\documentclass[twocolumn]{aastex63}

\usepackage[T1]{fontenc}
\usepackage{lmodern}
\usepackage{microtype} 
\usepackage{amsmath}
\usepackage{graphicx}           
\usepackage{latexsym}           
\usepackage{bm}                 
\usepackage[english]{babel}
\usepackage{color}              
\usepackage{scrextend}
\deffootnote{4pt}{0em}{\textsuperscript{\thefootnotemark}\,}  

\newcommand{\code}[1]{\texttt{#1}}
\newcommand{\drizzlepac}{\code{DrizzlePac}}%
\newcommand{\astrodrizzle}{\code{AstroDrizzle}}%
\newcommand{\tweakreg}{\code{TweakReg}}%
\newcommand{\photutils}{\code{Photutils}}%
\newcommand{\sextractor}{\code{SourceExtractor}}%
\newcommand{\EAZY}{\code{EAZY}}%
\newcommand{\CIGALE}{\code{CIGALE}}%
\newcommand{\IDL}{\code{IDL}}%

\newcommand{\teal}[1]{{#1}}
\newcommand{\tealnb}[1]{{#1}}
\newcommand{\mAB}{\ensuremath{m_{\rm AB}}}%
\newcommand{\mTF}{\ensuremath{m_{\rm 0.24}}}
\newcommand{\Dmagerr}{\ensuremath{\sigma_{\Delta m}}}

\definecolor{grey}{rgb}{0.5,0.5,0.5}%
\newcommand{\gsim}{\ensuremath{\gtrsim}}%
\newcommand{\lsim}{\ensuremath{\lesssim}}%

\newcommand{\eg}        {e.g.,}%
\newcommand{\ie}        {i.e.,}%
\newcommand{\etal}      {\hbox{et~al.}}%

\newcommand{\HST}       {HST}%
\newcommand{\JWST}      {JWST}%
\newcommand{\Gaia}      {Gaia}%
\newcommand{\Chandra}   {Chandra}%
\newcommand{\NuSTAR}    {NuSTAR}%

\newcommand{\tsim}      {\ensuremath{\sim}}%
\newcommand{\ttimes}    {\ensuremath{\times}}%
\newcommand{\tpm}       {\ensuremath{\pm}}%
\newcommand{\tdeg}      {\ensuremath{^{\circ}}}%
\newlength{\txw}
\newlength{\txh}

{\relax}%
{\relax}%

\received{January 9th, 2024}
\revised{March 11th, 2024}
\accepted{March 22nd, 2024}
\submitjournal{ApJS}

\graphicspath{{./}{figures/}}

\begin{document}

\title{TREASUREHUNT: Transients and Variability Discovered with \HST\ in the \JWST\ North Ecliptic Pole Time Domain Field}

\author[0000-0003-3351-0878]{Rosalia O'Brien}
\affiliation{School of Earth and Space Exploration, Arizona State University, Tempe, AZ 85287-1404, USA}

\author[0000-0003-1268-5230]{Rolf A.~Jansen}
\affiliation{School of Earth and Space Exploration, Arizona State University, Tempe, AZ 85287-1404, USA}

\author[0000-0001-9440-8872]{Norman A.~Grogin}
\affiliation{Space Telescope Science Institute, 3700 San Martin Drive, Baltimore, MD 21218, USA}

\author[0000-0003-3329-1337]{Seth H.~Cohen}
\affiliation{School of Earth and Space Exploration, Arizona State University, Tempe, AZ 85287-1404, USA}

\author[0000-0002-0648-1699]{Brent M.~Smith}
\affiliation{School of Earth and Space Exploration, Arizona State University, Tempe, AZ 85287-1404, USA}

\author[0000-0001-6564-0517]{Ross M.~Silver}
\affiliation{Astrophysics Science Division, NASA Goddard Space Flight Center, Greenbelt, MD 20771, USA}

\author[0000-0002-2203-7889]{W.P.~Maksym, III}
\affiliation{Center for Astrophysics $|$ Harvard \& Smithsonian, Cambridge, MA 02138, USA}
\affiliation{NASA Marshall Space Flight Center, Huntsville, AL 35812, USA}

\author[0000-0001-8156-6281]{Rogier A.~Windhorst}
\affiliation{School of Earth and Space Exploration, Arizona State University, Tempe, AZ 85287-1404, USA}

\author[0000-0001-6650-2853]{Timothy Carleton}
\affiliation{School of Earth and Space Exploration, Arizona State University, Tempe, AZ 85287-1404, USA}

\author[0000-0002-6610-2048]{Anton M.~Koekemoer}
\affiliation{Space Telescope Science Institute, 3700 San Martin Drive, Baltimore, MD 21218, USA}

\author[0000-0001-6145-5090]{Nimish P.~Hathi}
\affiliation{Space Telescope Science Institute, 3700 San Martin Drive, Baltimore, MD 21218, USA}

\author[0000-0001-9262-9997]{Christopher N.A.~Willmer}
\affiliation{Department of Astronomy\,/\,Steward Observatory, University of Arizona, Tucson, AZ 85721, USA}

\author[0000-0003-1625-8009]{Brenda L.~Frye}   
\affiliation{Department of Astronomy\,/\,Steward Observatory, University of Arizona, Tucson, AZ 85721, USA}


\author[0000-0003-0321-1033]{M.~Alpaslan}
\affiliation{Center for Cosmology and Particle Physics, Department of Physics, New York University, NY 10003, USA}

\author[0000-0002-3993-0745]{M.L.N.~Ashby}
\affiliation{Center for Astrophysics $|$ Harvard \& Smithsonian, Cambridge, MA 02138, USA}

\author[0000-0003-4439-6003]{T.A.~Ashcraft}
\affiliation{School of Earth and Space Exploration, Arizona State University, Tempe, AZ 85287-1404, USA}
\affiliation{Department of Physics, Eckerd College, St. Petersburg, FL USA}

\author{S.~Bonoli}
\affiliation{Donostia International Physics Center, Paseo Manuel de Lardizabal 4, E-20118 Donostia-San Sebasti{\'{a}}n, Spain}
\affiliation{Ikerbasque, Basque Foundation for Science, E-48013 Bilbao, Spain}

\author{W.~Brisken}
\affiliation{National Radio Astronomy Observatory\,/\,Associated Universties, Inc., Charlottesville, VA 22903, USA}

\author[0000-0002-1697-186X]{N.~Cappelluti}
\affiliation{Physics Department, University of Miami, Miami, FL 33124, USA}

\author{F.~Civano}
\affiliation{Center for Astrophysics $|$ Harvard \& Smithsonian, Cambridge, MA 02138, USA}
\affiliation{Astrophysics Science Division, NASA Goddard Space Flight Center, Greenbelt, MD 20771, USA}

\author[0000-0003-1949-7638]{C.J.~Conselice}
\affiliation{Jodrell Bank Centre for Astrophysics, The University of Manchester, Manchester, M13\,9PL, U.K.}

\author[0000-0003-4236-9642]{V.S.~Dhillon}
\affiliation{Department of Physics \& Astronomy, The University of Sheffield, Sheffield, S3\,7RH, U.K.}
\affiliation{Instituto de Astrof\'{i}sica de Canarias, E-38205 La Laguna, Tenerife, Spain}

\author[0000-0001-9491-7327]{S.P.~Driver}
\affiliation{International Centre for Radio Astronomy Research, The University  of Western Australia, Crawley, WA\,6009, Australia}

\author[0000-0001-6889-8388]{K.J.~Duncan}
\affiliation{Institute for Astronomy, The University of Edinburgh, Edinburgh, EH9\,3HJ, U.K.}

\author{R.~Dupke}
\affiliation{Observat{\'o}rio Nacional, Rio de Janeiro, Brazil}
\affiliation{Eureka Scientific, Inc., Oakland, CA 94602, USA}

\author[0000-0001-5060-1398]{M.~Elvis}
\affiliation{Center for Astrophysics $|$ Harvard \& Smithsonian, Cambridge, MA 02138, USA}

\author[0000-0002-0670-0708]{G.G.~Fazio}
\affiliation{Center for Astrophysics $|$ Harvard \& Smithsonian, Cambridge, MA 02138, USA}

\author[0000-0001-8519-1130]{S.L.~Finkelstein}
\affiliation{Department of Astronomy, The University of Texas at Austin, Austin, TX 78712, USA}

\author[0000-0003-1436-7658]{H.B.~Gim}
\affiliation{Department of Physics, Montana State University, Bozeman, MT 59717, US}

\author[0000-0003-1880-3509]{A.~Griffiths}
\affiliation{School of Physics and Astronomy, The University of Nottingham, Nottingham, NG7\,2RD, U.K.}

\author[0000-0001-8751-3463]{H.B.~Hammel}
\affiliation{Association of Universities for Research in Astronomy, Washington, D.C. 20004, USA}

\author[0000-0003-4738-4251]{M.~Hyun}
\affiliation{Korea Astronomy and Space Science Institute, Yuseong-gu, Daejeon 34055, Republic of Korea}

\author[0000-0002-8537-6714]{M.~Im}
\affiliation{SNU Astronomy Research Center, Dept. of Physics \& Astronomy, Seoul 08826, Republic of Korea}

\author[0000-0003-4665-8521]{V.R.~Jones}
\affiliation{School of Earth and Space Exploration, Arizona State University, Tempe, AZ 85287-1404, USA}

\author[0000-0001-5120-0158]{D. Kim}
\affiliation{Department of Astronomy and Space Science, Chungnam National University, Yuseong-gu, Daejeon 34134, Republic of Korea}

\author{B.~Ladjelate}
\affiliation{Instituto de Radioastronomia Millim{\'e}trica, E\,18012 Granada, Spain}

\author[0000-0003-2366-8858]{R.L.~Larson}
\affiliation{Department of Astronomy, The University of Texas at Austin, Austin, TX 78712, USA}
\affiliation{School of Physics and Astronomy, Rochester Institute of Technology, Rochester, NY 14623, USA}

\author[0000-0002-9226-5350]{S.~Malhotra}
\affiliation{Astrophysics Science Division, NASA Goddard Space Flight Center, Greenbelt, MD 20771, USA}

\author[0000-0001-6434-7845]{M.A.~Marshall}
\affiliation{National Research Council of Canada, Herzberg Astronomy \&
Astrophysics Research Centre, Victoria, British Columbia V9E\,2E7, Canada}
\affiliation{ARC Centre of Excellence for All Sky Astrophysics in 3 Dimensions
(ASTRO 3D), Australia}

\author[0000-0001-7694-4129]{S.N.~Milam}
\affiliation{Astrochemistry Laboratory, NASA Goddard Space Flight Center, Greenbelt, MD 20771, USA}

\author[0000-0002-2361-7201]{J.D.R.~Pierel}
\affiliation{Space Telescope Science Institute, 3700 San Martin Drive, Baltimore, MD 21218, USA}

\author[0000-0002-1501-454X]{J.E.~Rhoads}
\affiliation{Astrophysics Science Division, NASA Goddard Space Flight Center, Greenbelt, MD 20771, USA}

\author[0000-0003-1947-687X]{S.A.~Rodney}
\affiliation{Department of Physics \& Astronomy, University of South Carolina, Columbia, SC 29208, USA}

\author[0000-0001-8887-2257]{H.J.A.~R{\"o}ttgering}
\affiliation{Leiden Observatory, Leiden University, P.O.~Box 9513, 2300 RA Leiden, The Netherlands}

\author[0000-0003-3527-1428]{M.J.~Rutkowski}
\affiliation{Dept. of Physics and Astronomy, Minnesota State University-Mankato, Mankato, MN 56001, USA}

\author[0000-0003-0894-1588]{R.E.~Ryan,~Jr.}
\affiliation{Space Telescope Science Institute, 3700 San Martin Drive, Baltimore, MD 21218, USA}

\author[0000-0003-1810-0889]{M.J.~Ward}
\affiliation{Centre for Extragalactic Astronomy, Dept. of Physics, Durham University, Durham DH1\,3LE, U.K.}

\author[0000-0002-0486-0222]{C.W.~White}
\affiliation{School of Earth and Space Exploration, Arizona State University, Tempe, AZ 85287-1404, USA}

\author[0000-0002-0587-1660]{R.J.~van\,Weeren}
\affiliation{Leiden Observatory, Leiden University, P.O.~Box 9513, 2300 RA Leiden, The Netherlands}

\author[0000-0002-7791-3671]{X.~Zhao}
\affiliation{Center for Astrophysics $|$ Harvard \& Smithsonian, Cambridge, MA 02138, USA}


\author[0000-0002-7265-7920]{J.~Summers}
\affiliation{School of Earth and Space Exploration, Arizona State University,
Tempe, AZ 85287-1404, USA}

\author[0000-0002-9816-1931]{J.C.J.~D'Silva}
\affiliation{International Centre for Radio Astronomy Research, The University  of Western Australia, Crawley, WA\,6009, Australia}
\affiliation{ARC Centre of Excellence for All Sky Astrophysics in 3 Dimensions
(ASTRO 3D), Australia}

\author[0000-0002-6150-833X]{R.~Ortiz.~III}
\affiliation{School of Earth and Space Exploration, Arizona State University, Tempe, AZ 85287-1404, USA}

\author[0000-0003-0429-3579]{A.S.G.~Robotham}
\affiliation{International Centre for Radio Astronomy Research, The University  of Western Australia, Crawley, WA\,6009, Australia}

\author[0000-0001-7410-7669]{D.~Coe}
\affiliation{Space Telescope Science Institute, 3700 San Martin Drive, Baltimore, MD 21218, USA}
\affiliation{Association of Universities for Research in Astronomy (AURA) for the European Space Agency (ESA), STScI, Baltimore, MD 21218, USA}
\affiliation{Center for Astrophysical Sciences, Department of Physics and Astronomy, The Johns Hopkins University, Baltimore, MD 21218, USA}

\author[0000-0001-6342-9662]{M.~Nonino}
\affiliation{INAF-Osservatorio Astronomico di Trieste, Via Bazzoni 2, 34124 Trieste, Italy}

\author[0000-0003-3382-5941]{N.~Pirzkal}
\affiliation{Space Telescope Science Institute, 3700 San Martin Drive, Baltimore, MD 21218, USA}

\author[0000-0001-7592-7714]{H.~Yan}
\affiliation{Department of Physics and Astronomy, University of Missouri,
Columbia, MO 65211, USA}


\author{T.~Acharya}
\affiliation{School of Earth and Space Exploration, Arizona State University, Tempe, AZ 85287-1404, USA}


\correspondingauthor{Rosalia O'Brien, Rolf A.~Jansen}
\email{robrien5@asu.edu, rolf.jansen@asu.edu}


\shortauthors{O'Brien, R., \etal}
\shorttitle{Transients and Variability in the \JWST\ NEP Time Domain Field}

\begin{abstract}
The \JWST\ North Ecliptic Pole (NEP) Time Domain Field (TDF) is a $>$14\arcmin\ diameter field optimized for multi-wavelength time-domain science with {\JWST}. It has been observed across the electromagnetic spectrum both from the ground and from space, including with the Hubble Space Telescope (\HST).
As part of HST observations over 3 cycles (the ``TREASUREHUNT'' program), deep images were obtained with ACS/WFC in F435W and F606W that cover almost the entire \JWST\ NEP TDF. Many of the individual pointings of these programs partially overlap, allowing an initial assessment of the potential of this field for time-domain science with \HST\ and {\JWST}. The cumulative area of overlapping pointings is \tsim88 arcmin$^2$, with time intervals between individual epochs that range between 1 day and 4$+$ years. \teal{To a depth of \mAB\ $\simeq$ 29.5 mag (F606W), we present the discovery of 12 transients and 190 variable candidates. For the variable candidates, we demonstrate that Gaussian statistics are applicable, and estimate that \tsim80 are false positives}.
The majority of the transients will be supernovae, although at least two are likely quasars. Most variable candidates are AGN, where we find \teal{0.42\%} of the general $z$ \lsim\ 6 field galaxy population to vary at the $\tsim3\sigma$ level. Based on a 5-year timeframe, this translates into a random supernova areal density of up to \tsim0.07 transients per arcmin$^2$ (\tsim245 deg$^{-2}$) per epoch, and a variable AGN areal density of \teal{\tsim1.25} variables per arcmin$^2$ (\teal{\tsim4500} deg$^{-2}$) to these depths.
\end{abstract}



\section{Introduction} \label{sec:intro}

With the successful launch, commissioning, and first year of science observations, the James Webb Space Telescope (\JWST) has opened new opportunities for the study of the faint (potentially \mAB\,\gsim\,30 mag) variable Universe.
After discovering time-varying phenomena (i.e., transients and objects that vary in brightness or position), one would need the capacity to monitor these objects at their astrophysically relevant timescales, to determine the nature of their variability. For example, rapid follow-up of supernovae (SNe) is crucial to determine their types and distances. Flexible follow-up of active galactic nuclei (AGN) allows the timescale of AGN variability (which can range from several days to several years) to be measured, which can be linked to properties of the supermassive black hole (SMBH) at the center of its host galaxy \citep[e.g.,][]{Vanden_Berk_2004}. However, transient follow-up and variability monitoring with \JWST\ become difficult when Sun angle restrictions, power generation, and micrometeoroid mitigation limit when any particular area of the sky can be observed. The Continuous Viewing Zones (CVZs) centered on the ecliptic poles are the only locations where \JWST\ can observe at any time of the year and avoid these constraints.

\citet{jansen_2018} selected a particularly suitable area within \JWST's northern CVZ, the \JWST\ North Ecliptic Pole (NEP) Time Domain Field (TDF), to enable the exploration of time-varying phenomena with \JWST\ at high redshifts, but also within the halo of our own Galaxy, and in the extreme outer Solar System at high ecliptic latitudes.
The \JWST\ NEP TDF is a $>$14\arcmin\ diameter field centered on (RA, Dec)$_{\text{J2000}}$ = (17:22:47.896, +65:49:21.54), that was carefully chosen to minimize foregrounds, to be devoid of stars brighter than $m_K$ \tsim\ 15.5 mag, and maximize the observing efficiency for time-domain science with {\JWST}.
Its location near the NEP allows for follow-up of transients and variable sources at any cadence and at any time of the year. In addition, the NEP suffers minimal Zodiacal foregrounds, with minimal variation in the course of a year, naturally allowing for more sensitive observations per unit time.
While the ecliptic latitude of the \JWST\ NEP TDF is very high ($b_{\text{ecl}}$ $\simeq$ 86.2\tdeg), it is located at intermediate Galactic latitude ($b^{\rm II}$ $\simeq$ 33.6\tdeg), providing a clear but relatively long sight-line through the halo of our Galaxy. This makes the field useful for deep Galactic time-domain science with
\JWST\ as well.

This field serves as an important testbed for advancing our understanding of the time-varying universe at fainter limits than can be reached from the ground at optical and near-IR wavelengths, where transients and variability can hold cosmological significance (\eg\ dark energy, cosmic star formation rate, and the evolution of supermassive blackholes).
Type\,Ia SNe are often used as standard candles to provide critical constraints on the Hubble constant, the mass density of the universe, the cosmological constant, the deceleration parameter, and the age of the Universe \citep[e.g.,][]{riess_1998}.
Core Collapse (CC) SNe are equally important, as they not only provide most of the heavy elements in our universe \citep[e.g.,][]{matteucci_1986}, but also should trace the rate of instantaneous massive star formation. CC SNe rates and their connection to cosmic star formation rates remains poorly understood, as there seems to be a significant mismatch between observed CC SNe rates and what is expected \citep[see, e.g.,][]{cappellaro_2014}.

In addition to studying transient events like SNe, we can uncover new insights into the nature of astrophysical processes that lead to AGN variability. AGN are powered by matter accreting onto a SMBH \citep{Lynden-Bell_1969}. All massive galaxies host central SMBHs \citep[e.g.,][]{Magorrian_1998, Kormendy_2013}, and irregular or varying rates of infall, as well as turbulence and temperature fluctuations, can cause variations in brightness over time. The study of these variable AGN can provide a better understanding of the complex, unstable processes occurring around supermassive black holes \citep[e.g.,][]{Shakura_1976, Ulrich_1997, Kawaguchi_1999}. As AGN are linked to the evolution of their host galaxy \citep[e.g.,][]{Gebhardt_2000, Ferrarese_2000, Kormendy_2013}, this would also allow a better understanding of the co-evolution of galaxies and their central black holes.

AGN are often identified via X-ray or mid-IR emission, colors, or spectroscopic signatures, however, these methods can miss AGN that are intrinsically faint or lack X-ray emission. Variability in particular provides a unique way to identify AGN that might be missed via these other methods \citep[e.g.,][]{boutsia_2009, Pouliasis_2019}. \teal{\citet{Lyu_2022} explore various AGN selection methods in GOODS-S, including those based on X-ray emission, UV to mid-IR spectral energy distributions (SEDs), mid-IR colors, radio emission, and variability, and find \tsim10\% of the AGN in this field to exhibit optical variability\footnote{The optical variable sample is selected from \citet{Pouliasis_2019} (10 year baselines, and 5 epochs per object).}. We refer to \citet{Lyu_2022} and \citet{Lyu_2023} for a broader treatment of the role of various AGN selection methods.}

Variability proves especially promising for identifying faint AGN, as there appears to exist an anti-correlation between AGN brightness and variability amplitude \citep{Hook_1994, Trevese_1994, Cristiani_1997, Giveon_1999, Vanden_Berk_2004, Wilhite_2008, Zuo_2012}. In addition, the amplitude or timescale of AGN variability is correlated accretion disk size, and therefore black hole mass \citep[e.g.,][]{Xie_2005, Wold_2007, Li_2008, Wilhite_2008, Burke_2021} and redshift \citep{Cristiani_1990, Hook_1994, Trevese_1994, Vanden_Berk_2004}. A variability survey can thus provide a more complete census of the AGN population and its evolution over cosmic time. In addition, AGN identified through variability may be spectroscopically confirmed at any later time.

\teal{A large number of studies focusing on AGN variability have been published in the last few years \citep[for a recent review, see][]{Lyu_2022_review}.} \teal{At optical wavelengths, AGN have been identified} via their variability on scales from weeks to years \citep{cohen_2006, Morokuma_2008, villiforth_2010, sarajedini_2003, sarajedini_2011}.


As part of Hubble Space Telescope (HST) programs GO\,15278 (PI: R.~Jansen) and GO\,16252/16793 (TREASUREHUNT; PIs: R.~Jansen \& N.~Grogin) UV--Visible imaging of the JWST NEP TDF was secured with the F275W (0.272\,\micron), F435W (0.433\,\micron), and F606W (0.592\,\micron) filters. Areas of partial overlap between individual visits of the observations allow a first look at 2- to 4-epoch object variability and transients in this field.
For a detailed description of the observational design of these HST surveys, data reduction, stacked images, and source photometry, we refer the reader to Jansen et~al.\ (2024a,b; in prep.).

While previous HST variability studies like \citet{sarajedini_2003} (Hubble Deep Field North) and \citet{cohen_2006} (Hubble Ultra Deep Field) achieved impressive depths (29.0 and 30.5 mag, respectively), they are limited to approximately \tsim6.5 arcmin$^2$. \citet{villiforth_2010}, on the other hand, leveraged the expansive area (\gsim500 arcmin$^2$) of the Great Observatories Origins Deep Survey (GOODS), but to a relatively shallow depth (\tsim26.0 mag). At the same exquisite angular resolution (full width half max \lsim\ 0\farcs09), TREASUREHUNT observations provide a unique combination of depth (\mAB\ \lsim\ 29.5 mag) and area (\tsim88 arcmin$^2$) for a variability analysis.

In this work, we provide the first time-domain study of the JWST NEP TDF at visible wavelengths, presenting 12 transients (supernova candidates) and \tsim100 variable candidates (AGN candidates) identified using TREASUREHUNT data. In \S~\ref{sec:data}, we briefly summarize the TREASUREHUNT HST data, and data processing specific to the detection of transients and variable objects. \S~\ref{sec:methods} explains our methods for identifying transients and variability in the NEP TDF. In \S~\ref{sec:transient_results} and \S~\ref{sec:variability_results} we showcase our supernova candidate detections and our variable AGN candidates, respectively. All magnitudes are in AB units\footnote{Defined as $\mAB = -2.5 \log(F_{\nu}) + 8.90$ mag, where the flux density $F_{\nu}$ is in Jy.} \citep{Oke_1983}. Where relevant, we adopt a flat $\Lambda$CDM cosmology with $H_0 = 68$ km s$^{-1}$ Mpc$^{-1}$, $\Omega_m = 0.32$, and $\Omega_{\Lambda} = 0.68$ \citep{PlanckCollaboration2016, PlanckCollaboration2018}.

\begin{figure*}[ht]
    \centering
    \includegraphics[width=\textwidth]{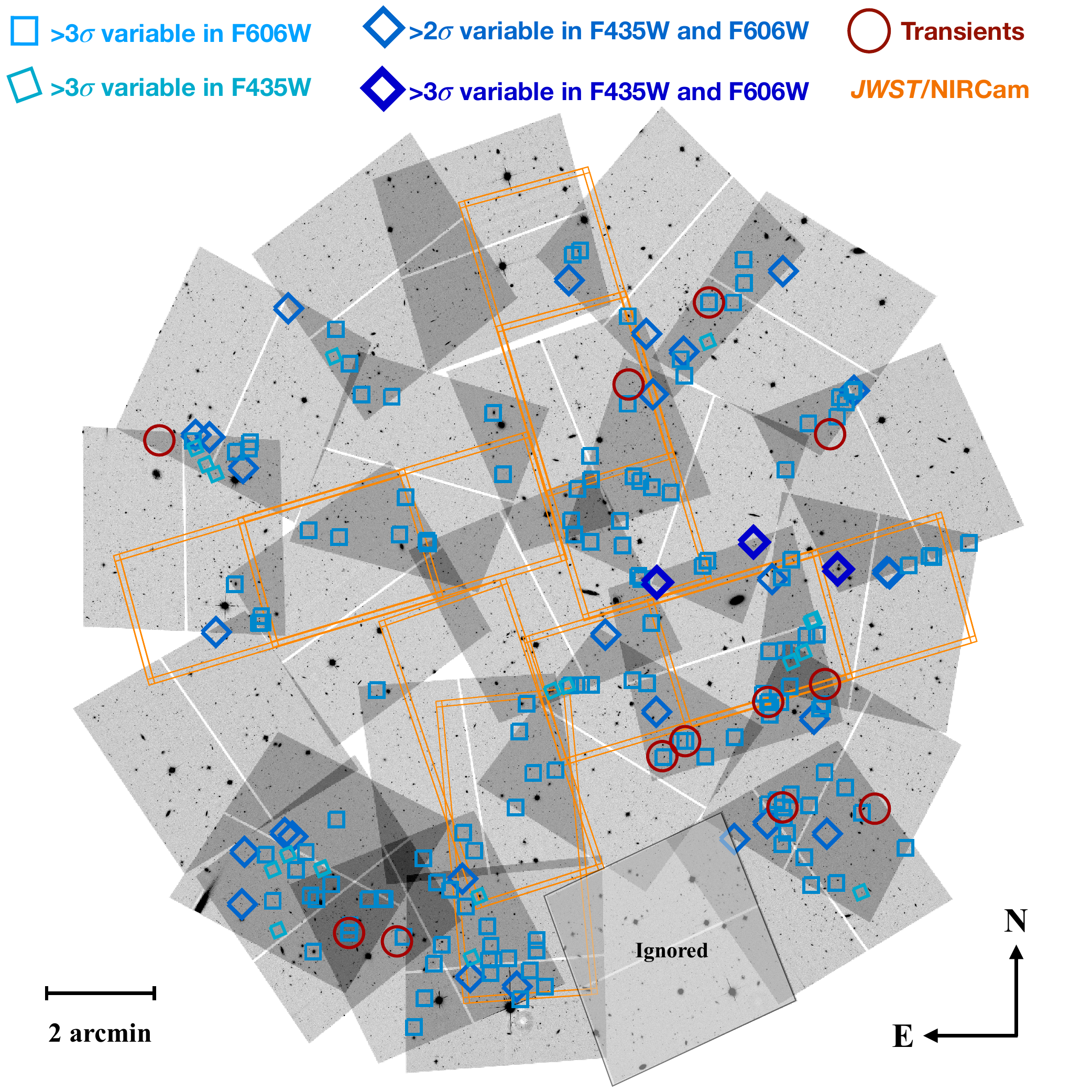}
    \caption{Full extent of \HST\ ACS/WFC coverage of the JWST NEP TDF. Regions of overlap between visits in the mosaic are highlighted in progressively darker shades of grey for areas of 2-, 3-, and 4-epoch overlap. This paper focuses on these regions of overlap, with a total area of 87.94 arcmin$^2$, to investigate transients and variable sources. Red circles mark the positions of the 12 identified transients, while blue squares and diamonds mark the 190 unique variable candidates discovered in this study. We expect \tsim80 of these sources to be false positives (see \S~\ref{sec:variable_methods}). Symbol shapes and colors identify the samples from which individual variables were selected, as indicated by the legend at the top of the figure. For comparison, the JWST/NIRCam coverage of this field is outlined in orange. We estimate photometric redshifts (\S~\ref{sec:redshift_methods}) for objects with both HST and JWST photometry. One visit (\texttt{jeex02}; at bottom right) with a bad astrometric solution was ignored in this work.}
    \label{fig:f606w_mosaic}
\end{figure*}

\section{Data and Processing} \label{sec:data}

\subsection{HST Observations and Data Processing}

Observations from \HST\ GO\,15278 were taken between 2017 October 1 and 2019 February 9, and those from GO\,16252+16793 (TREASUREHUNT\footnote{We will loosely refer to the observations and data of the GO\,15278 program as also part of TREASUREHUNT hereafter, as the intent and survey strategy were identical.}) between 2020 September 25 and 2022 October 31. The former program targeted the central ($r$ $\lsim$ 5\arcmin) portion of the JWST NEP TDF, and the latter an outer annulus to $r$ $\sim$ 7$\farcm$8. Together, these programs provide near-contiguous coverage with the Wide Field Camera~3 (WFC3/UVIS) in the F275W filter ($\lambda_c$ \tsim\ 0.272 \micron) and with the Advanced Camera for Surveys (ACS/WFC) in the F435W and F606W filters ($\lambda_c$ \tsim\ 0.433 and \tsim0.592\,\micron), with a total area of \tsim194 arcmin$^2$ (ACS/WFC). The full \HST\ coverage of the JWST NEP TDF is shown in Figure~\ref{fig:f606w_mosaic}.

As described in more detail in Jansen et~al.\ (2024a,b; in prep.), both programs used 4-orbit CVZ and pseudo-CVZ\footnote{True CVZ orbits imply Earth Limb angles \gsim15.5\tdeg, but uninterrupted observations can be implemented at smaller angles down to \tsim7.6\tdeg\ away from the Earth limb when phased to execute when that limb is dark.  We term these possible but non-standard opportunities ``pseudo-CVZ''.} visits to reach 2-$\sigma$ limiting depths of \mAB\ $\simeq$ 28.0, 28.6, and 29.5 mag in F275W, F435W and F606W, respectively. These limiting depths were determined using the \sextractor\ \citep{bertin_1996} \texttt{MAG\_AUTO} aperture, within areas with single-epoch coverage (lighter grey in Figure~\ref{fig:f606w_mosaic}), and are the adopted nominal survey depths. 

The TREASUREHUNT data were retrieved from MAST and post-processed (after the standard ACS and WFC3 calibration pipeline steps of flatfielding and correction for Charge Transfer Efficiency (CTE) effects) to mask satellite trails, remove residual CTE trails, and fit and subtract the background (Jansen et~al.\ 2024a,b; in prep.).  The latter was needed because the observing strategy deliberately used the full duration of CVZ orbits, as well as pseudo-CVZ orbits where \HST\ was allowed to dip closer to the Earth limb than the nominal limit for CVZ orbits. This resulted in excess background light that helped fill charge traps in the F275W exposures with their exceedingly dark natural sky background \citep[e.g.,][]{OBrien_2023}, but also rendered the observed background level meaningless.

The background-subtracted images were corrected for geometric distortion and aligned to the \Gaia/DR3 \citep{GaiaMission, GaiaDR3} astrometric reference grid using the \drizzlepac{} tasks \tweakreg{} and \astrodrizzle{} \citep[see][]{hoffmann_2021}. \teal{The \astrodrizzle{} software also flags cosmic-rays, a crucial step when identifying faint variability, where cosmic ray induced signals could cause spurious variability if not properly removed. We expect, and tested (as discussed in more detail in the Appendix), that all cosmic ray and other image artifacts were largely removed through the \astrodrizzle{} step, as each single 4-orbit visit resulted in 8--9 exposures per filter that were stacked together.}

All 8--9 exposures per visit were taken at the same pointing with small $\le$0.2\arcsec\ dither offsets in RA and Dec. For each visit, we drizzle-combined the individual post-processed, rectified and aligned exposures per filter (Jansen et~al.\ 2024a,b; in prep.). For ACS/WFC F606W and F435W, this resulted in 24 rectified and aligned science images (and 24 associated weight maps), and for WFC3/UVIS F275W in 23 images and associated weight maps, each at a pixel scale of 0\farcs030\,pix$^{-1}$ and centered on the nominal field center of the JWST NEP TDF.
We also use the full mosaics of all drizzle-combined images per filter, generated with the same image dimensions and pixel scale, and aligned to \Gaia/DR3 to within 0\farcs008 in RA and Dec.

The footprints of individual ACS/WFC (and to a lesser extent WFC3/UVIS) visits partially overlap one another, with a cumulative area of overlap of \tsim88 arcmin$^2$. These overlapping regions are key to searching for transients and variability within this field. Figure~\ref{fig:f606w_mosaic} showcases the overlapping regions in progressively darker shades of grey for areas of 2-, 3- and 4-epoch overlap. The time interval between observations in these regions of overlap range between 1 day and 4$+$ years.

We note that while the very small number of epochs and sparse time sampling can catch transients \teal{like SNe and variable sources like AGN,} they cannot provide a light curve to trace their evolution. \teal{Further epochs would provide a more robust identification of variable sources, where we may characterize the scatter in the measured brightness \citep[e.g.,][]{Pouliasis_2019, Sokolovsky_2017}}.

While our primary focus is the identification of time variable objects within the TREASUREHUNT HST data, we opt to incorporate ancillary data to better understand certain objects of interest. Specifically, we incorporate \JWST/NIRCam (0.8--5\,\micron), \NuSTAR\ and XMM-Newton, \Chandra, and MMT/Binospec data.

\subsection{JWST Observations and Data Processing}

\JWST/NIRCam observations within the JWST NEP TDF were obtained between 2022 August 25 and 2023 May 30 as part of \JWST\ GTO\,2738 (PIs: R.~Windhorst \& H.~Hammel) in 7 broadband filters (F090W through F444W) and 1 mediumband filter (F410M) as part of the PEARLS program \citep{windhorst_2023}. \teal{Figure \ref{fig:f606w_mosaic} shows the JWST observations outlined in orange, highlighting the 4 epochs, corresponding to the four orthogonal ``spokes''. Each spoke consists of two NIRCam pointings that partially overlap in the center. Each pointing includes a total exposure time of 3.5 hours across all 8 filters, or about 3000 sec per filter. Each spoke has an area of 13\farcm7, totaling 54\farcm7 of JWST observations.}

The data were retrieved from MAST and post-processed by the PEARLS team using their custom pipeline to mitigate $1/f$ noise, alleviate wisps in the NIRCam/SW filters, mask snowball artifacts, and flatten the background across read-out amplifier boundaries. The individual post-processed images where rectified and aligned to \Gaia/DR3 independent of the \HST\ images, but using nearly identical methods. \teal{Full mosaics of the field were created for each filter with a 0\farcs03 platescale. The 5 sigma point-source limit is typically between 28.0 and 29.1 mag, depending on the filter.} For more details on the reduction, calibration and post-processing, we refer to \citet{windhorst_2023}, \citet{Robotham_2023}, and Jansen et~al.\ 2024c (in prep.). 
In \S~\ref{sec:redshift_methods}, we will use PEARLS aperture photometry of transient and varying objects discovered in areas with both \HST\ and \JWST\ coverage to fit SEDs.

\subsection{X-Ray Observations and Data Processing}

The JWST NEP TDF has also been observed extensively in X-rays with \NuSTAR, XMM-Newton, and {\Chandra}. \NuSTAR\ has monitored the field since Cycle 5 (PID 5192, 6218, 8180, and 9267; PI: F.~Civano). For this work we use the Cycle 5+6 data and catalog published in \citet[]{zhao_2021cat, zhao_2021, Zhao_2024}. The 21 \NuSTAR\ observations took place over 28 months (between 2019 September and 2022 January), while the three XMM-Newton observations (part of the \NuSTAR\ cycle 6 program) were taken over 15 months (between 2020 October and 2022 January). The total \NuSTAR\ exposure from Cycle 5+6 is 1.5\,Ms, reaching a flux limit of 3.3\ttimes10$^{-14}$\,erg\,cm$^{-2}$\,s$^{-1}$ in the 3--24 keV band. We note that this is the deepest \NuSTAR\ field taken thus far. The total XMM-Newton exposure time is 62\,ks, reaching a limit of 8.7\ttimes10$^{-16}$\,erg\,cm$^{-2}$\,s$^{-1}$ in the 0.5--2 keV band.

We also use the first 1.3\,Ms of \Chandra\ observations of the JWST NEP TDF (PID 19900666, 20900658, 21900294 and 22900038; PI: W.P.~Maksym) spanning cycles 19--23. \Chandra\ observed the field 46 times between 2018 April 12 and 2022 September 25 (a span of 4.6 years), with exposures ranging from 9 to 59.9\,ks. Beginning 2021 August 29, these occurred in 90\,ks epochs every 3 months, typically broken into 2 or more individual visits (OBSIDs) spanning no more than 30 days per epoch.  We reprocessed and reduced the data using {\tt ciao} \citep{Fruscione_2006}, corrected the cross-observations using source catalogs generated by {\tt wavdetect}, and astrometrically registered them to Pan-STARRS \citep{Chambers_2016}.  We then used {\tt merge\_obs} to generate combined event files and exposure maps. \teal{The faintest 3$\sigma$ detection in the catalog is $1.35\times10^{-16}$ erg s$^{-1}$ cm$^{-2}$ in the full 0.5--7 keV band assuming galactic absorption and a $\Gamma=1.4$ power law.}

\subsection{Ground Based Spectroscopy}

MMT/Binospec \citep{Binospec} observations of the JWST NEP TDF in 2019 September (PI: C.~Willmer) used a 270 lines/mm grating and a 1\arcsec\ slit width to yield spectra spanning \tsim3900--9240\AA\ sampled at 1.30\AA/pix and with a resolution of \tsim4.9\AA\ (full width half max) at 6500{\AA}. The exposure time was 1 hr (4\ttimes900\,s), reaching $S/N$ \tsim\ 5 at \mAB\ \tsim\ 22.5 mag for quiescent galaxies, and a magnitude deeper for galaxies with significant emission lines.


\section{Methods to Identify Transients and Variability} \label{sec:methods}

In this section, we detail our methods for identifying transients and variability in the 55 areas of overlap (shown as darker shades of grey in Figure~\ref{fig:f606w_mosaic}) with a combined area of \tsim88 arcmin$^2$.

Transients are usually defined as objects that only appear for a short duration. For this work, where we expect most transients to be SNe, we consider relatively bright point sources that are present in one epoch, yet do not appear in another. We searched for transients using difference images, and considered any source detected therein a transient candidate.  In contrast, variable sources tend to show less extreme differences in brightness and are not as evident in difference images. Variable sources are therefore identified after measuring the source flux and its associated uncertainty within a small aperture. Consequently, some sources we identify as transients may be variable sources with extreme amplitudes of variability (e.g., quasars), and some variable sources may be dim transients or transients caught well away from peak brightness (e.g., SNe). Nonetheless, we assume most transients we discover will be SNe, while most variable sources will be weak AGN. However, the methods presented here may identify a variety of objects, including variable stars, tidal disruption events, gamma ray bursts, or quasars. We did not identify any objects that showed appreciable movement between visits. For all objects discovered here, further analysis is required to determine their true nature.

\subsection{Transient Search} \label{sec:transient_methods}

Transients were identified by visual inspection of difference images, each of which was created as follows. First, we bring the drizzled images of individual visits to the same pixel grid using their \Gaia-aligned WCS and the Astropy \citep{astropy:2013, astropy:2018, astropy:2022} affiliated package \texttt{reproject}. Then, we generate two distinct difference images for each overlapping region in the deeper F606W images, i.e., 110 difference images for the 55 overlapping areas.  For instance, if Visit 1 overlaps with Visit 2, we generate a difference image by subtracting Visit 2 from Visit 1, and another difference image by subtracting Visit 1 from Visit 2. In each  difference image we then visually identify sources with positive flux that are consistent in appearance with bearing the imprint of the point spread function (PSF). We did not use the F435W images for our transient search, because they are \tsim0.9 mag shallower.

\teal{Although rare}, bright cosmic ray (CR) induced signals can sometimes survive in final drizzled products and subsequently appear in difference images. We inspected the individual pipeline-calibrated, flat-fielded exposures at the corresponding position of each transient candidate to ensure they appear in each exposure, bear the imprint of the PSF, and are not (residuals of) CRs or other spurious noise. We originally found 20 transient candidates, of which 8 were confirmed to be due to CRs and were removed from further consideration.

We measured the apparent F606W magnitudes of each transient using the difference images.
To measure transient magnitudes, we centered a circular aperture with a radius of 8 pixels on the transient in the difference image, and measured its magnitude within that aperture. These apertures are \gsim5\ttimes\ the full width half max (FWHM) of the PSF and encompass the transient signal. We used the Python photometry package \photutils\ \citep{bradley_2022} and specified the exact center of each aperture. We created a difference error map by adding the weight (WHT) maps from the individual visits in quadrature, then taking ERR = $1/\sqrt{\text{WHT}}$. We direct this error map into \photutils\ to derive an uncertainty on the reported flux. Magnitude errors are then calculated as $\Delta m = (2.5/\ln{10}) \cdot \sigma_F/F$, with $F$ the measured flux, and $\sigma_F$ the uncertainty thereon, both as reported by \photutils.

For each transient, we also performed matched-aperture photometry on a subset of F435W difference images, generated for those regions of overlap that contained a transient.

\subsection{Variability Search} \label{sec:variable_methods}

To detect variability in galaxies, we first constructed source catalogs using \sextractor, with object positions and aperture photometry for each visit. For initial object detection, as summarized in Table~\ref{tab:sext_params}, we require an object to be detected at the 1.5$\sigma$ level above the local sky level (given a gain value corresponding to the median exposure time of the individual exposures in that visit and filter) after smoothing with a Gaussian kernel with a FWHM of 3 pixels, and require a minimum number of 5 contiguous pixels to minimize detection of spurious sources. For the same reason, we also generously masked the pixels along the detector borders, where fewer contributing exposures result in excess noise and imperfect rejection of CR signal.

 %
\begin{table}[t!]
\begin{center}
   \caption{Relevant \sextractor\ parameters for variability identification and aperture photometry.\label{tab:sext_params}}
   \vspace*{-6pt}
   \begin{tabular}{lll}
\hline\hline
Parameter                 & Variability Aperture      & Full Aperture       \\[4pt]
\hline
Aperture Size             & 0\farcs24                 & \texttt{FLUX\_AUTO} \\
\texttt{DETECT\_MINAREA}  & 5                         & 5                   \\
\texttt{DETECT\_THRESH}   & 1.5                       & 1.5                 \\
\texttt{ANALYSIS\_THRESH} & 1.5                       & 1.5                 \\
\texttt{DEBLEND\_MINCONT} & 0.001                     & 0.06                \\
\texttt{FILTER\_NAME}     & gauss\_3.0                & gauss\_3.0          \\
\texttt{GAIN}             & \multicolumn{2}{c}{\itshape median exposure time}        \\
\texttt{BACK\_SIZE}       & 128                       & 128                 \\
\texttt{BACK\_FILTERSIZE} & 3                         & 3                   \\
\texttt{BACKPHOTO\_THICK} & 50                        & 50                  \\
\hline
   \end{tabular}
\end{center}
\begin{minipage}{0.485\textwidth}{\small
\textbf{Note.} For each \sextractor\ parameter in the first column, the second an third columns list its value for identifying variable sources and for full aperture photometry, respectively. We used a circular aperture with a diameter of 0\farcs24 diameter for variability identification, specified using the \texttt{PHOT\_APERTURES} parameter, and the \texttt{FLUX\_AUTO} aperture for full aperture photometry. We set the \texttt{GAIN} value to the median exposure time of the individual exposures combined per visit.}
\end{minipage}
\end{table}

We compared brightness measurements across distinct visits to identify potential variables. We specifically focus on detecting variability in the F435W and F606W filter observations, as the F275W filter magnitude limit is substantially brighter, the total area of overlap between individual F275W visits is substantially smaller, and the WFC3 F275W observations are not contemporaneous with the ACS/WFC F435W and F606W ones.

\subsubsection{Photometry} \label{sec:photometry}

\begin{figure}[ht]
    \centering
    \includegraphics[scale = 0.375]{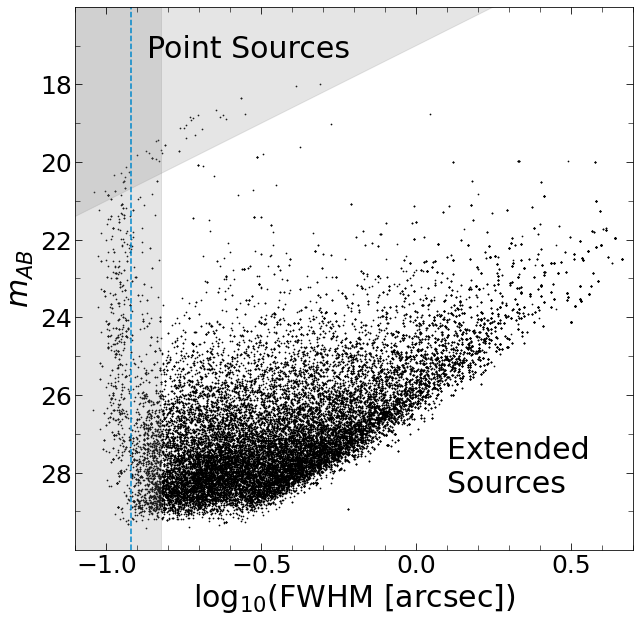}
    \caption{Star-galaxy separation, following \citet{windhorst_2011}, using the \sextractor\ \texttt{MAG\_AUTO} magnitudes in F606W versus full width at half maximum (FWHM). Objects within grey regions, with FWHM $<$ 0\farcs15 or \mAB\ $<$ 17 $-$ 4\,$\log_{10}(\textrm{FWHM})$, are flagged as stars and are not considered in this study. We also show the radius (0\farcs12) of the aperture used to identify variability as a vertical blue dashed line.}
    \label{fig:elephants_trunk}
\end{figure}

Our goal is to leverage HST's high resolution to isolate the potentially varying nuclei within galaxies and SNe near their core while ignoring the non-varying extended component. To achieve this, we first set the minimum contrast for deblending in \sextractor\ (\texttt{DEBLEND\_MINCONT}) to 0.001. This value is lower than is typical, but allows isolating the nuclei of galaxies \textit{and} potentially locate variability in bright regions that may not be in the very center of a galaxy. In addition to a low deblending threshold, we opted to use apertures close to the FWHM of stellar objects. For our resolution of \lsim0\farcs09, we used a circular aperture with a diameter of 0\farcs24 (8 pixels). For apertures much larger than this the surrounding galaxy tends to dilute any variability of a compact central source. We henceforth refer to this small aperture as the 0\farcs24 aperture, and to the corresponding magnitudes as \mTF. The radius of this aperture is shown as the dashed blue vertical line in Figure~\ref{fig:elephants_trunk} to facilitate comparison with detected stellar objects.

With \texttt{DEBLEND\_MINCONT} = 0.001, the measured variability sometimes appears offset from the centers of galaxies. This could be due to off-center SNe or SMBHs that have yet to settle onto the centers of their host galaxies. To measure host galaxy properties for these offset events, we also ran \sextractor\ with \texttt{DEBLEND\_MINCONT} = 0.06, which reduces deblending of galaxies into individual clumps. We thus have two types of aperture photometry for each source: 1) within the small circular 0\farcs24 diameter aperture used to identify variability, and 2) for the full Kron aperture, as identified by \sextractor. Table~\ref{tab:sext_params} summarizes the relevant \sextractor\ parameters used for these two types of measurements. For our matched aperture F435W photometry, we used \sextractor\ in dual image mode, employing the full F606W mosaic as the detection image.

To identify variability, we need to first carefully match the objects from one visit catalog to another. We did so using the Astropy \texttt{match\_coordinates\_sky} function, which finds the nearest object in the catalog from one visit to a given object in a catalog from another. To ensure slight differences in position do not contribute to the measured variability, we only considered two sources to be a match across two different visits if they are within one pixel of each other (0\farcs03). We increased this threshold to 0\farcs08 when matching the F606W catalog to the F435W catalog, to account for the fact that objects may have different PSF and surface brightness distributions at different wavelengths. Matches between the \texttt{DEBLEND\_MINCONT} = 0.001 and \texttt{DEBLEND\_MINCONT} = 0.06 catalogs were identified using a search radius of 3 arcsec, since the highly deblended catalog may find variability in the outskirts of host galaxies, and thus the positions may be offset by much more than our astrometric uncertainties and pixel size.

We opted to ignore stars in our variability analysis due to uncertainties in the ACS PSF \citep[e.g.,][]{villiforth_2010}. We flagged as stars all objects with a FWHMt less than 0\farcs15 or \mAB\ $<$ 17 $-$ 4\,$\log_{10}(\textrm{FWHM})$ mag. Grey shading in Figure~\ref{fig:elephants_trunk} indicates the stars and faint point-like sources that were removed from our analysis. In addition, we ignored objects with a \sextractor\ \texttt{FLAGS}\footnote{\url{https://sextractor.readthedocs.io/en/latest/Flagging.html}} $>$ 2, including objects with saturated pixels, truncated footprints, corrupted apertures/footprints, or other issues. Last, we also ignored all objects within the \texttt{jeex02} visit in the F606W filter, as it has a poor astrometric solution that hampers reliable alignment and variability measurements (Jansen et~al.\ 2024a,b; in prep.).

\teal{Some variable candidates, like variable stars and ultra-bright AGN like quasars, will be point-like and excluded by our methods. Missing quasars in particular may bias our final AGN estimates. Nonetheless, these objects are relatively rare. In addition, our methods would miss any AGN that does not vary on the timescales we probe. Variability identified in this paper is therefore the \emph{minimum} amount of variability expected in this field.}

\subsubsection{Variable Source Selection} \label{sec:source_selection}

Sources were identified as variable if they varied by more than 3$\times$ their combined photometric uncertainty. Therefore, careful consideration of uncertainties is crucial to ensure variable sources are robustly identified.
Combining and resampling multiple lower-resolution images into a higher-resolution composite image (using, e.g., the drizzle algorithm) causes both the signal and the noise in adjacent pixels to become correlated. Photometric uncertainties are known to be underestimated when correlated pixel noise is not taken into account \citep[e.g.,][]{casertano_2000, labbe_2003, blanc_2008, whitaker_2011, papovich_2016}. In addition, the ACS PSF is known to vary due to telescope breathing \citep[e.g.,][]{villiforth_2010}, which could potentially contaminate variability identification performed within small apertures. \teal{Finally, cosmic rays may cause changes in brightness if not filtered out properly.} We employed an empirical approach to statistically account for correlated pixel noise and PSF changes, as well as other sources of error, regardless of their origin.

\begin{figure*}[ht]
    \centering
    \includegraphics[width=\textwidth]{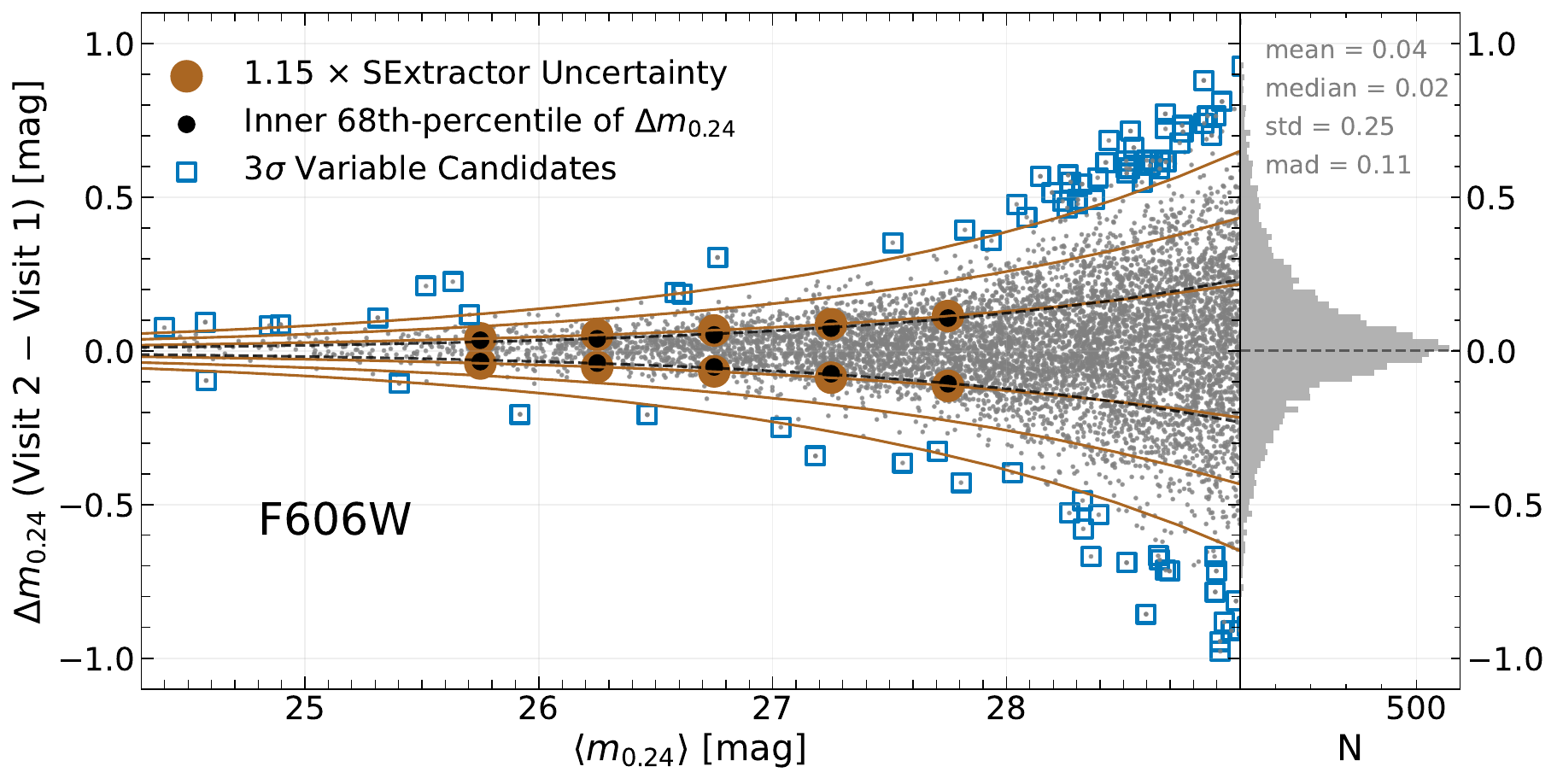}
    \caption{Calibration of photometric uncertainties to identify variable sources. We plot the difference in measured F606W magnitude, $\Delta \mTF$, versus the average measured F606W magnitude, $\mTF$, for objects observed more than once in regions of overlapping HST coverage, where $\Delta \mTF$ is defined in the sense Visit\,2 $-$ Visit\,1. Each small grey point is an individual object, as detected and measured with \sextractor. The large brown circles depict \sextractor\ photometric errors, scaled  by a factor 1.15 to match the observed distribution of magnitude differences (shown as small black circles) measured in 0.5 mag wide bins for 25.5 $<$ \mTF\ $\le$ 28.0 mag such that \teal{\tsim68.3\% of all data points fall between them.} The dashed black curves represent the inner \teal{\tsim68.3\%} range of the distribution at each magnitude and are polynomial fits to the black points. The brown curves therefore represent \teal{\textit{median}} 1, 2 and 3$\sigma$ ranges as fit to the brown circles. Blue squares identify objects that vary at \teal{$\ge$3$\Dmagerr$} in F606W, \teal{where $\Dmagerr$ is defined in Equation \ref{eq:magerr}}. The histogram on the right-hand side illustrates the distribution of magnitude differences for all objects detected in the field. In grey, we list mean, median, standard deviation, and median absolute deviation of the y-axis distribution.}
    \label{fig:variable_candidates}
\end{figure*}

Figure~\ref{fig:variable_candidates} highlights our method to best calibrate photometric uncertainties, which is based on the variability search methods of \citet{cohen_2006}. Measured magnitudes were compared for all objects that appear in more than one visit. For each such object, Figure~\ref{fig:variable_candidates} shows the magnitude difference, $\Delta \mTF$, versus the mean magnitude, \mTF. We used the distribution of these magnitude differences to calibrate the empirical (true) photometric scatter, where the inner \teal{brown curves} contain \tsim68.26\% of the data points in each magnitude bin (shown as black circles and curves). To account for correlated pixel noise, a varying PSF, and other unaccounted-for sources of error, we multiplied all \sextractor\ magnitude errors by a fixed scale factor (1.15 for F606W and 1.35 for F435W) to best match the observed distribution of photometric scatter. The scaled result is represented by the large brown circles. Magnitude difference uncertainties \teal{(for F606W)} were calculated as:
\begin{equation}\label{eq:magerr}
    \Dmagerr = \tealnb{1.15 \times}(2.5\,N/\ln{10})\,\sqrt{(\sigma_1/F_1)^2 + (\sigma_2/F_2)^2},
\end{equation}

\noindent with $N$ the number of sigma, $F_1$ ($F_2$) the measured flux for Visit 1 (Visit 2), and $\sigma_1$ ($\sigma_2$) the corresponding scaled flux error. 

\teal{We emphasize that Figure~\ref{fig:variable_candidates} is used to calibrate photometric uncertainties and \textit{not} to identify variable candidates.} The blue squares represent objects flagged as variable at \teal{\teal{$\ge$3$\Dmagerr$}} in F606W. \teal{These objects are determined to be those which vary by more than 3$\times$ their individual photometric uncertainty. In general, these sources fall outside the brown 3$\sigma$ curves in Figure~\ref{fig:variable_candidates}, but this isn't always the case. Some objects outside the brown 3$\sigma$ curves are not flagged as variable if their individual photometric uncertainties are larger than the median. Similarly, some objects inside the brown 3$\sigma$ curves are  flagged as variable if their individual photometric uncertainties are smaller than the median.}

In general, we do not expect zeropoint offsets between visits to be a concern, as changes in zeropoint over the duration of the TREASUREHUNT program are $<$1\% for ACS/WFC \citep{bohlin_2020, OBrien_2023}. The histogram on the right side of Figure~\ref{fig:variable_candidates} is centered at $\Delta \mTF$ = 0.02 mag (with a standard deviation of 0.25 mag that is dominated by the photometric uncertainties of faint sources), and therefore shows that the zeropoints between detectors and different regions on the same detector as calibrated by the standard HST pipeline are correct to \lsim0.02 mag.

We only considered objects brighter than $\mTF \simeq 28.6$ mag in F435W or $\mTF \simeq 29.5$ mag in F606W  (corresponding to the predicted 2$\sigma$ magnitude limits for sources in the field) in at least one epoch.
We manually sorted through variable candidates, and removed 63 sources that are either contaminated by diffraction spikes from a bright star, or are close to the edge of a detector. In a vast majority of these cases a diffraction spike overlapped the object in one visit but not in another, observed at a different orientation.

We created a variability catalog for objects that vary more than \teal{$3\Dmagerr$} in F435W, and similarly a variability catalog for objects that vary more than \teal{$3\Dmagerr$} and \teal{$5\Dmagerr$} in F606W. In addition, we created catalogs with sources that vary (in the same direction) in \textit{both} filters with \teal{$2\Dmagerr$} and \teal{$3\Dmagerr$} significance. The significance of the catalogs for objects that vary in both filters to \teal{2$\Dmagerr$} is $\sqrt{2}\,(2\Dmagerr) \simeq 2.83\Dmagerr$, and to \teal{$3\Dmagerr$} is $\sqrt{2}\,(3\Dmagerr) \simeq 4.24\Dmagerr$. We consider the latter sample (variability detected in both filters to \teal{$3\Dmagerr$}) to be our most conservative, robust sample, as objects that vary in both filters are most likely to be true variables, although the shallower F435W images severely limit their number. 

\begin{figure}
    \centering
    \includegraphics[width=0.473\txw]{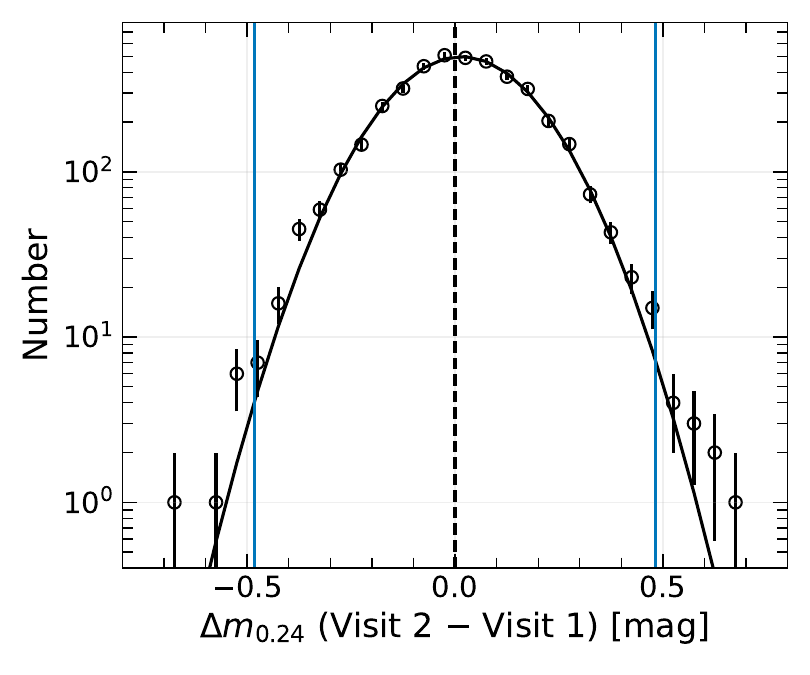}
    \caption{\teal{Histogram of the distribution of magnitude differences measured between two epochs ($\Delta\mTF$) for all galaxies (4,059 total) between $28.0 < \mAB < 28.5$ mag. The black circles show the number of galaxies ($N$) for each $\Delta\mTF$ bin, where the error bar is $\sqrt{N}$. The solid black curve is a Gaussian fit to the data. The vertical blue lines represent the 3$\sigma$ thresholds based on the Gaussian fit. This proves that the distribution of $\Delta\mTF$ follows a Gaussian distribution, such that the false detections can be assumed to follow Gaussian statistics (see Section \ref{sec:false_detections}).}}
    \label{fig:gaussian}
\end{figure}

\begin{figure}
    \centering
    \includegraphics[width=0.473\txw]{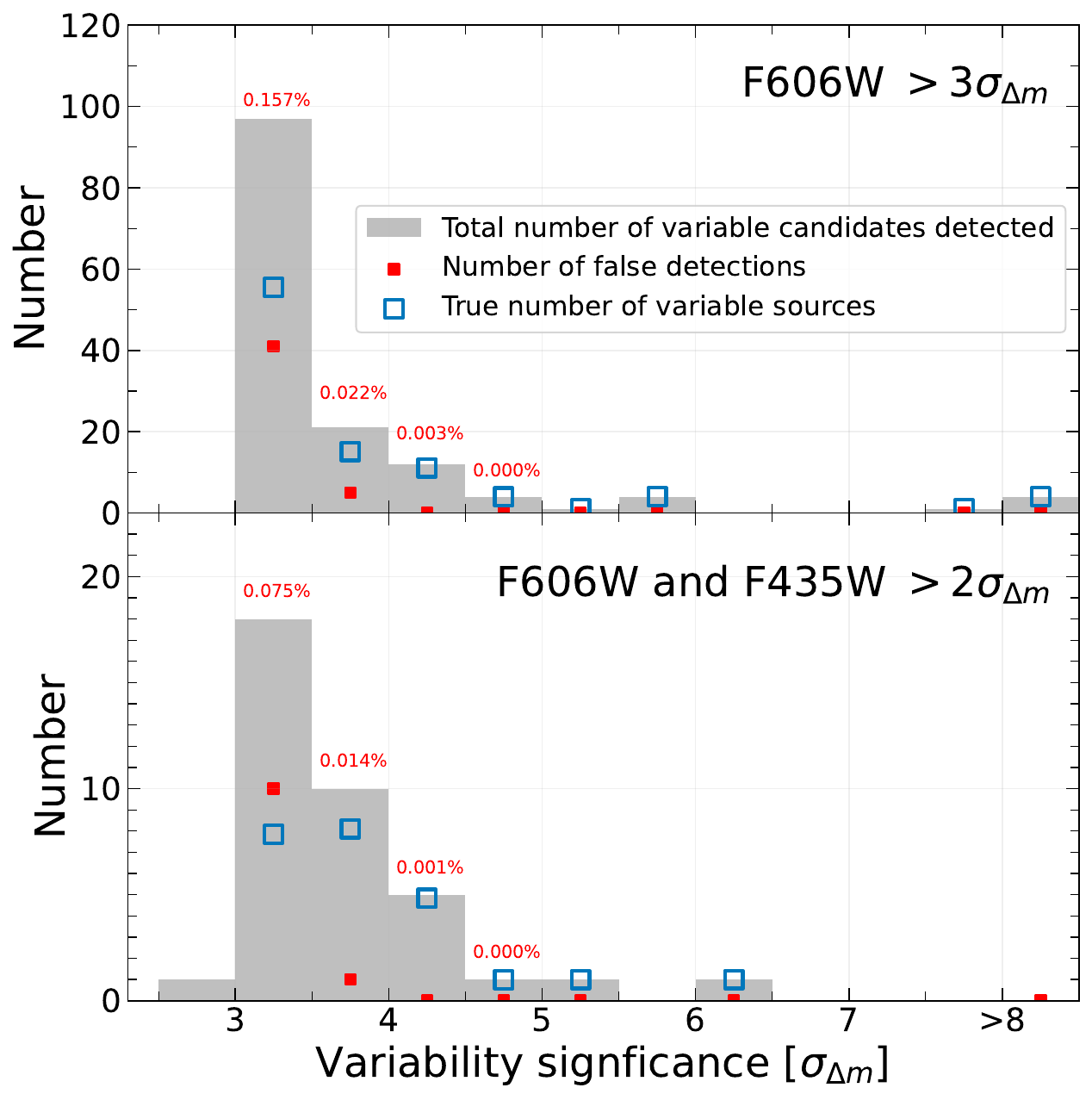}
    \caption{Histograms of the number of variable candidates (light grey) and estimated numbers of false positives (red) and genuine variable sources (blue) as a function of the significance (in units of \teal{$\Dmagerr$}) of the variability detected.  The number of genuine variables and false positives add up to the total number of candidates in each bin. The red labels above each bin show the percentage of all objects in the field that are expected to be false positives for that \teal{$\Dmagerr$-bin}. The top panel is for sources that varied by more than \teal{3$\Dmagerr$} in F606W, and the bottom panel for sources that varied by more than \teal{2$\Dmagerr$} in both F606W and F435W, and did so in the same sense.  Hence, assuming Gaussian statistics, the upper panel reflects a two-sided test, and the lower panel a one-sided test. \teal{In total, we estimate \tsim80 false positives in the sample of sources that varied by more than \teal{3$\Dmagerr$} in F606W, the sample of sources that varied by more than \teal{3$\Dmagerr$} in F435W, and the sample of sources that varied by more than \teal{2$\Dmagerr$} in both F606W and F435W.}}
    \label{fig:false_detections}
\end{figure}

\subsubsection{Accounting for False Detections} \label{sec:false_detections}

Since variable sources are identified using \teal{2$\Dmagerr$ or 3$\Dmagerr$} significance levels, we can expect false detections (e.g., noise peaks). \teal{We herein refer to the individual variable sources identified in this work as variable candidates, since we cannot determine which individual sources exhibit genuine variability and which are spurious detections.}

\teal{We predict the number of false detections using Gaussian statistics. As demonstrated in Figure \ref{fig:gaussian}, we observe a Gaussian distribution in $\Delta\mTF$ for all 4,059 galaxies with $28.0 < \mAB < 28.5$ mag. We confirm that brightness bins of $25.0 < \mAB < 27.0$ (2,550 galaxies), $27.0 < \mAB < 28.0$ (4,892 galaxies), and $28.5 < \mAB < 29.0$ mag (5,336 galaxies) similarly follow a Gaussian distribution.}

Given a Gaussian distribution, we expect a false detection rate of 0.27\% (assuming a two-tailed test, where the difference in magnitude can either be positive or negative) for variable candidates identified at \teal{3$\Dmagerr$} significance. That rate falls rapidly for levels higher than \teal{3$\Dmagerr$}. We also identify variable candidates that vary in \textit{both} filters (in the same direction) with \teal{2$\Dmagerr$ and $3\Dmagerr$} significance. Since we implement the requirement that the variability must be of the same sign in both filters, we estimate the amount of false detections using a one-tailed test (i.e., 0.135\% \teal{false detections at 3$\Dmagerr$}) for these samples.

To demonstrate the methodology by which we assessed the number of false positives as a function of the significance of the detected variability, Figure~\ref{fig:false_detections} shows the number of detected variable candidates, the expected number of false positives (and percentage of the total number of objects in the parent sample), and the number of genuine variable sources in \teal{0.5$\Dmagerr$} wide bins for the F606W sample that varies by more than \teal{$3\Dmagerr$} and the F435W$+$F606W sample that varies by more than \teal{2$\Dmagerr$}.

\teal{In addition to false detections due to Gaussian statistics, image artifacts like cosmic rays, hot pixels, bias stripping, amplifier offsets, CTE trails, dust motes, optical ghosts, and cross-talk, may cause artificial variability to be detected. Of these, cosmic rays are the brightest and most prevalent, yet will not affect our results due to robust flagging and having 8 exposures (per pointing, filter and epoch) to stack using the \astrodrizzle{} software. We explore this in more detail in the Appendix. The other image artifacts are largely removed by the standard HST pipeline. Therefore, we do not expect and find no evidence for image artifacts to bias our results. Nonetheless, confirmation of which individual sources genuinely vary will require follow-up observations.}

 

\section{Results}

\begin{figure*}[ht]
    \centering
    \includegraphics[width=0.995\txw]{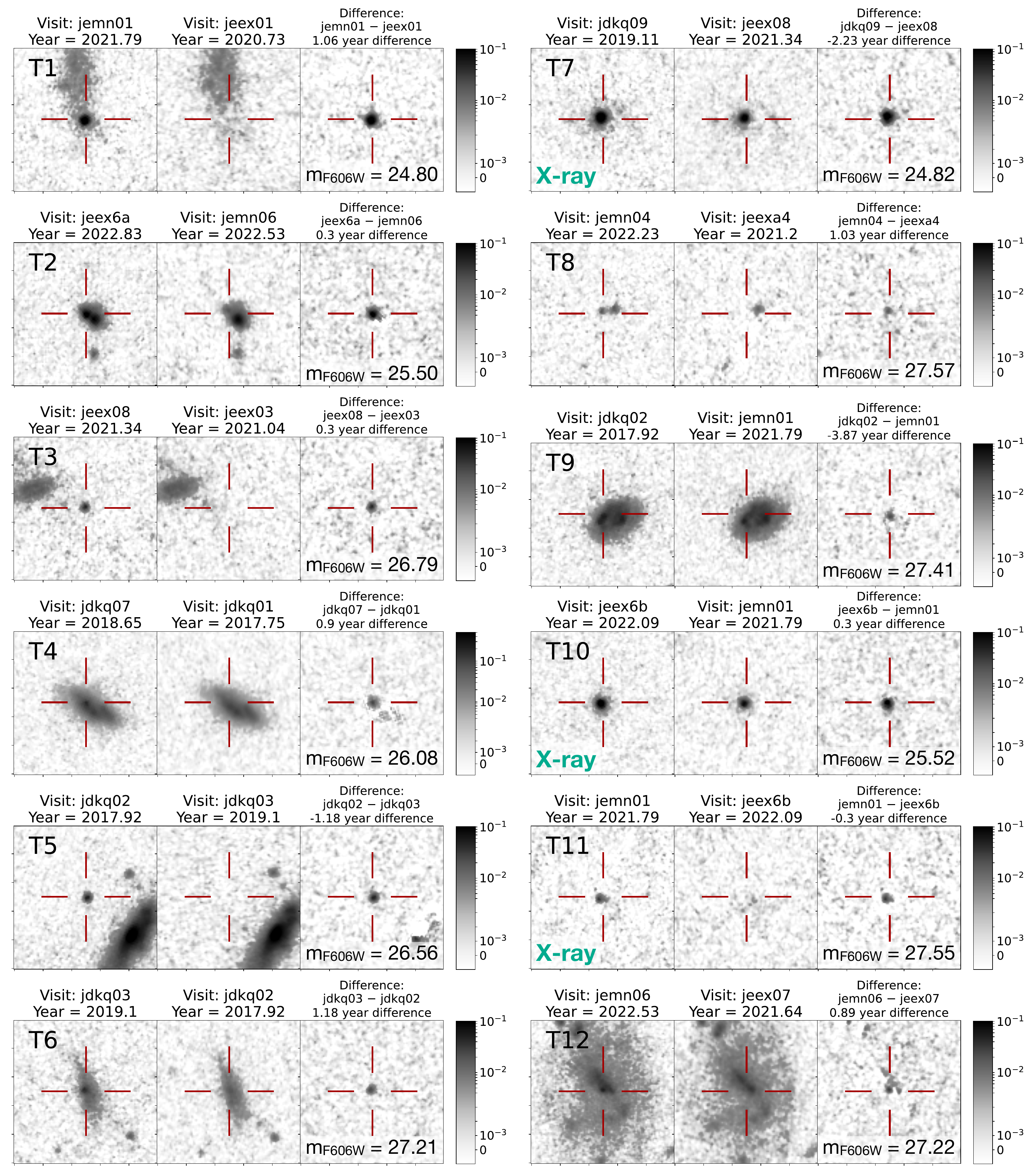}
    \caption{F606W images of each of the 12 transients identified within the JWST NEP TDF. For each transient we show a 3-panel composite of 100\ttimes100 pixel (3\arcsec\ttimes3\arcsec) cutouts from the image in which the transient is present (left), the image in which the transient is absent (middle), and the difference image (right). Labels give the transient ID, the visit IDs and dates of observation, the time interval between epochs, and the apparent F606W magnitude of the transient. Transients T7, T10 and T11 were also detected in X-rays (see Table \ref{tab:xray}), with T7 and T10 likely quasars. Grey-scale bars to the right of each 3-panel set are in units of $e^-$/s.\label{fig:transient_cutouts}}
\end{figure*}

\begin{table*}[ht!]
\centering
    \caption{Transients identified in HST TREASUREHUNT observations of the JWST NEP TDF\label{tab:final_transients}}
    \vspace*{-6pt}
    \begin{tabular}{lllllllrcc}
    \hline\hline
    \multicolumn{1}{l}{ID}&\multicolumn{1}{c}{RA}&\multicolumn{1}{c}{Dec}&\multicolumn{1}{l}{Visit\,1}&\multicolumn{1}{l}{Visit\,2}&\multicolumn{1}{c}{$t_1$}&\multicolumn{1}{c}{$t_2$}&\multicolumn{1}{c}{$\Delta t$}&\multicolumn{1}{c}{$m_{F435W}$}&\multicolumn{1}{c}{$m_{F606W}$}\\
       &\multicolumn{1}{c}{[deg]}&\multicolumn{1}{c}{[deg]}&  &  &\multicolumn{1}{c}{[yr]}&\multicolumn{1}{c}{[yr]}&\multicolumn{1}{c}{[yr]}&\multicolumn{1}{c}{[mag]}&\multicolumn{1}{c}{[mag]}\\[4pt]
    \hline
    T1 & 260.505468 & 65.798221 & jemn01 & jeex01 & 2021.789 & 2020.733 &   1.056 &       \nodata        &         24.80 (0.05) \\
    T2 & 260.585670 & 65.906841 & jeex6a & jemn06 & 2022.830 & 2022.530 &   0.300 &         25.89 (0.29) &         25.50 (0.09) \\
    T3 & 260.835152 & 65.727517 & jeex08 & jeex03 & 2021.344 & 2021.035 &   0.309 &       \nodata        &         26.79 (0.33) \\
    T4 & 260.641524 & 65.883517 & jdkq07 & jdkq01 & 2018.647 & 2017.749 &   0.898 &         26.49 (0.59) &         26.08 (0.16) \\
    T5 & 260.618390 & 65.777831 & jdkq02 & jdkq03 & 2017.918 & 2019.101 &$-$1.183 &         26.45 (0.52) &         26.56 (0.26) \\
    T6 & 260.602593 & 65.782095 & jdkq03 & jdkq02 & 2019.101 & 2017.918 &   1.183 &       \nodata        &         27.21 (0.45) \\
    T7 & 260.802040 & 65.725119 & jdkq09 & jeex08 & 2019.106 & 2021.344 &$-$2.238 &         24.44 (0.08) &         24.82 (0.05) \\
    T8 & 260.967884 & 65.867384 & jemn04 & jeexa4 & 2022.227 & 2021.204 &   1.023 &       \nodata        &         27.57 (0.67) \\
    T9 & 260.544912 & 65.793210 & jdkq02 & jemn01 & 2017.918 & 2021.789 &$-$3.871 &       \nodata        &         27.41 (0.53) \\
   T10 & 260.471298 & 65.762740 & jeex6b & jemn01 & 2022.093 & 2021.789 &   0.304 &         25.59 (0.21) &         25.52 (0.09) \\
   T11 & 260.535175 & 65.763209 & jemn01 & jeex6b & 2021.789 & 2022.093 &$-$0.304 &         27.75 (1.70) &         27.55 (0.66) \\
   T12 & 260.501418 & 65.869263 & jemn06 & jeex07 & 2022.530 & 2021.637 &   0.893 &        \nodata        &         27.22 (0.44) \\[2pt]
        \hline\\[-6pt]
        \end{tabular}
    \begin{minipage}{\textwidth}{\small
    \textbf{Notes:} Table columns (1) through (8) list the transient ID, the celestial coordinates (identified by eye) of the transient in decimal degrees, the root names of the visit where the transient is present (Visit 1) and where it is absent (Visit 2), the corresponding dates of observation, and the time interval $\Delta t$ between Visits 1 and 2. The two final columns list the measured F435W and F606W magnitudes of the transient (\mAB\ at $t_1$ relative to $t_2$, measured within a circular aperture centered on the transient with a radius of 8 pixels), where the magnitude uncertainties are given in parentheses.}
    \end{minipage}
\end{table*}

In this section, we present our findings regarding the identification and characterization of transients and variable objects in our data set. We detected a total of 12 transients, including two quasar candidates, and 190 unique galaxies exhibiting variability (where \tsim 80 are expected to be false positives). Figure~\ref{fig:f606w_mosaic} provides a visual representation of the locations of both the transient (red circles) and variable candidates (blue squares and diamonds). Variable candidates are identified in almost all areas of overlap, except those with insufficient astrometric precision or excess noise.

We note that the techniques used to identify transients and variable candidates (as described in \S\ref{sec:methods}) cause some objects detected as variable candidates to also be classified as transient candidates (e.g., the quasars). Rather than appearing and then disappearing, these actually vary in brightness over time with a sufficiently large amplitude that they were detectable in our difference images.

\subsection{Transients in the JWST NEP Time Domain Field} \label{sec:transient_results}

We identified a total of 12 transients, for each of which we show 3\arcsec\ttimes3\arcsec\ cutouts of the F606W images from the relevant visits and of the corresponding difference image in Figure~\ref{fig:transient_cutouts}. For each of these transients, Table~\ref{tab:final_transients} lists the coordinates, dataset IDs and date of observation of each visit, time interval between epochs, as well as the measured apparent brightness (difference in magnitude between the two epochs) in F606W. We also report the F435W magnitude, if the matched-aperture flux exceeded the F435W detection limit.  We call the epoch in which the transient signal appears ``Visit\,1''.
The transients range in brightness from \mAB\ \tsim\ 24.8 mag to \tsim27.5 mag in F606W. This may not correspond to the peak brightness of the transient, as the measured brightness depends on when the transient event image was taken during its respective light-curve. At magnitudes fainter than \mAB\ \tsim\ 27.5 mag, the difference images become dominated by image noise, preventing reliable identification of transients. 


In all, we detected \tsim0.14 transients per arcmin$^2$ of overlapping area, or \tsim491 deg$^{-2}$. Since two epochs are needed to detect a transient, the areal density per epoch is therefore \tsim0.07 transients per arcmin$^2$ (\tsim245 deg$^{-2}$), or \tsim0.77 per ACS/WFC footprint per epoch. For comparison, \citet{Dahlen_2012} searched for SNe in the GOODS fields (using the ACS/F850LP filter), and identified 118 SNe in a total area of 1 deg$^2$ to $m_{F850LP}$ \tsim\ 26 mag. If we only consider the 4 transients with $m_{F606W}$ \lsim\ 26.0 mag, we find an areal density of \tsim164 deg$^{-2}$, consistent with \citet{Dahlen_2012} when factoring in differences in detection filter, SNe colors and redshift distribution, and small number statistics. We also note that it is likely that not all our transients are SNe.


\subsubsection{Transients with Significant X-ray Detections}

Next, we checked whether any of the transients were detected in X-rays within XMM-Newton, \NuSTAR, or \Chandra\ observations of the field.  We matched the transient positions with published and preliminary X-ray source catalogs, and found that 3 of the transients are noticeable X-ray emitters.  While an in-depth analysis of the X-ray observations of these sources is beyond the scope of the present paper and is deferred to future work, we present the measured XMM-Newton, \NuSTAR, and \Chandra\ fluxes in Table~\ref{tab:xray}.

For the XMM-Newton and \NuSTAR\ catalogs, we used a matching radius of 5\arcsec\ and 30\arcsec, respectively, and found a likely match for transients T7, T10, and T11.
We inspected the \Chandra\ 0.5--7 keV event-file images for obvious sources, and here also identified sources consistent with T7 and T10.
For the other transients, we extracted source counts from a region with a radius of 2\arcsec\ and background counts from a nearby source-free region with a radius of 10\arcsec, but found no evidence for any X-ray counterparts above 1$\sigma$ significance. \Chandra\ upper limits vary across the field due to detector geometry, as well as vignetting and PSF spreading for off-axis sources, but we expect typical 3$\sigma$ upper limits of \tsim1.4\ttimes10$^{-5}$ ct\,s$^{-1}$ for the full 1.3\,Ms exposure.  For T7 and T10, we then extracted source fluxes using the \texttt{ciao} tool \texttt{srcflux} on the cross-registered event files for each OBSID. To convert count rates to fluxes, we used a model with assumed Galactic absorption and a $\Gamma$ = 1.4 power law.  At $r$ $\simeq$ 6.5\arcmin\ from the field center (characteristic of T7 and T10), this corresponds to 3$\sigma$ upper limits of 3.2\ttimes10$^{-16}$ erg\,s$^{-1}$\,cm$^{-2}$.  Note, however, that the epochs of individual X-ray observations do not necessarily match or overlap with specific HST visits in which a transient was observed or absent.

\begin{table}[t]
    \begin{center}
    \caption{Transients identified in HST observations of the JWST NEP TDF with significant X-ray detections\label{tab:xray}}
    \vspace*{-6pt}
    \begin{tabular}{llrr}
        \hline\hline
        Transient&Detected by&\multicolumn{1}{c}{Flux}&\multicolumn{1}{c}{Band}\\
                 &            &\multicolumn{1}{c}{[erg\,s$^{-1}$\,cm$^{-2}$]}&\multicolumn{1}{c}{[keV]}\\[4pt]
        \hline
        T7  & Chandra    & (5.36$\pm$0.64)$\times$10$^{-15}$ & 0.5--7\\
        T7  & XMM-Newton & (1.41$\pm$0.41)$\times$10$^{-15}$ & 0.5--2\\
        T7  & XMM-Newton & $<$9.03\,$\times$10$^{-15}$ & 2--10\\[2pt]
        \hline
        T10 & Chandra    & (2.68$\pm$0.53)$\times$10$^{-15}$ & 0.5--7\\
        T10 & XMM-Newton & (1.26$\pm$0.42)$\times$10$^{-15}$ & 0.5--2\\
        T10 & XMM-Newton & $<$4.97\,$\times$10$^{-15}$ & 2--10\\[2pt]
        \hline
        T11 & NuSTAR     & (1.14$\pm$0.41)$\times$10$^{-14}$ & 3--8\\[2pt]
        \hline
    \end{tabular}
    \end{center}
    \begin{minipage}{0.485\textwidth}{\small
    \textbf{Note.} Table columns list (1) the transient ID, (2) the X-ray observatory that also detected the transient source, (3) the measured X-ray flux, and (4) the energy band in which that flux was measured. The XMM-Newton 2--10 keV upper limits correspond to the 90\% confidence level.}
    \end{minipage}
\end{table}

\subsubsection{Notes on Individual Transients}

\emph{Transient T1} is detected only in F606W (although it is faintly discernible also in F435W) and appears in or superimposed on the outskirts of a diffuse, possibly clumpy disk galaxy, \tsim1\farcs08 S and 0\farcs16 W of the estimated galaxy center. It has a red (\textsl{F435W}$-$\textsl{F606W}) color compared to the galaxy in a color composite image, and its point-like morphology and complete absence in Visit\,2 (1.06 yr earlier) suggest it is a supernova (SN) within this $m_{\rm F606W}$ \tsim\ 23.9 mag host galaxy.

\emph{Transient T2} is detected in both F606W and F435W and appears off-center in a centrally concentrated galaxy (possibly an elliptical galaxy) or in its fainter and partially overlapping apparent companion galaxy. It is located \tsim0\farcs14 E and 0\farcs09 N of the core of the brighter ($m_{\rm F606W}$ \tsim\ 24.5 mag) galaxy and closer in projected distance (\tsim0\farcs07 W and 0\farcs10 S) to the center of the fainter one (not separately photometered by \sextractor). Its point-like morphology and complete absence in Visit\,2 (0.30 yr earlier) suggest it is a SN within either galaxy, with a visible color similar to that of its host.

\emph{Transient T3} is detected only in F606W. Its very red color is similar to that of a nearby extended disk galaxy or grouping of galaxies that is barely visible in F435W. Its point-like morphology and complete absence in Visit\,2 (0.31 yr earlier) suggest it is a SN associated with this $m_{\rm F606W}$ \tsim 24.0 mag host galaxy, appearing \tsim1\farcs11 W and 0\farcs32 S of the center of the brightest region, although the relatively large separation from the main body of this potential host may allow different interpretations.

\emph{Transient T4} is detected in both F606W and F435W within a $m_{\rm F606W}$ \tsim\ 23.0 mag disk galaxy with knots of apparent star formation and a secure MMT/Binospec spectroscopic redshift of $z$ = 0.615{\,}. It appears \tsim0\farcs10 N and 0\farcs10 E (a projected distance of \tsim1.0 kpc) of the estimated galaxy center, and its color is similar to that of the host. T4 appears point-like and is completely absent in Visit\,2 (0.90 yr earlier), but is much fainter than its host galaxy, suggesting a SN caught either before or significantly past its peak brightness.  T4 is presently the only transient with a spectroscopic redshift.

\emph{Transient T5} is detected in both F606W and F435W and appears well outside a nearby, highly inclined disk galaxy. It is located \tsim0\farcs10 E and 0\farcs01 N of the core of that bright $m_{\rm F606W}$ \tsim\ 22.5 mag galaxy, with a color that is similar or slightly bluer.  No hint of this point-like source is discernible in Visit\,2 (1.18 yr later), suggesting that T5 may be a SN associated with the disk galaxy, although the relatively large separation from this potential host allows other interpretations.

\emph{Transient T6} is detected only in F606W (with perhaps a hint discernible in F435W) within a red and diffuse disk galaxy, \tsim0\farcs09 E and 0\farcs09 N of the estimated galaxy center. It has a red color consistent with that of the galaxy, and its point-like morphology and complete absence in Visit\,2 (1.18 yr earlier) suggest it is a SN within this $m_{\rm F606W}$ \tsim\ 24.4 mag host galaxy.

\emph{Transient T7} is detected in both F606W and F435W, appears point-like and isolated. It lacks any nearby potential host galaxy, although there may be a hint of signal \tsim0\farcs59 due E and due W, and \tsim0\farcs23 due S, as well as some ``fuzz'' \tsim0\farcs42 to the NW in both Visits that could be due to a very faint potential host or galactic companions. Unlike most transients reported here, T7 is also clearly detected in Visit\,2 (2.24 yr later) and  has significant X-ray detections by both \Chandra\ and XMM-Newton (see Table~\ref{tab:xray}). This transient is thus unlikely to be a SN, but rather a quasar that exhibits significant variation in brightness.

\emph{Transient T8} is detected only in F606W and is our faintest transient at $m_{\rm F606W}$ = 27.57 mag. It appears \tsim0\farcs27 E and 0\farcs04 S of a faint ($m_{\rm F606W}$ \tsim\ 27.3 mag) and small nearby galaxy that may be similar in color, although possibly not quite as red. This point source is absent in Visit\,2 (1.02 yr earlier), and the proximity to a potential host galaxy suggests that it could be a SN.

\emph{Transient T9} is detected only in F606W (although it is faintly discernible also in F435W) and appears in what looks like either a spiral arm of a relatively face-on disk galaxy or in an interacting galaxy pair\footnote{A configuration resembling, e.g., NGC\,5278/79 (Arp\,239) or NGC\,6050/IC\,1179 (Arp\,272).}, \tsim0\farcs20 E and 0\farcs08 N of the estimated galaxy center. Its color is redder than that of an adjacent clump in the same spiral arm, but is consistent with the color of most of the disk. Its point-like morphology, projected location within a spiral arm, and absence in Visit\,2 (3.87 yr later) suggest that T9 is a Type\,II SN associated with massive star formation within this $m_{\rm F606W}$ \tsim\ 23.4 mag host galaxy.

\emph{Transient T10} is detected in both F606W and F435W, appears point-like and is isolated, lacking any nearby potential host galaxy. Unlike most transients reported here, T10 is also clearly detected in Visit\,2 (0.30 yr earlier) and has significant X-ray detections by both \Chandra\ and XMM-Newton (Table~\ref{tab:xray}). This transient is thus unlikely to be a SNe, but rather a quasar that exhibits significant variation in brightness.

\emph{Transient T11} is detected in both F606W and F435W, appears point-like and is isolated. The nearest galaxies are \tsim2\farcs3 to the S and \tsim2\farcs6 to the N, both $>$3 galaxy diameters away. Its color is relatively blue, allowing it to be clearly visible in F435W though $m_{\rm F606W}$ = 27.55 mag. This point source is absent in Visit\,2 (0.30 yr later). Interestingly, T11 was detected with \NuSTAR\ in the 3--8 keV band, and thus is unlikely to be a SNe.

\emph{Transient T12} is detected only in F606W and appears adjacent to a stellar bar across the center of an extended face-on spiral disk galaxy, \tsim0\farcs12 S and 0\farcs02 E of the estimated galaxy center. Its color is redder than that bar, but consistent with the color of other portions of the faint disk. Its point-like morphology, projected location within the disk close to the central portions, and absence in Visit\,2 (0.89 yr earlier) suggest that T12 is a SN within this $m_{\rm F606W}$ \tsim\ 22.5 mag host galaxy.


The four brightest transients, T1, T2, T7, and T10, all of which are brighter than 25.6 mag, may be bright SNe or quasars. The significant X-ray detections of T7 and T10 specifically argue for the latter. In contrast, one of the faintest transients in this study, T11 with \mAB\ \tsim\ 27.5 mag, is especially interesting as it is the only transient without a potential host detected in F606W, yet with noticeable X-ray emission.

\begin{table*}[ht!]
\centering
  \caption{Number of variable candidates identified and inferred in F606W and F435W.\label{tab:variability_stats}}
\hspace*{-0.75in}\begin{tabular}{llrrrr}
\hline\hline
\multicolumn{1}{c}{Sample} & All Objects & Apparent Variables & False Positives & Inferred Variables & \% variable\\[4pt]
\hline
F606W \teal{(3$\Dmagerr$)}           & 26,468 & 145 &   48 & 97 & 0.37\%\\
F606W \teal{(5$\Dmagerr$)}           & 26,468 &  10 &    0 & 10 & 0.04\%\\
F435W \teal{(3$\Dmagerr$)}           & 13,574 &  23 &   18 &  5 & 0.04\%\\
F606W and F435W \teal{(2$\Dmagerr$)} & 13,574 &  38 &   13 & 25 & 0.18\%\\
F606W and F435W \teal{(3$\Dmagerr$)} & 13,574 &   3 & $<1$ &  3 & 0.02\%\\[2pt]
\hline\\[-6pt]
  \end{tabular}
\begin{minipage}{\textwidth}{\small
\textbf{Note.} Table columns list (1) the sample of variable candidates, with the significance limit given between parentheses, (2) the total number of sources in the parent sample, limited to the F435W parent sample in the case of a combined F606W+F435W selection, (3) the number of objects satisfying the criteria at face value, (4) the number of false positive detections expected, (5) the number of genuine variable sources after statistical correction for false positives, and (6) the inferred percentage of sources in the parent sample that are variable.}
\end{minipage}
\end{table*}

\begin{figure*}[ht]
    \centering
    \includegraphics[scale = 0.46]{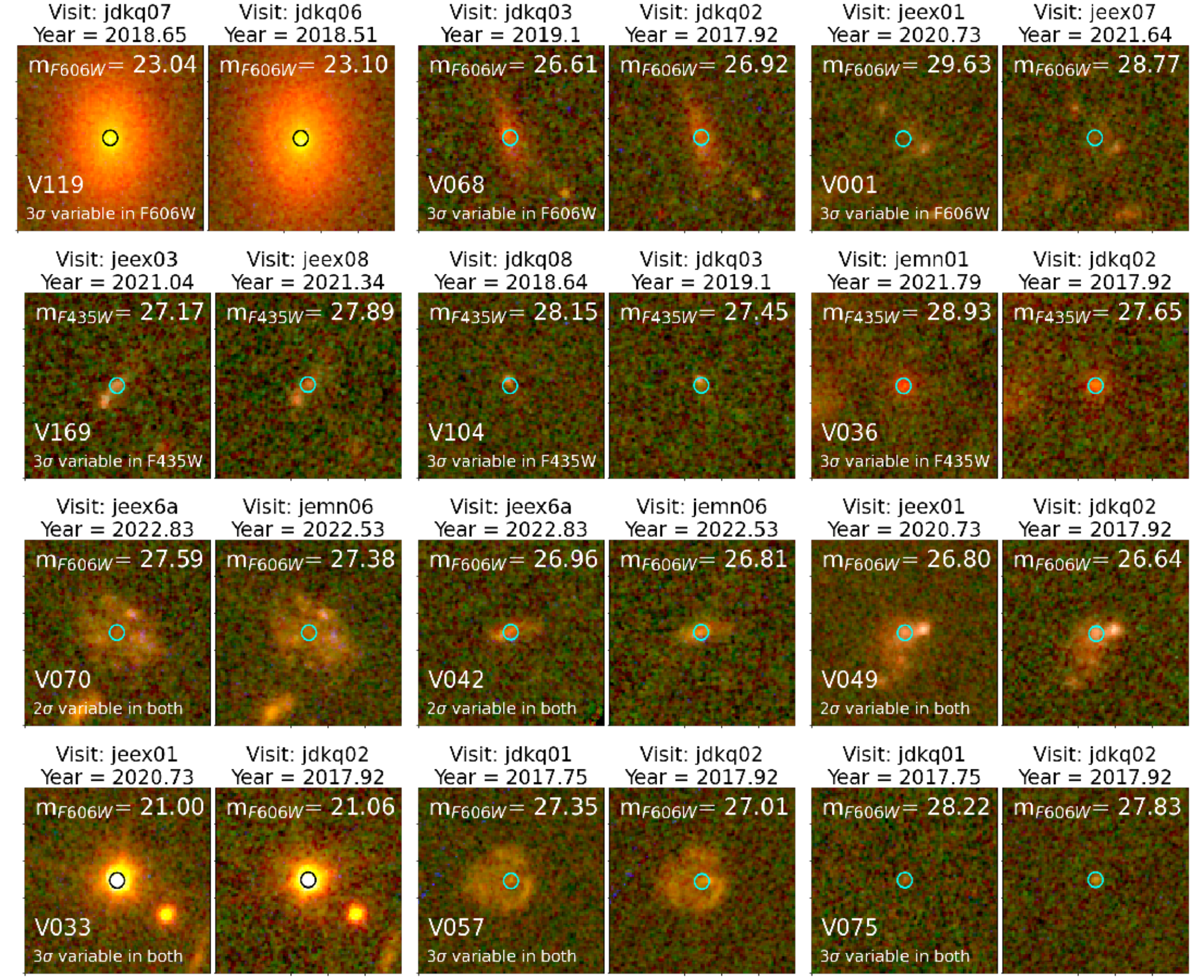}
    \caption{Color composites of 12 representative variable candidates, with F606W, F435W, and F275W shown in red, green, and blue hues, respectively. Each row of 100$\times$100 pixel (3\arcsec$\times$3\arcsec) cutouts correspond to a different sample: 1) F606W-only \teal{3$\Dmagerr$}, 2) F435W-only \teal{3$\Dmagerr$}, 3) F435W and F606W \teal{2$\Dmagerr$}, and 4) F435W and F606W \teal{3$\Dmagerr$} variability. A cyan or black circle at the center of each cutout represents the 0\farcs24 aperture in which variability was detected. Above each cutout, we note the visit identifier and decimal year of observation. At the top of each cutout, we include the measured \mTF\ magnitude for that visit. In the left panel of each pair of cutouts, at the bottom, we list the variable candidate identification number and whether it is variable in the F606W filter, the F435W filter, or both. Most of the variable candidates will be AGN, although some will be faint or obscured SNe.}\label{fig:variability_cutouts}
\end{figure*}

\subsection{Variability in the JWST NEP Time Domain Field} \label{sec:variability_results}

We identified 190 unique candidate variable candidates that meet the selection criteria of various samples discussed in \S~\ref{sec:variable_methods}, of which we estimate \tsim80 to be false positives (the exact number depends on how many false detections appeared in multiple variability samples). Figure~\ref{fig:f606w_mosaic} shows the locations of all 190 variable candidates. Symbol shapes (squares and diamonds) and hues identify the sample in which each variable was identified. Table~\ref{tab:variable_table} (at the end of this paper) provides a full list of IDs, celestial coordinates, dataset IDs and date of observation of each visit, as well as brightest measured apparent magnitude for each variable candidate. We also list the change in brightness between visits (in the sense Visit\,2 $-$ Visit\,1), and the significance of variability in units of \teal{$\Dmagerr$}, for both the F435W and F606W filters.

\teal{We emphasize that we can not claim any individual variable candidate varying at $<5\Dmagerr$ significance to be a genuine variable.  We can only place strong constraints on the overall number of variables in the area samples in a  statistical sense. Conversely, we also} do not exclude any specific sources as false detections from this final variability catalog, because we can only account for false positives in a statistical sense. We acknowledge that when an individual source is of interest, the source should be carefully analyzed and may require additional observations to ensure it is a genuine variable.

More relevant for population statistics, Table~\ref{tab:variability_stats} lists the inferred number of genuine variable candidates, factoring in false detections, for different significance levels in either the F606W or F435W filters, or both. Figure~\ref{fig:variability_cutouts} shows 3-color composite cutouts for a representative sample of 12 variable candidates. The bottom row shows our most conservative sample of 3 galaxies that vary in both F435W and F606W at \teal{$\ge$3$\Dmagerr$} significance. Examples of variable sources drawn from the less restrictive samples are shown in the first three rows.

\teal{We estimate the total number} of variable sources using the 1st, 3rd, and 4th rows of Table~\ref{tab:variability_stats}, since variables in the 2nd and 5th rows have more stringent selection criteria and are already included within these less restrictive samples. \teal{There are 206 objects listed in these specific rows of Table~\ref{tab:variability_stats}, where 16 are duplicates  (i.e., matching more than one of these sample selection criteria), leaving 190 unique variable candidates. With an estimated \tsim80 false positives, this results in \tsim110 unique variable sources, or 0.42\% of all 26,468 detected sources in the field. This is the maximum variable source fraction inferred in this work, as these samples are not necessarily independent.} Our most conservative sample of \teal{$\ge$3$\Dmagerr$} variable objects in both F435W and F606W only makes up 0.02\% of the parent population.

We \teal{also} estimate the areal density of variable sources using the total number of inferred genuine variables. These \teal{\tsim110 unique variable candidates} translate to \teal{\tsim1.25} variables per arcmin$^2$ (\teal{\tsim4500} deg$^{-2}$) of overlapping area. We can therefore expect \teal{\tsim14} new variable sources per additional fully overlapping ACS/WFC footprint (\tsim11.3 arcmin$^2$), assuming F606W imaging to similar depths as the TREASUREHUNT observations.

The \teal{3$\Dmagerr$} threshold, which determines by what magnitude a galaxy must vary to be flagged as variable, is key to putting our findings in context with other work. Higher noise levels naturally result in fewer detected galaxies, so the threshold helps to normalize these variations. The solid brown curves in Figure~\ref{fig:variable_candidates} show the \teal{1$\Dmagerr$, 2$\Dmagerr$, and 3$\Dmagerr$} thresholds (for the F606W filter) as a function of magnitude. For all sources with $\mTF\sim26.0$ mag (a 1.0 mag wide bin centered on 26.0 mag), the median \teal{3$\Dmagerr$} threshold is $\Delta \mTF=0.14$ mag. In other words, a typical 26.0 mag source would need to vary by at least 0.14 mag to be flagged as variable for this specific data set. For bins centered on 27.0 and 28.0 mag, the \teal{3$\Dmagerr$} thresholds are 0.24 and 0.41 mag, respectively. For the F435W filter, the \teal{3$\Dmagerr$} thresholds for bins centered on 26.0, 27.0, and 28.0 mag are 0.26, 0.44, and 0.82 mag, respectively.

Although our variability identification pipeline was developed to identify variable AGN, we still detect 6 out of the 12 transients presented in Section \ref{sec:transient_results}. Specifically, it identifies transients T2, T3, T5, T6, T9, and T11. Transients T7 and T10 were originally identified to exhibit variability, but were flagged and removed as star candidates due to their small FWHM. Transients T1, T4, T8 and T12 did not satisfy the \teal{3$\Dmagerr$} threshold due their relatively large photometric uncertainties.

\begin{figure*}[ht]
    \centering
    \includegraphics[width=\txw]{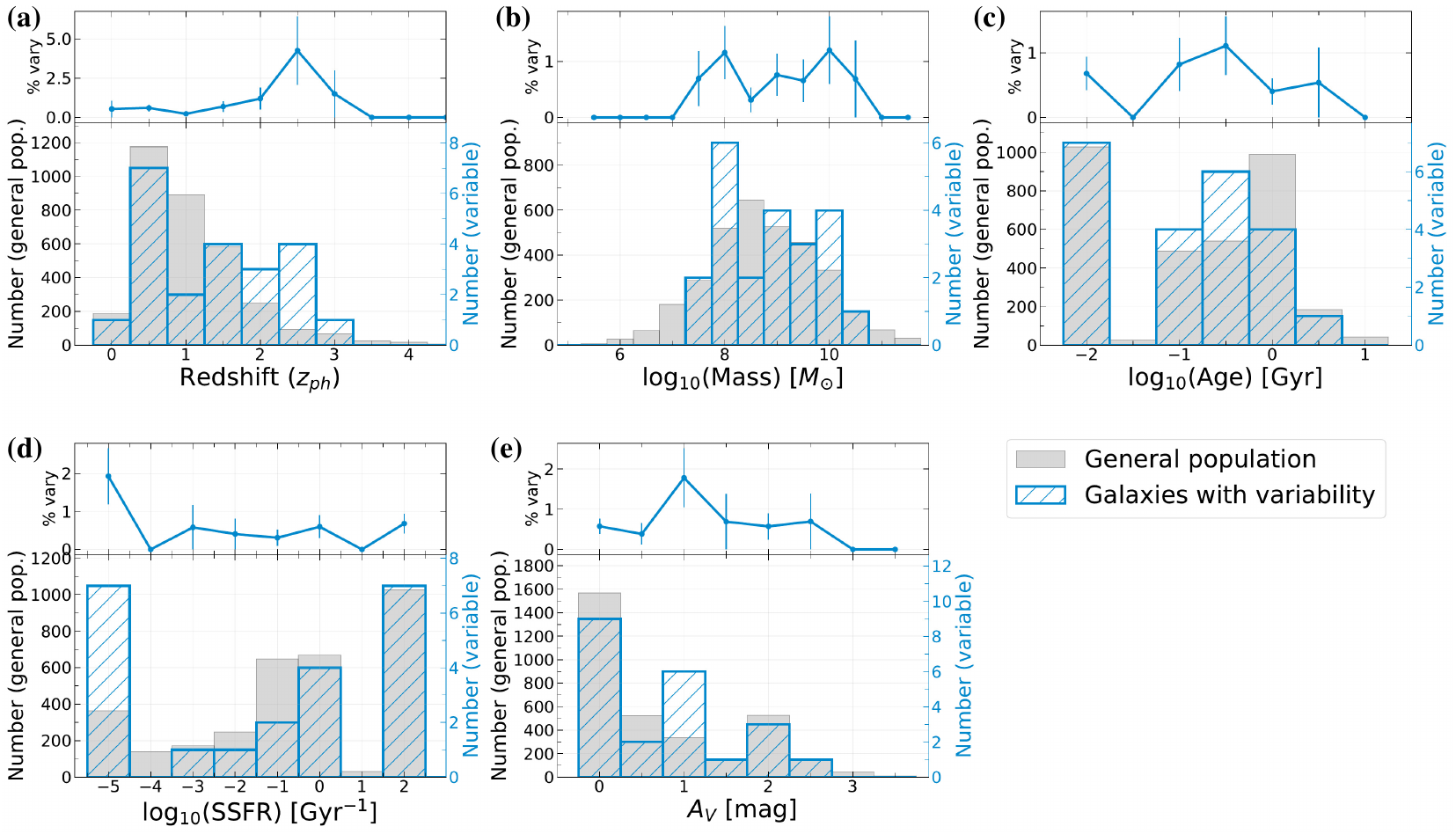}
    \caption{Comparison of galaxy properties in the general galaxy population and in galaxies that exhibit variability. Shown are distributions of (\emph{a}) photometric redshifts ($z_{ph}$), (\emph{b}) stellar mass, (\emph{c}) stellar population age, (\emph{d}) specific star formation rate (sSFR), and (\emph{e}) extinction by dust ($A_V$), for the general population (solid grey) and galaxies with variability (blue hatched). The y-axis labels on the left-hand side of each plot represents the number of galaxies in the general population, and those on the right-hand side (in blue) the number of galaxies with variability. Above each histogram panel, we
    plot the fraction of galaxies with variability (100\% \ttimes\ $N_{\rm var}/N_{\rm general}$) for each bin. The uncertainties assume Poisson statistics with $\sigma_N\propto\sqrt{N}$.This comparison only includes the subset of galaxies already flagged as reliable in \S~\ref{sec:photometry}, with both \HST\ and \JWST\ imaging (where photometric redshifts and SEDs can be fit), with $z_{ph}$ $<$ 5.5, $\chi^2$ $<$ 10, and where the 2$\sigma$ confidence interval in $p(z)$ spans $<$1 redshift
    unit.}\label{fig:galaxy_properties_hist}
\end{figure*}

\subsection{Photometric Redshift Estimation and SED Fits} \label{sec:redshift_methods}

To gain a better understanding of the transients and variable candidates in the field, we estimated photometric redshifts using \EAZY\ \citep{Brammer_2008}. This specifically allows us to understand the distances of the variable candidates, and to explore various properties of them, such as their masses, ages, dust extinction, and specific star formation rates (sSFRs).

Since that would be impossible with just the three HST filters, we also used PEARLS JWST/NIRCam images (see \S~\ref{sec:data}). Consequently, we estimate redshifts only for the subset of galaxies that fall within the overlapping coverage of both HST and JWST. We opted to not use the HST F275W filter for photometric redshift estimations, because most objects remain undetected at sufficient significance. We thus employed a total of 10 filters for estimating photometric redshifts: HST/ACS F435W and F606W, and JWST/NIRCam F090W, F115W, F150W, F200W, F277W, F356W, F410M, and F444W. To ensure the HST and JWST images are on the same pixel scale, we reprojected each HST image onto the PEARLS F444W image pixel grid using \texttt{reproject}. In each filter, we set all pixels without coverage in all 10 filters to NaN values, facilitating flagging of objects at image edges.

We created photometric catalogs for input into \EAZY\ using \sextractor. Recognizing that some galaxies exhibit variability that may bias a redshift estimation, we ran \sextractor\ on HST drizzled images of each individual visit instead of the full HST mosaic, and use the measurement with the fainter flux. We assume the fainter measurement represents the photometry if no variability were present ---galaxies may temporarily brighten, but they cannot get dimmer on human time scales. For the JWST data, we used the full mosaics, since only a single epoch yet exists at any given location within the field. The NIRCam F444W image served as the \sextractor\ detection image in all cases. For HST images, we set \texttt{MAG\_ZEROPOINT} to 26.49 mag for F606W and to 25.65 mag for F435W. For our JWST images, \texttt{MAG\_ZEROPOINT} was set to 28.0865 mag for all filters, as appropriate for their 0\farcs030 platescale, to convert from MJy\,sr$^{-1}$ \citep[see][]{windhorst_2023}. Other relevant parameters are the same for HST and JWST, and are as listed in the right-most column of Table~\ref{tab:sext_params}.

\EAZY(v1.0) was run on the resulting 10-band photometry. Following the approach outlined in Figure~\ref{fig:variable_candidates}, we scale \sextractor\ magnitude uncertainties by a factor of 3 upon input into \EAZY. This factor of 3 is larger than the factor of 1.15 used for scaling the 0\farcs24 aperture magnitudes in \S\ref{sec:variable_methods}, primarily due to the larger aperture sizes used here. We used the \texttt{eazy\_v1.0} templates on a redshift grid spanning 0.01 $\le$ $z$ $\le$ 15, allowing only single templates to be fit. We adopted \texttt{z\_ml} as the redshift estimate and did not include any priors.

We used the \EAZY\ redshifts as input to a custom \IDL\ spectral energy distribution (SED) fitting code that was also recently used by \citet{Meinke_2021, Meinke_2023}. This code starts with \citeauthor{Bruzual_2003} (\citeyear{Bruzual_2003}; BC03) SED models assuming a Salpeter IMF and an  exponentially decaying star-formation history specified by the decay scale, $\tau$, which ranges from 0.01 to 100 Gyr in 16 logarithmically spaced steps. Extinction by dust is specified on a grid of 0.0 $\le$ $A_V$ $\le$ 4.0 mag in steps of 0.2 mag assuming a \citet{Calzetti_2000} extinction law. For a given redshift, the age, $T$, is not allowed to exceed the age of the Universe (assuming a \citet{PlanckCollaboration2018} cosmology). The best-fit model is chosen by minimizing the $\chi^2$ and scaling for the stellar mass, $M$.
The (unweighted) present-day value of the star formation rate, $\Psi(T)$, is computed using the best-fit stellar mass, decay scale ($\tau$), and age ($T$) from $\Psi(T)=\Psi_0 e^{-T/\tau}$, where $\Psi_0$ is the star formation rate at time $T=0$. Given that $M=\int_0^T \Psi(t)dt$, we can solve $\Psi_0$ analytically from $\Psi_0=(M/\tau)/(1-e^{-T/\tau})$.

With 10-band, 0.4--5\,\micron\ photometry, we required $\chi^2$ $<$ 10 and $z$ $<$ 5.5. Since all objects of interest are detected in F606W, they must be at $z$ $<$ 5.5. However, these cuts alone will not eliminate all bad fits, especially if the photometric uncertainties in some crucial filters are large. We added an additional condition that the limits on the 2$\sigma$ confidence interval in \teal{redshift probability distribution, $p(z)$,} as computed by \EAZY\ (namely \texttt{l95} and \texttt{u95}) must not differ by more than 1 redshift unit. These conditions result in a sample with reliable photometric redshift estimates that includes 22 galaxies exhibiting variability (28\% of the original sample of 79 variables that fall within the \HST\ and \JWST\ footprints) and 3296 normal galaxies (31\% of the 10,664 galaxies that fall within the \HST\ and \JWST\ footprints). We only considered galaxies that were already marked as reliable in \S~\ref{sec:photometry} (the 26,468 sources in Table \ref{tab:variability_stats}).

Due to the necessarily limited grid of SED model parameters, any galaxy with an inferred sSFR $<$ 10$^{-5}$ yr$^{-1}$ is categorized as ``quiescent'' and its sSFR is set to 10$^{-5}$ yr$^{-1}$. Our templates also cannot distinguish stellar population ages younger than 10 Myr, so we set all apparent age solutions $<$0.01 Gyr to 0.01 Gyr. Given that age is inversely correlated with the inferred sSFR, where $\log_{10}($Age$) \sim -2$ corresponds to $\log_{10}($sSFR$)\sim 2$, we also set all $\log_{10}$(sSFR)$>2$ to 2. This adjustment is solely made to assess collective trends in galaxies with variability. 

In Figure~\ref{fig:galaxy_properties_hist}, we compare photometric redshift ($z_{ph}$),  stellar mass, stellar population age, specific star formation rate (sSFR), and dust extinction ($A_V$) of galaxies that exhibit variability to typical galaxies in the general population. Overall, we sample a broad range of redshifts, masses, ages, star formation histories, and dust extinction. The most prominent peaks in the fraction of galaxies with variability with respect to the general galaxy population (expressed in \%) occur at a redshift of $z_{ph}$ $\simeq$ 2.5, a stellar mass of \tsim10$^{8}$ and \tsim10$^{10}$ $M_{\odot}$, an age of \tsim10$^{-0.5}$ Gyr, a sSFR of \tsim10$^{-5}$ Gyr$^{-1}$, and an attenuation $A_V$ $\simeq$ 1.0 mag. The median redshift of our variability sample is $z_{ph}$ = 1.3, but we sample variability up to $z_{ph}$ = 2.9{\,}. Similarly, our galaxies with variability sample masses $M$ = (0.02--20)\ttimes10$^9$ $M_{\odot}$, ages $T$ = (0.01--2) Gyr, sSFR = (10$^{-5}$--10$^2$) Gyr$^{-1}$, and $A_V$ = (0.0--2.4) mag.

Three transients fell within an area that also has PEARLS 8-filter JWST/NIRCam coverage, yet all three had \EAZY\ photometric redshifts with a $\chi^2 > 10$. This could be due to a significant non-thermal component from an AGN, or due to the presence of strong emission lines that could mimic a continuum break. As transient T4 has a secure MMT/Binospec spectroscopic redshift of $z$ = 0.615, we used the Code Investigating GALaxy Emission (\CIGALE) \citep{Boquien_2019} to better fit the SED of its host galaxy. \CIGALE\ has the benefit of incorporating emission lines into its models, while the previously mentioned SED fitting algorithm does not and resulted in poorer fits for this transient host.
We used a delayed burst-quench star-formation history model \citep{Ciesla_2017}, since the standard exponentially declining model can struggle to reproduce higher SFRs, and its morphology suggests that the transient host likely underwent a recent star-formation episode. The $e$-folding times tested were 1, 3, 5, and 8 Gyr, the main stellar population spanned 1--8 Gyr in increments of 1 Gyr, the burst/quench age ranged from 0.1, 0.5, and 1 Gyr, and the possible SFR ratios before and after these burst ages were 0.1, 0.3, and 0.6. Our stellar component used the BC03 library using the Salpeter IMF, testing metallicities of 0.004, 0.02, and 0.05 $Z_{\odot}$ with young stellar ages being either 10 or 100 Myr. We included a nebular component to account for emission lines using all the default parameters, where the nebular component is self-consistent with the stellar population in the sense that its strength is determined by the Lyman continuum of the total SED. To account for dust attenuation, we used \citet{Calzetti_2000} extinction with possible modified power-law slopes of $-$0.4, 0.2, and 0. For dust \textit{emission}, we used the \citet{Dale_2014} model with $\alpha$ slopes of 1.0, 2.0, and 3.0{\,}.

\begin{figure}[t]
    \centering\vspace*{6pt}
    \includegraphics[scale=0.47]{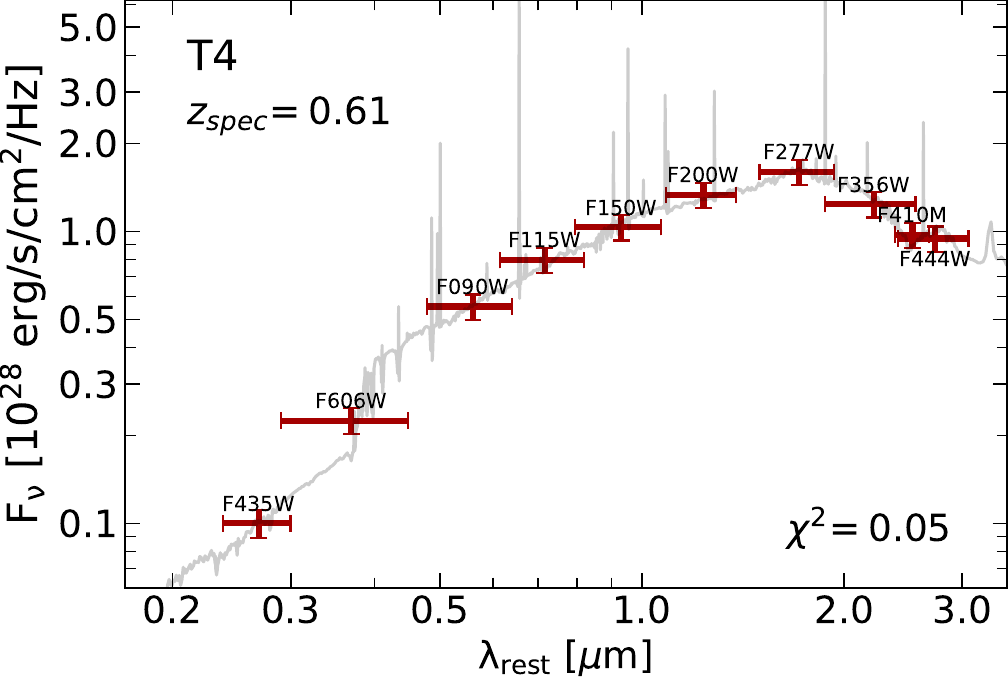}
    \caption{SED fit of transient T4 with \CIGALE, adopting the MMT/Binospec spectroscopic redshift of 0.615. The red points with vertical error bars show the measured fluxes of the transient host and their associated uncertainties. The horizontal bars on each data point represent the width of the bandpass. The light grey spectrum represents the best \CIGALE\ fit, with the reduced $\chi^2$ value shown.\\ }
    \label{fig:transient_seds}
\end{figure}

The best fit SED for the T4 host galaxy is shown in Figure~\ref{fig:transient_seds} and corresponds to a stellar population age of 3.6\,Gyr, $A_V$ = 0.5 mag of attenuation by dust, and a stellar mass of 1.2$\times$10$^{10}$ $M_{\odot}$. Its sSFR is 0.5 $M_{\odot}$ Gyr$^{-1}$. The reported best-fit value of 0.05 of the reduced $\chi^2$ indicates that the (conservative) photometric uncertainties upon input to \CIGALE\ were overestimating the actual ones somewhat.


\vfill

\section{Discussion} \label{sec:discussion}

An analysis of the detailed properties of individual transients and variable candidates is beyond the scope of this paper.  We can, however, make some general statements about the properties of these source populations, using their observed optical light distributions (morphology) and \EAZY\ SED fits. From the point-like morphology and off-center locations at small projected distances to likely host galaxies, we infer that the majority of the discovered transients are SNe.
The majority of the variable candidates, on the other hand, are found either in or very near the center of galaxies or are isolated without a discernible host, and are surmised to be AGN.  Most of these AGN are normal SMBHs in galaxies, with stochastically fluctuating accretion at lower luminosities than seen in quasars.
Nonetheless, it is important to consider that other phenomena, such as SNe from central starbursts, stochastic microlensing events, or tidal disruption events, could also contribute to the observed central variability. SNe that are heavily obscured and SNe that are caught at very early or at late times can contribute to variability detected in local peaks in surface brightness well away from host galaxy centers.

\subsection{Classification of Transients}

Based on their positions relative to likely host galaxies and the PSF-like appearance in the difference images, we propose that nine of the transients (T1, T2, T3, T4, T5, T6, T8, T9, and T12) are SNe. It is reasonable to expect that \tsim40--50\% of detected SNe are Type~Ia, as suggested by previous studies \citep{Dahlen_2012, Graur_2014, rodney_2014, cappellaro_2015}. If we assume that we have a minimum of nine SNe, we can anticipate a minimum of three Type~Ia SNe with the remainder being CC SNe.

The TREASUREHUNT observing strategy provided piecemeal UV--Visible imaging rather than time-domain monitoring. As a result, these SN candidates will have long faded and can no longer be followed up for spectroscopic identification. Nonetheless, now that this initial imaging exists, future monitoring observations could efficiently expand the present sample and would allow such spectroscopic follow-up. The sum of the areas of overlap used in this work samples \tsim88 arcmin$^{2}$, which is about half of the total \HST\ coverage of this field (\tsim194 arcmin$^2$). Therefore, we can expect about 2$\times$ as many transients if this field were observed again. Full overlapping coverage of this field could yield at least six new Type~Ia supernovae, and coupled with spectra and rapid follow-up, can provide essential constraints on fundamental cosmological parameters like the Hubble constant, the mass density of the universe, the cosmological constant, the deceleration parameter, and the age of the Universe \citep[e.g.,][]{riess_1998,Riess_2023}. In addition to Type Ia SNe, comprehensive HST coverage of this field could yield at least six CC SNe observations. Particularly when combined with ancillary data, this could contribute significantly to resolving the observed discrepancies in CC SN rates compared to theoretical models \citep{cappellaro_2014}. With a sufficiently large sample, the SN rate as a function of redshift could also be constrained. Furthermore, the increased volume of observations enhances the likelihood of capturing rare or unexpected events, potentially leading to groundbreaking discoveries in the field.

At least three of the transients are unlikely to be SNe. They either were detected in both epochs or they are isolated with no apparent host galaxy. When considering only the optical emission, transients T7 and T10 exhibit substantial changes in brightness between the two epochs, of $\Delta m$ = 0.83 $\tpm$ 0.06 mag and 0.82 $\tpm$ 0.09 mag, respectively. That would be equivalent to the presence of an additional source with \mAB\ \tsim\ 24.82 and 25.52 mag.  The observed centroid positions in both
epochs match to within 0\farcs001, such that the additional signal is indistinguishable from being due to a nuclear source that changed brightness. We therefore suggest that T7 and T10 may be quasars.  The fact that they both were detected in soft X-rays at flux levels $>$2\ttimes10$^{-15}$ erg\,s$^{-1}$\,cm$^{-2}$ lends additional evidence to this interpretation.

T11 is a unique transient with no visible host, yet was detected in hard X-rays by NuSTAR. If that X-ray emission is indeed associated with T11 then it could be a faint (\mAB\,\gsim\,29 mag), high-redshift quasar that briefly flared up to be detectable in F606W.  If the X-ray emission is unrelated, then we speculate that T11 may represent a SN located within an undetected host galaxy. That host could be a faint dwarf galaxy at $z$ \lsim\ 6, or a more massive host at $z$ \gsim\ 6 to explain the non-detection in F606W in the second visit. Unfortunately, T11 falls outside the area with \JWST\ coverage, so we cannot distinguish between these two scenarios at this time. \citet{chakrabarti_2018} have demonstrated that the rates of SNe in dwarf galaxies can be three times higher (per unit volume) than in typical spiral galaxies. Given the limited understanding of star formation in dwarf galaxies, a substantial number of SN candidates like T11 could offer a unique opportunity to identify otherwise undetected dwarf galaxies and gain more insights into their star formation rates.

Overall, determining the nature of each transient remains challenging without spectroscopic data secured close in time. Nonetheless, these detected events showcase the considerable potential and effectiveness of transient science within the JWST NEP TDF.

\subsection{Properties of Variable Candidates}

Our methods were developed to isolate variable candidates that are coincident with local maxima in surface brightness that can be isolated and photometered using \sextractor. This naturally optimizes detection of variable AGN, but faint SNe are also detected when located offset from the center of the host galaxy. In addition, some offset variability may also be due to accretion disks of SMBHs that have not yet settled in the center of the host galaxy \citep[e.g.,][]{Reines_2020}. Examples of offset variables are shown in Figure \ref{fig:variability_cutouts} (objects V001 and V169).

\teal{In general, the number of variable sources inferred here is expected to be fewer than the true number in the field, as some sources will vary on timescales that were not probed in this work. Constructing a sample of genuine variable sources in the field requires follow-up observations at various time sampling intervals, possibly extending over decades.}

Figure~\ref{fig:galaxy_properties_hist} shows that galaxies exhibiting variability span a broad range of stellar mass, stellar population age, star formation history, and dust extinction. This underscores that rest-frame UV--Visible variability can be a powerful
tactic to identify accreting AGN in a wide cross section of the galaxy population.
Especially notable is that AGN were detected through their variability even in dusty
galaxies with $A_V$ \gsim\ 2 mag.

The top portion of each panel in Figure~\ref{fig:galaxy_properties_hist}, of the fraction of galaxies with variability within each histogram bin for the quantity plotted, shows peaks in most of the panels, although the formal significance of these peaks (assuming Poisson counting statistics) is generally \lsim2$\sigma$ due to small number statistics in individual bins.

First, in Figure~\ref{fig:galaxy_properties_hist}\emph{a}, we find a relatively high fraction of galaxies with variability around $z$\,\tsim\,2.5, while at both lower and higher redshifts variability appears to be more rare. This is in contrast with both \citet{Villforth_2012}, who find optical--near-infrared variable AGN predominantly toward lower redshifts, and \citet{sarajedini_2011}, who find an increase in optically varying AGN with increasing redshift. \citet{Zhong_2022} find a similar number of optically variable AGN to peak at $z\sim 1$. We suspect that the number of variable sources as a function of redshift is influenced by cosmic variance of a relatively rare source population, and thus differs depending on the field observed. The fraction of variable sources may be flat over the 0\,\lsim\,$z$\,\lsim\,3 redshift range when averaged over an area of sufficient size.

The observed stellar mass distribution (Figure~\ref{fig:galaxy_properties_hist}\emph{b}) of galaxies exhibiting variability shows an apparent excess at both \tsim10$^8$ $M_{\odot}$ and \tsim10$^{10}$ $M_{\odot}$. We do not detect variability in the most massive galaxies, either due to small number statistics (most likely) or because their SMBHs are less likely to accrete at a sufficient rate to be detectable or have a longer period of relative quiescence between major accretion events. We also do not detect variability in galaxies with $M$\ \lsim\ 10$^7$ $M_{\odot}$, likely due to a combination of both small number statistics and larger photometric uncertainties in these faint systems, such that our \teal{3$\Dmagerr$} threshold is less likely to be met for a given amplitude of variability. If the relative distribution is taken at face values, then it could also hint at the presence of two separate populations of sources with variability: one associated with lower mass \tsim10$^8$ $M_{\odot}$ hosts, and another associated with higher mass hosts \tsim10$^{10}$ $M_{\odot}$. Such could be the case if variability associated with CC SNe favors lower mass host galaxies, and variability associated with SN\,Ia and AGN favors higher mass hosts.

In Figure~\ref{fig:galaxy_properties_hist}\emph{c} and \emph{d} we see that a significant fraction of the galaxies with variability are best fit with stellar population models characterized by active star formation (young ages \lsim30\,Myr and high sSFR \gsim30 Gyr$^{-1}$). Conversely, there is also a sizable fraction that is quiescent (sSFR \lsim\ 0.01 Gyr$^{-1}$) with stellar populations older than 1 Gyr. A possible excess appears for intermediate population ages (60--600 Myr).

Last, Figure~\ref{fig:galaxy_properties_hist}\emph{e} shows that while the galaxies exhibiting variability largely track the distribution of extinction seen in the general population, there may be an excess in the variability sample for $A_V$ \tsim\ 1 mag.

\subsection{Constraints on SMBH Mass, Accretion and Radiation Lifetimes}

\teal{The sparse and random time sampling of this field does not allow us to trace the evolution of faint AGN, but may still teach us various properties of them. AGN brightness varies on timescales corresponding to the light crossing time in the accretion disks and gas clouds surrounding SMBHs. For example, for a SMBH with a mass $M$ $\sim$ 10$^8$\,$M_{\odot}$, the timescale for variability cannot be shorter than $\sim$1 week in the rest-frame \citep[e.g.,][]{Xie_2005}.} For the majority of sources where the variability is due to AGN activity, that variability originates close to the central SMBH.  If the timescale we sample is the minimum timescale of variation, we can provide rough constraints on the SMBH masses following \citet{Xie_2005}. According to Schwarzschild Black Hole Theory (SBHT), the
mass of a black hole ($M_{\bullet}$ in $M_{\odot}$) is equal to the minimum timescale of variation ($\Delta t_{\rm min}$ in yr) as $M_{\bullet}$ = 4.29\ttimes10$^{11}$\,$\Delta t_{\rm min}$. Time intervals between observations in various areas of overlap
in the TREASUREHUNT data range from 1 day (0.0027 yr) to 4.78 yr.  At the median redshift of our variable candidate sample, $z_{ph} = 1.3$, the range in rest-frame timescales becomes 0.001--2.08 yr, giving SMBH masses of 5.1\ttimes10$^8$ \lsim\ $M_{\bullet}$ \lsim\ 8.9\ttimes10$^{11}$ $M_{\odot}$.  SMBH masses \gsim10$^{11}$ $M_{\odot}$ are unreasonable: objects within the region of overlap with a 4.78 yr interval between observations likely vary on much shorter timescales and we did not sample the minimum timescale of variation. Even a more typical time interval of \tsim1 yr between epochs will yield a \tsim10$^{11}$ $M_{\odot}$ SMBH mass. Taking the full range in redshifts sampled into account, we will therefore only claim that in select areas of overlap we can probe SMBHs with masses of $M_{\bullet}$\,\gsim\,(3--9)\ttimes10$^{8}$ $M_{\odot}$, but with no firm upper limit.

We can also frame our finding that \teal{a maximum of 0.42\%} of the general field galaxy population shows significant variability in the context of the accretion and radiation lifetimes of AGN. If we assume that: 1) all galaxies seen in this field with \HST\ have a central SMBH \citep[e.g.,][]{Kormendy_2013}, 2) the central SMBH is always visible when accreting, and 3) that a currently accreting AGN will be variable, then we would deduce that the AGN in our field \teal{are actively accreting for 0.42\% of the time.} Assumptions 2) and 3) are simplified, as some AGN will be hidden at rest-frame UV--Visible wavelengths by surrounding dust. This fraction could be as large as 2/3 of all AGN, if we use the \tsim30\% average Lyman continuum escape fraction of weak AGN \citep[]{smith_2018, Smith_2020, Smith_2024} as proxy for the fraction of AGN with direct unobscured sight lines to the observer.
Figure \ref{fig:galaxy_properties_hist}\emph{c} implies an average SED age of \tsim10$^{8}$ years (\tsim100 Myr) for the parent population, although the spread in age is wide (0.01--10 Gyr). A \teal{0.42\%} fraction of visibly variable AGN would correspond to an implied average AGN activity lifetime of \teal{0.42\%\ttimes10$^8$ $\sim$ 4\ttimes10$^5$ yr} (and a range of \teal{4\ttimes10$^4$--4\ttimes10$^7$ yr}). For comparison, \citet{Rawes_2015} estimated the optical synchrotron electron life times of AGN with visible jets observed with HST and Chandra at \tsim10$^4$ yr. However, these authors state that their estimated synchrotron lifetimes may be too short by at least a factor of two, and may be longer if synchrotron electrons are re-accelerated in the ambient magnetic field. On the other end of the activity scale, \citet{Jakobsen_etal2003} show that at least some quasars can remain active more or less continuously for \gsim10 Myr. With these assumptions and significant uncertainties, a visible AGN accretion time of \tsim10$^4$ to 10$^7$ yr could indeed result in a \teal{0.42\%} variability fraction in a galaxy population characterized by an average SED age of \tsim100 Myr (and a range of 0.01--10 Gyr) whose stars and gas feed that central engine.

Future work will need to secure and analyze larger samples of optically variable sources with deep X-ray imaging to better constrain UV--Visible synchrotron lifetimes, and wider and deeper HST+JWST images with spectroscopic (NIRISS or NIRSpec) redshifts to improve redshifts and the characterization of their stellar population.

  %
  %

\subsection{Caveats and Estimated Population Size}

It is important to note that with \HST\ data alone, distinguishing between variability caused by variable AGN and faint SNe is challenging. Variability at locations offset from the cores of galaxies could potentially indicate SN events or SMBHs that have not yet reached the galaxy center. In all four catalogs, we identify a total of 25 cases showing variability at locations other than the core, as identified through visual inspection. Identifying AGN will require complementing this work with ancillary observations of the field to determine whether galaxies exhibiting variability also emit mid-IR emission. Prolonged monitoring of variable objects within this field will contribute to a deeper understanding of SMBHs within galaxies, including their evolution over time and their influence on galactic physics.

We also note that the variable candidates presented here are a \emph{strict lower bound} to the number present in the field: many genuine variable sources, even if at $z$\,\lsim\,6 and detectable in F606W, were certainly missed because they did not vary sufficiently between specific individual visits to meet our \teal{3$\Dmagerr$} threshold, or if the timescale of their variability exceeded the duration of the TREASUREHUNT program. In the present study, the photometric uncertainties in each epoch of observation contributed equally to that \teal{3$\Dmagerr$} threshold.  Any future additional observations to similar depths would reduce the photometric uncertainty in the reference magnitudes, thus lowering the variability detection threshold. The magnitude and level of variability for all sources in this work is collected in Table~\ref{tab:variable_table}, so that they can be compared in future studies within the JWST NEP TDF.

Although significant \teal{(\ie\ $>$3$\Dmagerr$)} direct detections of variability to such exceeding faint limits through systematic monitoring campaigns with HST, JWST, or Roman would require a very large community investment, it is not outside the realm of what is possible.  Even at slightly shallower depths, and employing similar extrapolations, variability could account for a large fraction of all accreting SMBHs.


\section{Conclusion} \label{sec:conclusion}

This study highlights the JWST NEP TDF as an exceptional site for investigating time-variable astronomical phenomena, including SNe and (weak) variable AGN. We identified 12 transients and \tsim100 variable sources in ACS/WFC F606W and F435W images from the HST TREASUREHUNT program. We provide positional information, dates of observation, intervening time interval, and magnitudes for each of the 12 transients, which range in brightness from \tsim24.8 to \tsim27.5 mag. Their areal density is \tsim0.07 transients per arcmin$^2$ (\tsim245 per deg$^2$) per epoch. We argue that the vast majority of these transients are SNe. Three transients (T7, T10, and T11) were detected in X-rays, of which two (T7 and T10) appear isolated and point-like and are likely newly identified quasars. Transient T11 has no visible host galaxy, and its nature remains uncertain. We suspect it to be a faint quasar if the X-ray detection is indeed associated, or otherwise either part of a very faint dwarf galaxy at $z$\,\lsim\,6, or a transient in a more massive host at $z$\,\gsim\,6. One transient (T4) has a spectroscopic redshift $z$\,=\,0.615 from MMT/Binospec.

Our variability search revealed that \teal{0.42\%} of the general $z$\,\lsim\,6 field galaxy population exhibits variability at \teal{\gsim3$\Dmagerr$} significance to depths of 29.5 mag in F606W and 28.6 mag in F435W, with an areal density of \teal{\tsim1.25} variables per arcmin$^2$ (\teal{\tsim4500} per deg$^2$). We carefully identified variable candidates, using the measured distribution of magnitude differences for objects observed more than once to calibrate our photometric uncertainties, \teal{where the scaled photometric uncertainty is $\Dmagerr$, as defined in Equation \ref{eq:magerr}. Sources are flagged} as variable if they varied by more than \teal{$3\Dmagerr$} in F435W or F606W, or varied by more than \teal{2$\Dmagerr$} in both filters in the same sense. This revealed 190 unique variable candidates, of which we estimated \tsim80 to be false positives (using Gaussian statistics). Most of the variable candidates are coincident with the cores of galaxies, indicating potential AGN variability, while a smaller fraction appears associated with faint SNe. We also estimate photometric redshifts for our sample, from which we estimate masses, ages, dust extinction, and sSFRs. We briefly explore ways to estimate SMBH mass and AGN timescales using the time interval between overlapping visits and the observed rates of variability.


In conclusion, this work firmly establishes the JWST NEP TDF as a pivotal field for time-domain science with an initial harvest of transients and galaxies exhibiting variability. We emphasize the ability to identify AGN through their variability in the JWST NEP TDF. Future follow-up observations with HST or JWST could greatly increase the numbers of directly detected variables and to lower variability amplitudes.


\begin{acknowledgements}

We dedicate this paper to the memory of Ms. Susan Selkirk, whose expert graphics skills helped showcase our HST and JWST images of the NEP TDF at ASU, NASA, and elsewhere.

\teal{We thank the anonymous referee for comments that helped better clarify our results.}

RO, RAJ, RAW, SHC, AMK, NPH, BLF and CNAW acknowledge support from \HST\ grants HST-GO-15278.*, 16252.* and 16793.* from the Space Telescope Science Institute, which is operated by the Association of Universities for Research in Astronomy, Inc. (AURA) under contract NAS\,5-26555 from the National Aeronautics and Space Administration (NASA). RAJ, RAW and SHC also acknowledge support from NASA JWST Interdisciplinary Scientist grants NNX14AN10G, 80NSSC18K0200, NAG5-12460, and 21-SMDSS21-0013 from NASA Goddard Space Flight Center (GSFC). 

CNAW acknowledges support from the NIRCam Development Contract NAS5-02105 from GSFC to the University of Arizona. DK acknowledges support from the National Research Foundation of Korea (NRF) grant funded by the Korean government (MSIT) (No. NRF-2022R1C1C2004506). M.H.\ acknowledges the support from the Korea Astronomy and Space Science Institute grant funded by the Korean government (MSIT) (No.\ 2022183005) and the support from the Global Ph.D. Fellowship Program through the National Research Foundation of Korea (NRF) funded by the Ministry of Education (NRF-2013H1A2A1033110). S.B. acknowledges partial support from the project PID2021-124243NB-C21 funded by the Spanish Ministry of Science and Innovation. MI acknowledges the support from the National Research Foundation of Korea (NRF) grants, No. 2020R1A2C3011091, and No. 2021M3F7A1084525, funded by the Korea government (MSIT).

This work is based on observations associated with programs HST-GO-15278, 16252 and 16793 made with the NASA/ESA \emph{Hubble} Space Telescope. We thank our Program Coordinator, Tricia Royle, for her expert help scheduling this \HST\ program.

This work made use of observations associated with program JWST-GTO-2738 made with the NASA/ESA/CSA \emph{James Webb} Space Telescope.

\HST\ and \JWST\ data were obtained from the Mikulski Archive for Space Telescopes (MAST) at the Space Telescope Science Institute, which is operated by the Association of Universities for Research in Astronomy, Inc., under NASA contracts NAS\,5-26555 (\HST) and NAS5-03127 (\JWST). \teal{The specific observations analyzed can be accessed via \dataset[doi:10.17909/wv13-qc14]{https://doi.org/10.17909/wv13-qc14}.}

This work has also made use of data from the European Space Agency (ESA) mission \Gaia\ (\url{https://www.cosmos.esa.int/gaia}), processed by the \Gaia\ Data Processing and Analysis Consortium (DPAC, \url{https://www.cosmos.esa.int/web/gaia/dpac/consortium}). Funding for the DPAC has been provided by national institutions, in particular the institutions participating in the \Gaia\ Multilateral Agreement.

This work has also made use of data obtained from the Chandra Data Archive and the Chandra Source Catalog, and software provided by the Chandra X-ray Center (CXC) in the application packages CIAO and Sherpa.

This research has made use of NASA's Astrophysics Data System (ADS) bibliographic services (Kurtz et al. 2000).

We also acknowledge the indigenous peoples of Arizona, including the Akimel O'odham (Pima) and Pee Posh (Maricopa) Indian Communities, whose care and keeping of the land has enabled us to be at ASU's Tempe campus in the Salt River Valley, where this work was conducted.
\end{acknowledgements}

\facility{Based on observations with the NASA/ESA \emph{Hubble} Space Telescope (Wide Field Camera 3; Advanced Camera for Surveys), and on observations with the NASA/ESA/CSA \emph{James Webb} Space Telescope (NIRCam); Mikulski Archive for Space Telescopes,  \url{https://archive.stsci.edu}}

\software{Astropy \citep{astropy:2013, astropy:2018, astropy:2022}; Photutils \citep{bradley_2022}; MultiDrizzle/AstroDrizzle/DrizzlePac \citep{Koekemoer_2003, Fruchter_2009, hoffmann_2021}; \sextractor\ \citep{bertin_1996}; WCSTools 
\citep{Mink_2002, Mink_2011}; DS9 \citep{Joye_2003}.}

\clearpage
\bibliography{ref}{}
\bibliographystyle{aasjournal}

%
\begin{table*}[t!]
\movetableright=-0.75in
\caption{Variable Sources identified in HST TREASUREHUNT observations of the JWST NEP TDF\label{tab:variable_table}}
\setlength{\tabcolsep}{3pt}
{\scriptsize
\begin{tabular}{rrllllllrlrrlrr}
\hline\hline\\[-8pt]
\multicolumn{9}{c}{$\quad$} & \multicolumn{3}{c}{\hrulefill\ F435W \hrulefill}
  & \multicolumn{3}{c}{\hrulefill\ F606W \hrulefill}\\
\multicolumn{1}{c}{ID} & \multicolumn{1}{c}{CatID} &
   \multicolumn{1}{c}{RA} & \multicolumn{1}{c}{Dec} &
   \multicolumn{1}{c}{Visit1} & \multicolumn{1}{c}{Visit2} &
   \multicolumn{1}{c}{$t_1$} & \multicolumn{1}{c}{$t_2$} &
   \multicolumn{1}{c}{$\Delta t$} & \multicolumn{1}{c}{$m_{0.24}$} &
   \multicolumn{1}{c}{$\!\!\!\!\Delta m_{0.24}$} & \multicolumn{1}{c}{$\Dmagerr$} &
   \multicolumn{1}{c}{$m_{0.24}$} & \multicolumn{1}{c}{$\!\!\!\!\Delta m_{0.24}$} &
   \multicolumn{1}{c}{$\Dmagerr$}\\
  &   & \multicolumn{1}{c}{[deg]} & \multicolumn{1}{c}{[deg]} &   &   &
  \multicolumn{1}{c}{[yr]} & \multicolumn{1}{c}{[yr]} &
  \multicolumn{1}{c}{[yr]} & \multicolumn{1}{c}{[mag]} &
  \multicolumn{1}{c}{[mag]} &  & \multicolumn{1}{c}{[mag]} &
  \multicolumn{1}{c}{[mag]} &   \\
\multicolumn{1}{c}{(1)} & \multicolumn{1}{c}{(2)} & \multicolumn{1}{c}{(3)} &
  \multicolumn{1}{c}{(4)} & \multicolumn{1}{c}{(5)} & \multicolumn{1}{c}{(6)} &
  \multicolumn{1}{c}{(7)} & \multicolumn{1}{c}{(8)} & \multicolumn{1}{c}{(9)} &
  \multicolumn{1}{c}{(10)} & \multicolumn{1}{c}{(11)} &
  \multicolumn{1}{c}{(12)} & \multicolumn{1}{c}{(13)} &
  \multicolumn{1}{c}{(14)} & \multicolumn{1}{c}{(15)}\\[4pt]
\hline
V001 & 19364 & 260.4052851 & 65.8382044 & jeex01 & jeex07 & 2020.733 & 2021.637 & $-$0.904 & 28.86 (0.34) & $-$0.83 & 1.79 & 28.77 (0.21) & $-$0.86 & 3.64\\
V002 & 19815 & 260.4300229 & 65.8342205 & jeex01 & jeex07 & 2020.733 & 2021.637 & $-$0.904 & 28.84 (0.47) & $-$0.74 & 1.17 & 28.69 (0.33) &    1.15 & 3.04\\
V003 & 19785 & 260.4326188 & 65.8345545 & jeex01 & jeex07 & 2020.733 & 2021.637 & $-$0.904 & 27.93 (0.24) &    0.79 & 2.40 & 28.00 (0.15) &    0.55 & 3.23\\
V004 & 20003 & 260.4473218 & 65.8318503 & jeex01 & jeex07 & 2020.733 & 2021.637 & $-$0.904 & 28.65 (0.40) &    0.68 & 1.27 & 28.23 (0.17) &    0.59 & 3.04\\
V005 & 41428 & 260.4502146 & 65.7515390 & jemn01 & jeex6b & 2021.789 & 2022.093 & $-$0.304 & 29.10 (0.50) &    0.43 & 0.63 & 28.84 (0.33) &    1.13 & 3.00\\
V006 & 20130 & 260.4604344 & 65.8296368 & jeex01 & jeex07 & 2020.733 & 2021.637 & $-$0.904 & 28.33 (0.46) &    1.37 & 2.20 & 28.75 (0.23) &    0.62 & 2.36\\
V007 & 20104 & 260.4622684 & 65.8300200 & jeex01 & jeex07 & 2020.733 & 2021.637 & $-$0.904 & 27.49 (0.15) &    0.56 & 2.74 & 27.49 (0.09) &    0.28 & 2.84\\
V008 & 40135 & 260.4800396 & 65.7615899 & jemn01 & jeex6b & 2021.789 & 2022.093 & $-$0.304 & 28.97 (0.68) &    0.93 & 1.02 & 28.67 (0.30) &    1.05 & 3.10\\
V009 & 42426 & 260.4811656 & 65.7389668 & jemn01 & jeex6b & 2021.789 & 2022.093 & $-$0.304 & 24.68 (0.03) & $-$0.12 & 3.13 & 23.39 (0.01) & $-$0.04 & 2.97\\
V010 & 48837 & 260.4847004 & 65.8816131 & jemn06 & jeex07 & 2022.530 & 2021.637 &    0.893 & 27.92 (0.28) & $-$0.81 & 2.17 & 27.65 (0.10) & $-$0.26 & 2.14\\
V011 & 48800 & 260.4850911 & 65.8819144 & jemn06 & jeex07 & 2022.530 & 2021.637 &    0.893 & 27.17 (0.13) & $-$0.19 & 1.09 & 28.66 (0.36) & $-$1.29 & 3.12\\
V012 & 48831 & 260.4857008 & 65.8816066 & jemn06 & jeex07 & 2022.530 & 2021.637 &    0.893 & 27.92 (0.28) & $-$0.81 & 2.17 & 28.55 (0.32) &    1.23 & 3.33\\
V013 & 48977 & 260.4903910 & 65.8782280 & jemn06 & jeex07 & 2022.530 & 2021.637 &    0.893 & 28.33 (0.26) & $-$0.35 & 1.02 & 28.29 (0.19) &    0.77 & 3.52\\
V014 & 39334 & 260.4918589 & 65.7687739 & jemn01 & jeex6b & 2021.789 & 2022.093 & $-$0.304 & 28.75 (0.33) & $-$0.06 & 0.13 & 28.52 (0.22) &    0.81 & 3.24\\
V015 & 48934 & 260.4943736 & 65.8797585 & jemn06 & jeex07 & 2022.530 & 2021.637 &    0.893 & 28.53 (0.41) &    0.68 & 1.24 & 28.22 (0.17) &    0.59 & 3.07\\
V016 & 19975 & 260.4965078 & 65.8309014 & jeex01 & jeex07 & 2020.733 & 2021.637 & $-$0.904 & 19.18 (0.01) &    0.03 & 7.62 & 19.46 (0.03) &    0.13 & 3.42\\
V017 & 49160 & 260.4965104 & 65.8742828 & jemn06 & jeex07 & 2022.530 & 2021.637 &    0.893 & 28.43 (0.26) &    0.12 & 0.36 & 28.03 (0.13) &    0.47 & 3.14\\
V018 & 42276 & 260.4979360 & 65.7417349 & jemn01 & jeex6b & 2021.789 & 2022.093 & $-$0.304 & 28.77 (0.35) & $-$0.04 & 0.08 & 28.57 (0.23) & $-$0.81 & 3.03\\
V019 & 40970 & 260.5045403 & 65.7556663 & jemn01 & jeex6b & 2021.789 & 2022.093 & $-$0.304 & 28.12 (0.33) &    0.89 & 2.03 & 29.00 (0.34) &    0.92 & 2.37\\
V020 & 39013 & 260.5060240 & 65.7731007 & jemn01 & jeex6b & 2021.789 & 2022.093 & $-$0.304 & 28.86 (0.41) & $-$0.08 & 0.14 & 27.98 (0.13) &    0.57 & 3.71\\
V021 & 38120 & 260.5075136 & 65.7925038 & jemn01 & jdkq02 & 2021.789 & 2017.918 &    3.871 & 29.06 (0.64) & $-$0.83 & 0.96 & 28.74 (0.29) &    1.13 & 3.37\\
V022 & 38237 & 260.5090095 & 65.7913053 & jemn01 & jdkq02 & 2021.789 & 2017.918 &    3.871 & 28.63 (0.32) & $-$0.01 & 0.02 & 28.31 (0.17) &    0.60 & 3.05\\
V023 & 17632 & 260.5109164 & 65.8123441 & jeex01 & jdkq02 & 2020.733 & 2017.918 &    2.815 & 28.90 (0.43) & $-$0.45 & 0.76 & 28.61 (0.26) &    0.89 & 3.01\\
V024 & 38331 & 260.5132040 & 65.7883638 & jemn01 & jdkq02 & 2021.789 & 2017.918 &    3.871 & 28.16 (0.44) &    1.24 & 2.09 & 28.79 (0.24) &    0.62 & 2.22\\
V025 & 17466 & 260.5132587 & 65.8165789 & jeex01 & jdkq02 & 2020.733 & 2017.918 &    2.815 & 27.85 (0.28) & $-$1.16 & 3.07 & 29.40 (0.35) & $-$0.42 & 1.06\\
V026 & 17481 & 260.5139809 & 65.8162782 & jeex01 & jdkq02 & 2020.733 & 2017.918 &    2.815 & 27.85 (0.28) & $-$1.16 & 3.07 & 29.28 (0.37) &    0.71 & 1.66\\
V027 & 17482 & 260.5148620 & 65.8162530 & jeex01 & jdkq02 & 2020.733 & 2017.918 &    2.815 & 27.85 (0.28) & $-$1.16 & 3.07 & 28.13 (0.13) & $-$0.28 & 1.95\\
V028 & 42324 & 260.5156057 & 65.7410733 & jemn01 & jeex6b & 2021.789 & 2022.093 & $-$0.304 & 28.62 (0.34) &    0.16 & 0.36 & 27.34 (0.09) &    0.35 & 3.45\\
V029 & 49278 & 260.5165718 & 65.8723986 & jemn06 & jeex07 & 2022.530 & 2021.637 &    0.893 & 29.37 (0.51) &    0.01 & 0.02 & 28.17 (0.18) & $-$0.86 & 4.10\\
V030 & 39884 & 260.5168553 & 65.7637201 & jemn01 & jeex6b & 2021.789 & 2022.093 & $-$0.304 & 29.02 (0.59) &    0.66 & 0.83 & 28.04 (0.15) & $-$0.58 & 3.43\\
V031 & 17642 & 260.5181754 & 65.8119968 & jeex01 & jdkq02 & 2020.733 & 2017.918 &    2.815 & 25.73 (0.05) &    0.01 & 0.11 & 25.52 (0.03) &    0.23 & 5.72\\
V032 & 41667 & 260.5200938 & 65.7490569 & jemn01 & jeex6b & 2021.789 & 2022.093 & $-$0.304 & 28.80 (0.43) & $-$0.36 & 0.61 & 28.59 (0.28) &    1.02 & 3.14\\
V033 & 17861 & 260.5203993 & 65.8073115 & jeex01 & jdkq02 & 2020.733 & 2017.918 &    2.815 & 21.56 (0.01) & $-$0.04 & 4.18 & 21.00 (0.01) &    0.06 &14.72\\
V034 & 17017 & 260.5285670 & 65.8336403 & jeex01 & jdkq02 & 2020.733 & 2017.918 &    2.815 & 25.52 (0.04) & $-$0.01 & 0.15 & 24.85 (0.02) &    0.09 & 3.37\\
V035 & 17815 & 260.5286942 & 65.8079924 & jeex01 & jdkq02 & 2020.733 & 2017.918 &    2.815 & 28.89 (0.41) &    0.32 & 0.58 & 28.12 (0.15) &    0.61 & 3.58\\
V036 & 37564 & 260.5290447 & 65.8045199 & jemn01 & jdkq02 & 2021.789 & 2017.918 &    3.871 & 27.65 (0.29) & $-$1.28 & 3.28 & 26.37 (0.05) &    0.01 & 0.25\\
V037 & 39503 & 260.5295628 & 65.7668789 & jemn01 & jeex6b & 2021.789 & 2022.093 & $-$0.304 & 27.11 (0.12) & $-$0.24 & 1.47 & 28.53 (0.27) & $-$1.14 & 3.63\\
V038 & 37829 & 260.5301461 & 65.7975577 & jemn01 & jdkq02 & 2021.789 & 2017.918 &    3.871 & 28.69 (0.40) & $-$0.50 & 0.92 & 28.40 (0.19) &    0.68 & 3.10\\
V039 & 40938 & 260.5321327 & 65.7559458 & jemn01 & jeex6b & 2021.789 & 2022.093 & $-$0.304 & 28.72 (0.68) & $-$1.23 & 1.34 & 28.58 (0.25) & $-$0.90 & 3.18\\
V040 &  5015 & 260.5324664 & 65.8591989 & jdkq01 & jeex07 & 2017.749 & 2021.637 & $-$3.888 & 24.38 (0.03) &    0.07 & 2.13 & 22.35 (0.01) & $-$0.07 & 8.83\\
V041 & 37499 & 260.5337585 & 65.8079004 & jemn01 & jdkq02 & 2021.789 & 2017.918 &    3.871 & 24.84 (0.03) & $-$0.08 & 1.79 & 23.51 (0.01) &    0.08 & 5.92\\
V042 & 33111 & 260.5339889 & 65.9156836 & jeex6a & jemn06 & 2022.830 & 2022.530 &    0.300 & 27.20 (0.14) & $-$0.40 & 2.20 & 26.81 (0.06) & $-$0.15 & 2.09\\
V043 & 40338 & 260.5343798 & 65.7602481 & jemn01 & jeex6b & 2021.789 & 2022.093 & $-$0.304 & 29.16 (0.52) & $-$0.11 & 0.16 & 28.08 (0.14) & $-$0.49 & 3.07\\
V044 & 39949 & 260.5351364 & 65.7631956 & jemn01 & jeex6b & 2021.789 & 2022.093 & $-$0.304 & 27.97 (0.25) &    0.41 & 1.21 & 27.69 (0.19) &    1.27 & 5.95\\
V045 & 17127 & 260.5351702 & 65.8285103 & jeex01 & jdkq02 & 2020.733 & 2017.918 &    2.815 & 29.56 (0.59) & $-$0.11 & 0.14 & 28.62 (0.25) &    0.88 & 3.05\\
V046 & 41298 & 260.5354508 & 65.7527669 & jemn01 & jeex6b & 2021.789 & 2022.093 & $-$0.304 & 28.75 (0.43) & $-$0.42 & 0.73 & 27.86 (0.14) &    0.57 & 3.58\\
V047 & 38098 & 260.5366516 & 65.7930824 & jemn01 & jdkq02 & 2021.789 & 2017.918 &    3.871 & 28.25 (0.27) &    0.31 & 0.86 & 28.17 (0.16) &    0.72 & 3.86\\
V048 & 39683 & 260.5391875 & 65.7652426 & jemn01 & jeex6b & 2021.789 & 2022.093 & $-$0.304 & 28.13 (0.25) & $-$0.34 & 0.98 & 28.60 (0.28) & $-$1.05 & 3.25\\
V049 & 17165 & 260.5419552 & 65.8281206 & jeex01 & jdkq02 & 2020.733 & 2017.918 &    2.815 & 26.82 (0.10) & $-$0.34 & 2.63 & 26.64 (0.06) & $-$0.16 & 2.42\\
V050 & 38172 & 260.5436430 & 65.7915235 & jemn01 & jdkq02 & 2021.789 & 2017.918 &    3.871 & 28.01 (0.26) &    0.79 & 2.20 & 27.80 (0.11) &    0.48 & 3.79\\
V051 & 38266 & 260.5439886 & 65.7895398 & jemn01 & jdkq02 & 2021.789 & 2017.918 &    3.871 & 29.64 (0.73) & $-$0.22 & 0.23 & 28.69 (0.25) &    0.86 & 3.03\\
V052 & 17831 & 260.5441742 & 65.8076230 & jeex01 & jdkq02 & 2020.733 & 2017.918 &    2.815 & 28.48 (0.38) & $-$0.85 & 1.67 & 28.73 (0.33) &    1.16 & 3.06\\
V053 & 38122 & 260.5449020 & 65.7931821 & jemn01 & jdkq02 & 2021.789 & 2017.918 &    3.871 & 26.56 (0.08) & $-$0.03 & 0.26 & 25.82 (0.04) & $-$0.21 & 4.62\\
V054 & 39765 & 260.5451659 & 65.7641522 & jemn01 & jeex6b & 2021.789 & 2022.093 & $-$0.304 & 29.01 (0.54) &    0.58 & 0.80 & 28.54 (0.24) & $-$0.91 & 3.30\\
V055 & 40721 & 260.5457177 & 65.7587033 & jemn01 & jeex6b & 2021.789 & 2022.093 & $-$0.304 & 28.12 (0.32) & $-$0.87 & 2.01 & 28.31 (0.17) & $-$0.49 & 2.54\\
V056 & 37972 & 260.5485168 & 65.7954958 & jemn01 & jdkq02 & 2021.789 & 2017.918 &    3.871 & 28.55 (0.31) & $-$0.16 & 0.38 & 28.41 (0.30) &    1.42 & 4.11\\
V057 &    48 & 260.5548528 & 65.8386224 & jdkq01 & jdkq02 & 2017.749 & 2017.918 & $-$0.169 & 27.15 (0.12) & $-$0.49 & 3.00 & 27.01 (0.07) & $-$0.34 & 4.06\\
V058 & 33274 & 260.5610833 & 65.9123618 & jeex6a & jemn06 & 2022.830 & 2022.530 &    0.300 & 28.39 (0.41) &    0.82 & 1.47 & 28.21 (0.18) &    0.66 & 3.23\\
V059 & 32938 & 260.5613045 & 65.9189714 & jeex6a & jemn06 & 2022.830 & 2022.530 &    0.300 & 29.43 (0.64) &    0.16 & 0.19 & 28.50 (0.22) & $-$0.78 & 3.11\\
V060 &  6238 & 260.5683705 & 65.7831566 & jdkq03 & jdkq02 & 2019.101 & 2017.918 &    1.183 & 28.78 (0.46) &    0.52 & 0.84 & 27.37 (0.09) & $-$0.36 & 3.35\\
\hline\\[-8pt]
\end{tabular}
}
\end{table*}

\addtocounter{table}{-1}
\begin{table*}[t!]
\movetableright=-0.75in
\caption{(\textit{continued})}
\setlength{\tabcolsep}{3pt}
{\scriptsize
\begin{tabular}{rrllllllrlrrlrr}
\hline\hline\\[-8pt]
\multicolumn{9}{c}{$\quad$} & \multicolumn{3}{c}{\hrulefill\ F435W \hrulefill}
  & \multicolumn{3}{c}{\hrulefill\ F606W \hrulefill}\\
\multicolumn{1}{c}{ID} & \multicolumn{1}{c}{CatID} &
   \multicolumn{1}{c}{RA} & \multicolumn{1}{c}{Dec} &
   \multicolumn{1}{c}{Visit1} & \multicolumn{1}{c}{Visit2} &
   \multicolumn{1}{c}{$t_1$} & \multicolumn{1}{c}{$t_2$} &
   \multicolumn{1}{c}{$\Delta t$} & \multicolumn{1}{c}{$m_{0.24}$} &
   \multicolumn{1}{c}{$\!\!\!\!\Delta m_{0.24}$} & \multicolumn{1}{c}{$\Dmagerr$} &
   \multicolumn{1}{c}{$m_{0.24}$} & \multicolumn{1}{c}{$\!\!\!\!\Delta m_{0.24}$} &
   \multicolumn{1}{c}{$\Dmagerr$}\\
  &   & \multicolumn{1}{c}{[deg]} & \multicolumn{1}{c}{[deg]} &   &   &
  \multicolumn{1}{c}{[yr]} & \multicolumn{1}{c}{[yr]} &
  \multicolumn{1}{c}{[yr]} & \multicolumn{1}{c}{[mag]} &
  \multicolumn{1}{c}{[mag]} &  &
  \multicolumn{1}{c}{[mag]} & \multicolumn{1}{c}{[mag]} &   \\
\multicolumn{1}{c}{(1)} & \multicolumn{1}{c}{(2)} & \multicolumn{1}{c}{(3)} &
  \multicolumn{1}{c}{(4)} & \multicolumn{1}{c}{(5)} & \multicolumn{1}{c}{(6)} &
  \multicolumn{1}{c}{(7)} & \multicolumn{1}{c}{(8)} & \multicolumn{1}{c}{(9)} &
  \multicolumn{1}{c}{(10)} & \multicolumn{1}{c}{(11)} &
  \multicolumn{1}{c}{(12)} & \multicolumn{1}{c}{(13)} &
  \multicolumn{1}{c}{(14)} & \multicolumn{1}{c}{(15)}\\[4pt]
\hline
V061 & 41126 & 260.5686500 & 65.7542573 & jemn01 & jeex6b & 2021.789 & 2022.093 & $-$0.304 & 28.31 (0.42) &    1.15 & 2.01 & 28.27 (0.15) &    0.36 & 2.05\\
V062 & 33539 & 260.5688549 & 65.9067578 & jeex6a & jemn06 & 2022.830 & 2022.530 &    0.300 & 29.17 (0.57) & $-$0.40 & 0.52 & 28.29 (0.17) &    0.62 & 3.13\\
V063 & 33524 & 260.5855449 & 65.9068262 & jeex6a & jemn06 & 2022.830 & 2022.530 &    0.300 & 25.57 (0.05) &    0.16 & 2.27 & 25.41 (0.03) &    0.21 & 5.59\\
V064 & 34022 & 260.5865160 & 65.8956540 & jeex6a & jemn06 & 2022.830 & 2022.530 &    0.300 & 22.45 (0.01) & $-$0.07 & 4.92 & 29.03 (0.22) & $-$0.06 & 0.23\\
V065 &   234 & 260.5868385 & 65.8333537 & jdkq01 & jdkq02 & 2017.749 & 2017.918 & $-$0.169 & 28.37 (0.23) & $-$0.08 & 0.25 & 28.32 (0.18) & $-$0.71 & 3.40\\
V066 &  6629 & 260.5886084 & 65.7776747 & jdkq03 & jdkq02 & 2019.101 & 2017.918 &    1.183 & 28.05 (0.25) &    0.59 & 1.71 & 28.74 (0.24) &    0.89 & 3.17\\
V067 &   320 & 260.5901557 & 65.8318573 & jdkq01 & jdkq02 & 2017.749 & 2017.918 & $-$0.169 & 23.25 (0.01) &    0.02 & 1.05 & 21.70 (0.01) &    0.02 & 3.63\\
V068 &  6334 & 260.6025353 & 65.7820869 & jdkq03 & jdkq02 & 2019.101 & 2017.918 &    1.183 & 27.59 (0.15) & $-$0.05 & 0.23 & 26.61 (0.06) &    0.30 & 4.42\\
V069 & 34339 & 260.6027231 & 65.8860494 & jeex6a & jemn06 & 2022.830 & 2022.530 &    0.300 & 28.42 (0.38) &    0.98 & 1.89 & 28.79 (0.30) &    1.08 & 3.11\\
V070 & 34133 & 260.6032077 & 65.8928444 & jeex6a & jemn06 & 2022.830 & 2022.530 &    0.300 & 27.61 (0.18) & $-$0.55 & 2.28 & 27.38 (0.08) & $-$0.21 & 2.21\\
V071 & 34238 & 260.6051718 & 65.8906921 & jeex6a & jemn06 & 2022.830 & 2022.530 &    0.300 & 29.87 (0.82) &    0.07 & 0.07 & 28.70 (0.24) & $-$0.85 & 3.09\\
V072 &  6269 & 260.6061960 & 65.7821730 & jdkq03 & jdkq02 & 2019.101 & 2017.918 &    1.183 & 24.62 (0.03) & $-$0.02 & 0.50 & 23.86 (0.02) & $-$0.07 & 4.09\\
V073 &  3339 & 260.6121812 & 65.8528375 & jdkq01 & jdkq07 & 2017.749 & 2018.647 & $-$0.898 & 28.75 (0.32) &    0.21 & 0.48 & 28.12 (0.16) &    0.61 & 3.38\\
V074 &  6605 & 260.6177291 & 65.7775875 & jdkq03 & jdkq02 & 2019.101 & 2017.918 &    1.183 & 25.31 (0.04) & $-$0.06 & 1.01 & 24.53 (0.02) & $-$0.10 & 4.11\\
V075 &   663 & 260.6220866 & 65.8272493 & jdkq01 & jdkq02 & 2017.749 & 2017.918 & $-$0.169 & 27.68 (0.20) & $-$0.86 & 3.24 & 27.83 (0.11) & $-$0.40 & 3.01\\
V076 &  5684 & 260.6225709 & 65.7904307 & jdkq03 & jdkq02 & 2019.101 & 2017.918 &    1.183 & 27.92 (0.31) &    1.02 & 2.45 & 28.44 (0.17) &    0.46 & 2.42\\
V077 &  1294 & 260.6246197 & 65.8807391 & jdkq01 & jdkq07 & 2017.749 & 2018.647 & $-$0.898 & 27.98 (0.20) & $-$0.53 & 2.01 & 28.03 (0.12) & $-$0.40 & 2.81\\
V078 &  3134 & 260.6252980 & 65.8542599 & jdkq01 & jdkq07 & 2017.749 & 2018.647 & $-$0.898 & 28.44 (0.27) & $-$0.21 & 0.59 & 28.13 (0.15) & $-$0.53 & 3.05\\
V079 & 11949 & 260.6257482 & 65.8157473 & jdkq08 & jdkq02 & 2018.645 & 2017.918 &    0.727 & 28.87 (0.42) & $-$0.61 & 1.09 & 28.43 (0.22) & $-$0.98 & 3.89\\
V080 &  5284 & 260.6287600 & 65.7986047 & jdkq03 & jdkq02 & 2019.101 & 2017.918 &    1.183 & 29.42 (0.68) & $-$0.30 & 0.32 & 28.78 (0.31) & $-$1.18 & 3.33\\
V081 & 32136 & 260.6292399 & 65.8979164 & jeex6a & jdkq07 & 2022.830 & 2018.647 &    4.183 & 27.79 (0.21) &    0.76 & 2.67 & 26.81 (0.06) &    0.18 & 2.51\\
V082 & 11609 & 260.6327010 & 65.8281680 & jdkq08 & jdkq02 & 2018.645 & 2017.918 &    0.727 & 28.59 (0.27) & $-$0.36 & 0.99 & 28.56 (0.19) & $-$0.67 & 3.09\\
V083 &  2974 & 260.6335344 & 65.8561320 & jdkq01 & jdkq07 & 2017.749 & 2018.647 & $-$0.898 & 23.61 (0.02) & $-$0.01 & 0.60 & 21.91 (0.01) & $-$0.02 & 3.69\\
V084 &  4711 & 260.6350207 & 65.8290677 & jdkq01 & jdkq08 & 2017.749 & 2018.645 & $-$0.896 & 28.87 (0.36) & $-$0.32 & 0.67 & 28.61 (0.31) & $-$1.23 & 3.46\\
V085 &  2872 & 260.6381441 & 65.8573742 & jdkq01 & jdkq07 & 2017.749 & 2018.647 & $-$0.898 & 28.66 (0.35) & $-$0.57 & 1.21 & 28.37 (0.17) &    0.62 & 3.07\\
V086 &  5236 & 260.6391309 & 65.7993893 & jdkq03 & jdkq02 & 2019.101 & 2017.918 &    1.183 & 26.39 (0.07) &    0.04 & 0.37 & 24.53 (0.02) &    0.09 & 4.18\\
V087 & 31994 & 260.6421239 & 65.9029309 & jeex6a & jdkq07 & 2022.830 & 2018.647 &    4.183 & 23.99 (0.02) & $-$0.01 & 0.26 & 22.26 (0.01) &    0.03 & 4.12\\
V088 &  1426 & 260.6421779 & 65.8780433 & jdkq01 & jdkq07 & 2017.749 & 2018.647 & $-$0.898 & 28.76 (0.29) &    0.15 & 0.37 & 28.51 (0.20) &    0.76 & 3.24\\
V089 &  4463 & 260.6460086 & 65.8377478 & jdkq01 & jdkq08 & 2017.749 & 2018.645 & $-$0.896 & 28.77 (0.28) &    0.01 & 0.02 & 28.51 (0.24) & $-$0.91 & 3.29\\
V090 &  3860 & 260.6476671 & 65.8447077 & jdkq01 & jdkq07 & 2017.749 & 2018.647 & $-$0.898 & 29.00 (0.40) &    0.35 & 0.64 & 28.77 (0.35) & $-$1.20 & 3.01\\
V091 & 12341 & 260.6578804 & 65.8123865 & jdkq08 & jdkq03 & 2018.645 & 2019.101 & $-$0.456 & 28.23 (0.35) & $-$1.02 & 2.15 & 27.85 (0.11) & $-$0.28 & 2.22\\
V092 &  2939 & 260.6676957 & 65.8565418 & jdkq01 & jdkq07 & 2017.749 & 2018.647 & $-$0.898 & 28.10 (0.21) &    0.33 & 1.17 & 28.23 (0.17) &    0.58 & 3.03\\
V093 & 12870 & 260.6677285 & 65.7981739 & jdkq08 & jdkq03 & 2018.645 & 2019.101 & $-$0.456 & 28.88 (0.36) & $-$0.06 & 0.13 & 27.54 (0.09) & $-$0.33 & 3.14\\
V094 &  4423 & 260.6680216 & 65.8388153 & jdkq01 & jdkq08 & 2017.749 & 2018.645 & $-$0.896 & 27.10 (0.10) &    0.06 & 0.43 & 28.63 (0.25) &    0.90 & 3.13\\
V095 &  2469 & 260.6684303 & 65.8631755 & jdkq01 & jdkq07 & 2017.749 & 2018.647 & $-$0.898 & 29.23 (0.43) &    0.09 & 0.15 & 28.24 (0.33) & $-$1.57 & 4.18\\
V096 & 12868 & 260.6747247 & 65.7980539 & jdkq08 & jdkq03 & 2018.645 & 2019.101 & $-$0.456 & 28.62 (0.43) & $-$0.86 & 1.48 & 28.49 (0.24) & $-$0.88 & 3.22\\
V097 & 48064 & 260.6753285 & 65.9216637 & jemn05 & jeex6a & 2022.377 & 2022.830 & $-$0.453 & 28.21 (0.22) &    0.01 & 0.04 & 28.63 (0.24) & $-$0.89 & 3.16\\
V098 & 15034 & 260.6772860 & 65.8538229 & jdkq08 & jdkq07 & 2018.645 & 2018.647 & $-$0.002 & 28.04 (0.19) &    0.09 & 0.36 & 28.10 (0.15) &    0.68 & 3.84\\
V099 &  3932 & 260.6794726 & 65.8408931 & jdkq01 & jdkq07 & 2017.749 & 2018.647 & $-$0.898 & 29.21 (0.49) &    0.58 & 0.88 & 28.13 (0.14) &    0.49 & 3.09\\
V100 & 48124 & 260.6803062 & 65.9204980 & jemn05 & jeex6a & 2022.377 & 2022.830 & $-$0.453 & 28.10 (0.20) &    0.15 & 0.57 & 28.17 (0.16) & $-$0.69 & 3.82\\
V101 & 15545 & 260.6815935 & 65.8449714 & jdkq08 & jdkq07 & 2018.645 & 2018.647 & $-$0.002 & 28.85 (0.33) &    0.22 & 0.50 & 28.48 (0.21) &    0.76 & 3.23\\
V102 & 12867 & 260.6822444 & 65.7979657 & jdkq08 & jdkq03 & 2018.645 & 2019.101 & $-$0.456 & 26.89 (0.10) & $-$0.24 & 1.78 & 28.72 (0.26) &    0.94 & 3.17\\
V103 & 48474 & 260.6830765 & 65.9131667 & jemn05 & jeex6a & 2022.377 & 2022.830 & $-$0.453 & 27.95 (0.29) & $-$0.98 & 2.53 & 28.17 (0.14) & $-$0.42 & 2.51\\
V104 & 12872 & 260.6845500 & 65.7978732 & jdkq08 & jdkq03 & 2018.645 & 2019.101 & $-$0.456 & 27.45 (0.17) & $-$0.70 & 3.11 & 27.60 (0.09) & $-$0.09 & 0.85\\
V105 &  7585 & 260.6923363 & 65.7739049 & jdkq04 & jdkq03 & 2018.221 & 2019.101 & $-$0.880 & 28.17 (0.26) & $-$0.61 & 1.74 & 28.44 (0.21) & $-$0.95 & 3.88\\
V106 & 12912 & 260.6948845 & 65.7964272 & jdkq08 & jdkq03 & 2018.645 & 2019.101 & $-$0.456 & 23.61 (0.02) & $-$0.17 & 6.58 & 23.30 (0.01) &    0.01 & 0.88\\
V107 & 46007 & 260.6995940 & 65.7123076 & jemn02 & jdkq09 & 2021.940 & 2019.106 &    2.834 & 26.96 (0.09) & $-$0.01 & 0.06 & 24.36 (0.02) &    0.08 & 3.83\\
V108 & 44457 & 260.7055562 & 65.7256191 & jemn02 & jdkq09 & 2021.940 & 2019.106 &    2.834 & 27.12 (0.11) &    0.19 & 1.27 & 26.52 (0.05) &    0.19 & 3.21\\
V109 & 44621 & 260.7056172 & 65.7226130 & jemn02 & jdkq09 & 2021.940 & 2019.106 &    2.834 & 28.16 (0.40) & $-$1.14 & 2.11 & 28.63 (0.24) & $-$0.86 & 3.15\\
V110 &  7638 & 260.7078911 & 65.7730268 & jdkq04 & jdkq03 & 2018.221 & 2019.101 & $-$0.880 & 28.83 (0.55) & $-$1.05 & 1.43 & 28.64 (0.26) &    0.95 & 3.22\\
V111 & 45164 & 260.7102095 & 65.7169415 & jemn02 & jdkq09 & 2021.940 & 2019.106 &    2.834 & 28.80 (0.40) &    0.32 & 0.59 & 28.38 (0.27) &    1.36 & 4.32\\
V112 & 13412 & 260.7123985 & 65.7927245 & jdkq08 & jdkq04 & 2018.645 & 2018.221 &    0.424 & 27.84 (0.16) & $-$0.20 & 0.93 & 26.91 (0.06) & $-$0.25 & 3.47\\
V113 & 45918 & 260.7167982 & 65.7087560 & jemn02 & jdkq09 & 2021.940 & 2019.106 &    2.834 & 27.88 (0.21) & $-$0.16 & 0.60 & 26.49 (0.05) &    0.19 & 3.11\\
V114 &  7098 & 260.7175478 & 65.7848453 & jdkq04 & jdkq03 & 2018.221 & 2019.101 & $-$0.880 & 24.53 (0.03) & $-$0.01 & 0.41 & 22.81 (0.01) &    0.03 & 3.23\\
V115 & 45686 & 260.7192384 & 65.7123368 & jemn02 & jdkq09 & 2021.940 & 2019.106 &    2.834 & 27.88 (0.25) & $-$0.73 & 2.20 & 27.82 (0.12) & $-$0.37 & 2.78\\
V116 &  8197 & 260.7200109 & 65.7631919 & jdkq04 & jdkq03 & 2018.221 & 2019.101 & $-$0.880 & 24.17 (0.02) & $-$0.02 & 0.57 & 23.40 (0.01) &    0.06 & 4.30\\
V117 & 44851 & 260.7250941 & 65.7203820 & jemn02 & jdkq09 & 2021.940 & 2019.106 &    2.834 & 28.83 (0.40) & $-$0.05 & 0.10 & 28.32 (0.19) & $-$0.67 & 3.07\\
V118 & 11469 & 260.7287189 & 65.8579825 & jdkq07 & jdkq06 & 2018.647 & 2018.511 &    0.136 & 28.35 (0.34) &    0.89 & 1.94 & 28.38 (0.19) &    0.73 & 3.36\\
V119 & 10689 & 260.7358381 & 65.8754246 & jdkq07 & jdkq06 & 2018.647 & 2018.511 &    0.136 & 24.60 (0.03) &    0.04 & 0.91 & 23.04 (0.01) &    0.06 & 5.33\\
V120 & 44649 & 260.7372179 & 65.7239799 & jemn02 & jdkq09 & 2021.940 & 2019.106 &    2.834 & 28.27 (0.24) & $-$0.09 & 0.28 & 28.01 (0.15) & $-$0.53 & 3.07\\
\hline\\[-8pt]
\end{tabular}
}
\end{table*}

\addtocounter{table}{-1}
\begin{table*}[t!]
\movetableright=-0.75in
\caption{(\textit{continued})}
\setlength{\tabcolsep}{3pt}
{\scriptsize
\begin{tabular}{rrllllllrlrrlrr}
\hline\hline\\[-8pt]
\multicolumn{9}{c}{$\quad$} & \multicolumn{3}{c}{\hrulefill\ F435W \hrulefill}
  & \multicolumn{3}{c}{\hrulefill\ F606W \hrulefill}\\
\multicolumn{1}{c}{ID} & \multicolumn{1}{c}{CatID} &
   \multicolumn{1}{c}{RA} & \multicolumn{1}{c}{Dec} &
   \multicolumn{1}{c}{Visit1} & \multicolumn{1}{c}{Visit2} &
   \multicolumn{1}{c}{$t_1$} & \multicolumn{1}{c}{$t_2$} &
   \multicolumn{1}{c}{$\Delta t$} & \multicolumn{1}{c}{$m_{0.24}$} &
   \multicolumn{1}{c}{$\!\!\!\!\Delta m_{0.24}$} & \multicolumn{1}{c}{$\Dmagerr$} &
   \multicolumn{1}{c}{$m_{0.24}$} & \multicolumn{1}{c}{$\!\!\!\!\Delta m_{0.24}$} &
   \multicolumn{1}{c}{$\Dmagerr$}\\
  &   & \multicolumn{1}{c}{[deg]} & \multicolumn{1}{c}{[deg]} &   &   &
  \multicolumn{1}{c}{[yr]} & \multicolumn{1}{c}{[yr]} &
  \multicolumn{1}{c}{[yr]} & \multicolumn{1}{c}{[mag]} &
  \multicolumn{1}{c}{[mag]} &  &
  \multicolumn{1}{c}{[mag]} & \multicolumn{1}{c}{[mag]} &  \\
\multicolumn{1}{c}{(1)} & \multicolumn{1}{c}{(2)} & \multicolumn{1}{c}{(3)} &
  \multicolumn{1}{c}{(4)} & \multicolumn{1}{c}{(5)} & \multicolumn{1}{c}{(6)} &
  \multicolumn{1}{c}{(7)} & \multicolumn{1}{c}{(8)} & \multicolumn{1}{c}{(9)} &
  \multicolumn{1}{c}{(10)} & \multicolumn{1}{c}{(11)} &
  \multicolumn{1}{c}{(12)} & \multicolumn{1}{c}{(13)} &
  \multicolumn{1}{c}{(14)} & \multicolumn{1}{c}{(15)}\\[4pt]
\hline
V121 & 45250 & 260.7375391 & 65.7161188 & jemn02 & jdkq09 & 2021.940 & 2019.106 &    2.834 & 28.80 (0.50) & $-$0.64 & 0.96 & 28.77 (0.31) &    1.09 & 3.07\\
V122 & 44894 & 260.7383881 & 65.7200561 & jemn02 & jdkq09 & 2021.940 & 2019.106 &    2.834 & 28.63 (0.33) & $-$0.13 & 0.29 & 28.11 (0.16) &    0.56 & 3.16\\
V123 & 44048 & 260.7396362 & 65.7294908 & jemn02 & jdkq09 & 2021.940 & 2019.106 &    2.834 & 27.74 (0.18) & $-$0.21 & 0.86 & 28.57 (0.26) & $-$1.09 & 3.60\\
V124 & 43264 & 260.7454366 & 65.7380214 & jemn02 & jdkq09 & 2021.940 & 2019.106 &    2.834 & 19.25 (0.01) &    0.07 &13.92 & 19.50 (0.02) & $-$0.33 &13.27\\
V125 & 16264 & 260.7488596 & 65.7509596 & jdkq09 & jdkq04 & 2019.106 & 2018.221 &    0.885 & 28.41 (0.28) &    0.44 & 1.16 & 28.54 (0.21) & $-$0.72 & 3.03\\
V126 & 44831 & 260.7505188 & 65.7205403 & jemn02 & jdkq09 & 2021.940 & 2019.106 &    2.834 & 27.61 (0.29) & $-$1.18 & 3.01 & 27.89 (0.12) & $-$0.11 & 0.77\\
V127 & 45380 & 260.7514847 & 65.7150048 & jemn02 & jdkq09 & 2021.940 & 2019.106 &    2.834 & 27.73 (0.23) & $-$0.75 & 2.37 & 27.51 (0.10) & $-$0.24 & 2.15\\
V128 & 43642 & 260.7545349 & 65.7350399 & jemn02 & jdkq09 & 2021.940 & 2019.106 &    2.834 & 28.56 (0.32) &    0.29 & 0.67 & 28.30 (0.16) &    0.55 & 3.04\\
V129 & 15975 & 260.7564058 & 65.7560911 & jdkq09 & jdkq04 & 2019.106 & 2018.221 &    0.885 & 29.24 (0.46) &    0.07 & 0.11 & 28.32 (0.20) &    0.72 & 3.16\\
V130 & 42985 & 260.7569328 & 65.7430011 & jemn02 & jdkq09 & 2021.940 & 2019.106 &    2.834 & 27.87 (0.21) &    0.58 & 2.01 & 27.85 (0.11) &    0.34 & 2.67\\
V131 & 43199 & 260.7654001 & 65.7398784 & jemn02 & jdkq09 & 2021.940 & 2019.106 &    2.834 & 29.29 (0.59) &    0.21 & 0.26 & 28.63 (0.28) &    1.02 & 3.22\\
V132 & 44494 & 260.7709607 & 65.7242314 & jemn02 & jdkq09 & 2021.940 & 2019.106 &    2.834 & 28.30 (0.30) & $-$0.44 & 1.10 & 28.32 (0.18) & $-$0.68 & 3.33\\
V133 & 43060 & 260.7745589 & 65.7419473 & jemn02 & jdkq09 & 2021.940 & 2019.106 &    2.834 & 28.12 (0.28) &    0.72 & 1.90 & 28.35 (0.17) &    0.62 & 3.13\\
V134 & 45001 & 260.7764730 & 65.7187770 & jemn02 & jdkq09 & 2021.940 & 2019.106 &    2.834 & 26.10 (0.06) & $-$0.05 & 0.55 & 25.65 (0.03) &    0.12 & 3.04\\
V135 &  9852 & 260.7809939 & 65.8383387 & jdkq06 & jdkq05 & 2018.511 & 2018.363 &    0.148 & 27.94 (0.21) &    0.46 & 1.63 & 27.63 (0.10) &    0.39 & 3.45\\
V136 &  9811 & 260.7820403 & 65.8390785 & jdkq06 & jdkq05 & 2018.511 & 2018.363 &    0.148 & 27.51 (0.14) &    0.17 & 0.93 & 28.53 (0.19) &    0.70 & 3.18\\
V137 & 45900 & 260.7829845 & 65.7089929 & jemn02 & jdkq09 & 2021.940 & 2019.106 &    2.834 & 28.00 (0.26) &    0.64 & 1.83 & 28.06 (0.14) &    0.49 & 3.03\\
V138 & 16432 & 260.7838837 & 65.7488263 & jdkq09 & jdkq04 & 2019.106 & 2018.221 &    0.885 & 28.42 (0.38) & $-$0.86 & 1.70 & 28.48 (0.21) &    0.74 & 3.12\\
V139 & 46426 & 260.7901866 & 65.7007683 & jemn02 & jdkq09 & 2021.940 & 2019.106 &    2.834 & 28.16 (0.22) &    0.00 & 0.01 & 27.59 (0.10) & $-$0.43 & 3.74\\
V140 &  8862 & 260.7966543 & 65.8514172 & jdkq06 & jdkq05 & 2018.511 & 2018.363 &    0.148 & 25.51 (0.05) & $-$0.08 & 1.30 & 25.35 (0.03) & $-$0.11 & 3.06\\
V141 & 31209 & 260.7976797 & 65.7264133 & jeex08 & jdkq09 & 2021.344 & 2019.106 &    2.238 & 29.20 (0.55) & $-$0.57 & 0.76 & 28.40 (0.22) &    0.88 & 3.53\\
V142 &  9654 & 260.8007573 & 65.8408780 & jdkq06 & jdkq05 & 2018.511 & 2018.363 &    0.148 & 27.83 (0.20) &    0.48 & 1.82 & 28.65 (0.26) &    0.98 & 3.34\\
V143 & 28674 & 260.8063693 & 65.8799829 & jeex05 & jdkq06 & 2021.347 & 2018.511 &    2.836 & 28.27 (0.20) &    0.44 & 1.59 & 27.76 (0.09) &    0.36 & 3.44\\
V144 & 30070 & 260.8105090 & 65.7371529 & jeex08 & jdkq09 & 2021.344 & 2019.106 &    2.238 & 27.49 (0.15) & $-$0.18 & 0.88 & 28.05 (0.15) &    0.54 & 3.16\\
V145 & 30045 & 260.8122328 & 65.7373452 & jeex08 & jdkq09 & 2021.344 & 2019.106 &    2.238 & 28.47 (0.47) & $-$1.10 & 1.73 & 28.71 (0.35) &    1.23 & 3.04\\
V146 &  8477 & 260.8163753 & 65.7965563 & jdkq05 & jdkq04 & 2018.363 & 2018.221 &    0.142 & 26.81 (0.09) & $-$0.02 & 0.18 & 26.36 (0.05) & $-$0.21 & 3.53\\
V147 & 30059 & 260.8219253 & 65.7371863 & jeex08 & jdkq09 & 2021.344 & 2019.106 &    2.238 & 28.63 (0.45) &    0.76 & 1.25 & 28.27 (0.22) &    1.14 & 4.52\\
V148 & 28638 & 260.8274735 & 65.8806602 & jeex05 & jdkq06 & 2021.347 & 2018.511 &    2.836 & 28.37 (0.23) & $-$0.15 & 0.48 & 28.00 (0.14) &    0.49 & 3.09\\
V149 & 22712 & 260.8347912 & 65.7296116 & jeex03 & jdkq09 & 2021.035 & 2019.106 &    1.929 & 28.36 (0.27) &    0.06 & 0.16 & 28.32 (0.17) &    0.62 & 3.10\\
V150 & 27404 & 260.8352185 & 65.7275413 & jeex03 & jeex08 & 2021.035 & 2021.344 & $-$0.309 & 29.01 (0.70) & $-$0.69 & 0.73 & 27.38 (0.24) & $-$2.08 & 7.53\\
V151 & 47730 & 260.8358472 & 65.8893731 & jemn04 & jeex05 & 2022.227 & 2021.347 &    0.880 & 28.32 (0.30) &    0.86 & 2.10 & 28.52 (0.22) &    1.10 & 4.33\\
V152 &  9712 & 260.8429092 & 65.8400901 & jdkq06 & jdkq05 & 2018.511 & 2018.363 &    0.148 & 28.72 (0.34) & $-$0.39 & 0.85 & 28.68 (0.25) &    0.93 & 3.24\\
V153 & 23869 & 260.8442446 & 65.7597401 & jeex03 & jeex08 & 2021.035 & 2021.344 & $-$0.309 & 28.85 (0.39) &    0.30 & 0.57 & 28.69 (0.29) & $-$1.10 & 3.27\\
V154 & 47372 & 260.8451666 & 65.8991518 & jemn04 & jeex05 & 2022.227 & 2021.347 &    0.880 & 28.80 (0.68) & $-$0.83 & 0.90 & 28.77 (0.33) & $-$1.21 & 3.17\\
V155 & 47667 & 260.8468371 & 65.8914197 & jemn04 & jeex05 & 2022.227 & 2021.347 &    0.880 & 27.66 (0.25) &    1.03 & 3.01 & 27.96 (0.12) &    0.28 & 2.04\\
V156 & 22013 & 260.8477623 & 65.7413624 & jeex03 & jdkq09 & 2021.035 & 2019.106 &    1.929 & 28.30 (0.47) &    1.24 & 1.96 & 28.61 (0.25) &    0.92 & 3.23\\
V157 & 25243 & 260.8539353 & 65.7457044 & jeex03 & jeex08 & 2021.035 & 2021.344 & $-$0.309 & 27.75 (0.32) &    1.33 & 3.12 & 28.44 (0.17) & $-$0.06 & 0.33\\
V158 & 25240 & 260.8541929 & 65.7458702 & jeex03 & jeex08 & 2021.035 & 2021.344 & $-$0.309 & 27.75 (0.32) &    1.33 & 3.12 & 27.48 (0.09) & $-$0.06 & 0.59\\
V159 & 30067 & 260.8578555 & 65.7371669 & jeex08 & jdkq09 & 2021.344 & 2019.106 &    2.238 & 28.25 (0.26) & $-$0.48 & 1.37 & 28.07 (0.13) &    0.48 & 3.16\\
V160 & 27723 & 260.8600711 & 65.7221800 & jeex03 & jeex08 & 2021.035 & 2021.344 & $-$0.309 & 24.03 (0.02) & $-$0.03 & 1.06 & 21.98 (0.01) & $-$0.03 & 4.86\\
V161 & 22223 & 260.8619972 & 65.7380793 & jeex03 & jdkq09 & 2021.035 & 2019.106 &    1.929 & 28.21 (0.27) &    0.64 & 1.73 & 28.34 (0.16) & $-$0.72 & 3.79\\
V162 &  9531 & 260.8638246 & 65.8420120 & jdkq06 & jdkq05 & 2018.511 & 2018.363 &    0.148 & 29.46 (0.63) & $-$0.51 & 0.59 & 28.70 (0.27) &    1.01 & 3.25\\
V163 & 24676 & 260.8653220 & 65.7504766 & jeex03 & jeex08 & 2021.035 & 2021.344 & $-$0.309 & 28.42 (0.51) & $-$0.97 & 1.41 & 28.40 (0.20) &    0.71 & 3.09\\
V164 & 25310 & 260.8719766 & 65.7454975 & jeex03 & jeex08 & 2021.035 & 2021.344 & $-$0.309 & 28.54 (0.33) &    0.22 & 0.50 & 28.03 (0.16) & $-$0.67 & 3.68\\
V165 & 24248 & 260.8746031 & 65.7549045 & jeex03 & jeex08 & 2021.035 & 2021.344 & $-$0.309 & 27.52 (0.20) & $-$0.67 & 2.49 & 27.00 (0.07) & $-$0.22 & 2.64\\
V166 & 24737 & 260.8778166 & 65.7497988 & jeex03 & jeex08 & 2021.035 & 2021.344 & $-$0.309 & 26.33 (0.09) & $-$0.40 & 3.27 & 26.27 (0.05) & $-$0.07 & 1.31\\
V167 & 47239 & 260.8785658 & 65.9050944 & jemn04 & jeex05 & 2022.227 & 2021.347 &    0.880 & 27.33 (0.18) &    0.65 & 2.73 & 27.16 (0.08) &    0.22 & 2.40\\
V168 & 24124 & 260.8799575 & 65.7558947 & jeex03 & jeex08 & 2021.035 & 2021.344 & $-$0.309 & 27.77 (0.58) & $-$1.77 & 2.26 & 27.29 (0.08) & $-$0.24 & 2.55\\
V169 & 27313 & 260.8844831 & 65.7282510 & jeex03 & jeex08 & 2021.035 & 2021.344 & $-$0.309 & 27.17 (0.15) &    0.71 & 3.44 & 27.25 (0.08) &    0.11 & 1.29\\
V170 & 26382 & 260.8880858 & 65.7366434 & jeex03 & jeex08 & 2021.035 & 2021.344 & $-$0.309 & 28.41 (0.30) &    0.34 & 0.86 & 28.54 (0.23) &    0.93 & 3.44\\
V171 & 25315 & 260.8883667 & 65.7454453 & jeex03 & jeex08 & 2021.035 & 2021.344 & $-$0.309 & 27.37 (0.17) &    0.75 & 3.26 & 27.35 (0.08) &    0.08 & 0.92\\
V172 & 24771 & 260.8931970 & 65.7496619 & jeex03 & jeex08 & 2021.035 & 2021.344 & $-$0.309 & 29.00 (0.64) & $-$0.69 & 0.80 & 28.65 (0.27) & $-$1.03 & 3.32\\
V173 & 36203 & 260.8957973 & 65.8157113 & jeexa4 & jdkq05 & 2021.204 & 2018.363 &    2.841 & 28.61 (0.28) &    0.34 & 0.87 & 28.36 (0.16) &    0.59 & 3.32\\
V174 & 36152 & 260.8966926 & 65.8174593 & jeexa4 & jdkq05 & 2021.204 & 2018.363 &    2.841 & 26.22 (0.06) & $-$0.01 & 0.17 & 23.86 (0.01) &    0.21 &12.76\\
V175 & 36188 & 260.8977353 & 65.8156259 & jeexa4 & jdkq05 & 2021.204 & 2018.363 &    2.841 & 25.16 (0.04) & $-$0.02 & 0.37 & 24.81 (0.02) &    0.08 & 3.17\\
V176 & 36219 & 260.8981358 & 65.8157672 & jeexa4 & jdkq05 & 2021.204 & 2018.363 &    2.841 & 25.49 (0.05) & $-$0.04 & 0.65 & 25.26 (0.03) &    0.11 & 3.22\\
V177 & 36471 & 260.9048765 & 65.8670380 & jeexa4 & jemn04 & 2021.204 & 2022.227 & $-$1.023 & 27.99 (0.25) &    0.31 & 0.92 & 27.87 (0.12) &    0.44 & 3.23\\
V178 & 36634 & 260.9055490 & 65.8651822 & jeexa4 & jemn04 & 2021.204 & 2022.227 & $-$1.023 & 28.71 (0.46) & $-$0.79 & 1.29 & 28.20 (0.17) &    0.62 & 3.25\\
V179 & 24684 & 260.9077303 & 65.7505134 & jeex03 & jeex08 & 2021.035 & 2021.344 & $-$0.309 & 27.63 (0.20) &    0.76 & 2.75 & 27.82 (0.11) &    0.27 & 2.12\\
V180 & 26527 & 260.9094296 & 65.7355841 & jeex03 & jeex08 & 2021.035 & 2021.344 & $-$0.309 & 26.13 (0.07) &    0.19 & 2.07 & 25.04 (0.03) &    0.08 & 2.58\\
\hline\\[-8pt]
\end{tabular}
}
\end{table*}

\addtocounter{table}{-1}
\begin{table*}[t!]
\movetableright=-0.75in
\caption{(\textit{continued})}
\setlength{\tabcolsep}{3pt}
{\scriptsize
\begin{tabular}{rrllllllrlrrlrr}
\hline\hline\\[-8pt]
\multicolumn{9}{c}{$\quad$} & \multicolumn{3}{c}{\hrulefill\ F435W \hrulefill}
  & \multicolumn{3}{c}{\hrulefill\ F606W \hrulefill}\\
\multicolumn{1}{c}{ID} & \multicolumn{1}{c}{CatID} &
   \multicolumn{1}{c}{RA} & \multicolumn{1}{c}{Dec} &
   \multicolumn{1}{c}{Visit1} & \multicolumn{1}{c}{Visit2} &
   \multicolumn{1}{c}{$t_1$} & \multicolumn{1}{c}{$t_2$} &
   \multicolumn{1}{c}{$\Delta t$} & \multicolumn{1}{c}{$m_{0.24}$} &
   \multicolumn{1}{c}{$\!\!\!\!\Delta m_{0.24}$} & \multicolumn{1}{c}{$\Dmagerr$} &
   \multicolumn{1}{c}{$m_{0.24}$} & \multicolumn{1}{c}{$\!\!\!\!\Delta m_{0.24}$} &
   \multicolumn{1}{c}{$\Dmagerr$}\\
  &   & \multicolumn{1}{c}{[deg]} & \multicolumn{1}{c}{[deg]} &   &   &
  \multicolumn{1}{c}{[yr]} & \multicolumn{1}{c}{[yr]} &
  \multicolumn{1}{c}{[yr]} & \multicolumn{1}{c}{[mag]} &
  \multicolumn{1}{c}{[mag]} &  &
  \multicolumn{1}{c}{[mag]} & \multicolumn{1}{c}{[mag]} &  \\
\multicolumn{1}{c}{(1)} & \multicolumn{1}{c}{(2)} & \multicolumn{1}{c}{(3)} &
  \multicolumn{1}{c}{(4)} & \multicolumn{1}{c}{(5)} & \multicolumn{1}{c}{(6)} &
  \multicolumn{1}{c}{(7)} & \multicolumn{1}{c}{(8)} & \multicolumn{1}{c}{(9)} &
  \multicolumn{1}{c}{(10)} & \multicolumn{1}{c}{(11)} &
  \multicolumn{1}{c}{(12)} & \multicolumn{1}{c}{(13)} &
  \multicolumn{1}{c}{(14)} & \multicolumn{1}{c}{(15)}\\[4pt]
\hline
V181 & 37131 & 260.9097335 & 65.8596199 & jeexa4 & jemn04 & 2021.204 & 2022.227 & $-$1.023 & 28.03 (0.30) & $-$1.04 & 2.53 & 28.20 (0.15) & $-$0.48 & 2.72\\
V182 & 36713 & 260.9145485 & 65.8642884 & jeexa4 & jemn04 & 2021.204 & 2022.227 & $-$1.023 & 29.00 (0.85) &    0.78 & 0.68 & 27.94 (0.14) &    0.52 & 3.23\\
V183 & 35903 & 260.9149584 & 65.8265971 & jeexa4 & jdkq05 & 2021.204 & 2018.363 &    2.841 & 25.20 (0.04) & $-$0.04 & 0.62 & 23.07 (0.01) &    0.06 & 4.82\\
V184 & 47170 & 260.9280848 & 65.8129725 & jemn03 & jeexa4 & 2022.085 & 2021.204 &    0.881 & 27.47 (0.16) &    0.54 & 2.49 & 27.08 (0.08) &    0.21 & 2.45\\
V185 & 37073 & 260.9288557 & 65.8578725 & jeexa4 & jemn04 & 2021.204 & 2022.227 & $-$1.023 & 27.19 (0.15) & $-$0.63 & 3.12 & 27.47 (0.09) & $-$0.05 & 0.48\\
V186 & 36333 & 260.9330472 & 65.8683621 & jeexa4 & jemn04 & 2021.204 & 2022.227 & $-$1.023 & 24.58 (0.03) & $-$0.09 & 2.41 & 24.65 (0.02) & $-$0.06 & 2.32\\
V187 & 37009 & 260.9354158 & 65.8606655 & jeexa4 & jemn04 & 2021.204 & 2022.227 & $-$1.023 & 27.68 (0.30) & $-$1.30 & 3.17 & 28.08 (0.15) & $-$0.43 & 2.47\\
V188 & 36313 & 260.9427395 & 65.8689954 & jeexa4 & jemn04 & 2021.204 & 2022.227 & $-$1.023 & 27.81 (0.24) & $-$0.83 & 2.55 & 27.66 (0.10) & $-$0.26 & 2.20\\
V189 & 36623 & 260.9429745 & 65.8653180 & jeexa4 & jemn04 & 2021.204 & 2022.227 & $-$1.023 & 27.51 (0.24) & $-$1.14 & 3.57 & 27.59 (0.10) &    0.18 & 1.62\\
V190 & 36449 & 260.9444992 & 65.8672901 & jeexa4 & jemn04 & 2021.204 & 2022.227 & $-$1.023 & 28.37 (0.40) & $-$1.01 & 1.86 & 28.62 (0.34) &    1.22 & 3.11\\[2pt]
\hline\\[-8pt]
\end{tabular}

\begin{minipage}{\txw}{\small
\textbf{Note. ---} Columns (1) through (9) list the variable source ID, an
internal \sextractor\ catalog ID, the celestial coordinates (determined by
\sextractor\ for the position corresponding to $m_{0.24}$) of the variable
source in decimal degrees, the root names of the overlapping visits in which
the variable source was detected (Visit 1 and Visit 2), the corresponding UT
dates of observation, and the time interval $\Delta t$ between Visits 1 and 2.
Table columns (10) through (12) list for the F435W filter the brightest
$m_{0.24}$ of the two visits (with the magnitude uncertainty given in
parentheses), the change in brightness ($\Delta m_{0.24}$, using the same
definition as in Figure \ref{fig:variable_candidates}) between the two
visits, and the significance of the variability ($\sigma_{\rm var}$). Table
columns (13) through (15) list the same but for the F606W filter.
}
\end{minipage}
}
\end{table*}


\clearpage

\appendix

\section{\teal{Assessment of the Potential Impact of Cosmic Rays on Detected Variability}} \label{sec:cosmic_rays}

\tealnb{For variability searches with the two HST cameras --- WFC3 and ACS --- that have
been in the telescope for over 15-20 years it is prudent to consider the number of
cosmic rays (CRs) that will have hit our detectors and may have survived the
CR-clipping that happens during the drizzling process. } \tealnb{First, we consider how constant the CR-flux may be on HST's detectors. A summary of the best on-orbit data on the CR-flux 
is given by, \eg\ \citet[][their section 7.5]{Ryon2019} for ACS\footnote{see also
\url{https://www.stsci.edu/files/live/sites/www/files/home/hst/documentation/\_documents/acs/acs_ihb_cycle28.pdf}},
by \citet[][their section 5.4.10]{Dressel2016} for WFC3\footnote{see also 
\url{https://www.stsci.edu/files/live/sites/www/files/home/hst/documentation/\_documents/wfc3/wfc3_ihb_cycle24.pdf}}, 
and by \citet[][their Section 4.9]{McMaster2010} for WFPC2\footnote{see also
\url{https://www.stsci.edu/files/live/sites/www/files/home/hst/documentation/_documents/wfpc2/wfpc2_ihb_cycle17.pdf}
and \url{https://www.stsci.edu/files/live/sites/www/files/home/hst/documentation/_documents/wfpc2/wfpc2_dhb_v5.pdf}}.
Section 4.9 of \citet[][]{McMaster2010} suggest that the WFPC2 CR energy distribution follows a Weibull distribution,
which resembles a Poisson curve with a distinct low-energy cutoff. }

A good recent summary of over 1.2 billion CRs analyzed during the 30 year
operational lifetime of these four cameras inside HST is given by, e.g., 
\citet{Miles2021}. Their Table 6 shows that both ACS, WFPC2, and WFC3 have
a remarkably similar density of CR-hits of $\sim$1.1$\pm$0.1 hits/sec/cm$^2$ of
detector area. \cite{Miles2021} provide a better physically motivated Landau
distribution of the CR-energies for all HST cameras, which resembles the earlier
Weibull distribution of \cite{McMaster2010}. These CR energy distributions are
thus representative of the typical CR-distribution and frequency in typical HST
images, as long as one stays away from transitions through the South Atlantic
Anomaly (SAA), where the CR-flux can be temporarily much higher. Our NEP TDF images always avoid
the SAA transitions, and therefore, we will assume that the CR-flux is as given in these
respective Instrument Handbooks and \citet{Miles2021}.

Following the Instrument Handbooks, at
minimum four independent exposures of $\sim$1200 sec are needed to minimize
multiple CR impacts on the same object, and we use a much more conservative N=8
exposures per pointing in each filter. \citet{Windhorst1994a} investigate from
12-exposure pointings how much the CR-rejection improves if fewer than N=12
exposures are stacked, and find that for N$=$8 the sky-mean and sky-sigma
converge to the best achievable values (from the N=12 maximum exposure case) if
one clips at the level of 3--3.5$\sigma$, where $\sigma$ is the local sky
rms noise-level. So starting with CR-clipping on 8 exposures for every pointing
should be robust.

\tealnb{Nonetheless, we cannot rule out that the CR flux may occasionally be considerably higher than average, and so must account for the possibility that even while using N=8 exposures for each pointing in each filter, some CR's may have survived the clipping procedure upon drizzling and mimic variable objects at the 3.0$\sigma$ level. For this reason, we also performed additional tests, as described below.}

\tealnb{Cosmic rays are flagged in the Data Quality extension of ACS images with values of 4096 and 8192. For every variable object, we create a 8$\times$8 pixel box at the center of the measurement, corresponding closely to the 8-pixel diameter aperture used to identify variability (see Section \ref{sec:variable_methods}). We count the number of pixels within this box that may be affected by residual cosmic ray flux. We define cosmic-ray-affected pixels (CR-affected) to be those where a cosmic ray appears in that exact location in at least half of the stacked exposures (4 or more exposures). The location is determined by the WCS in the image header of each individual exposure. For every object, we do this for both epochs, and compare the number of cosmic rays in the two epochs. A source is determined to be CR-affected if it has at least one CR-affected pixel within the 8$\times$8 pixel box in either epoch. A CR-affected source, as defined here, \emph{may} be affected by comic rays, although it is likely not due to the stacking of 8 individual exposures with \astrodrizzle{}.}

\begin{figure}
    \centering
    \includegraphics[width=1\linewidth]{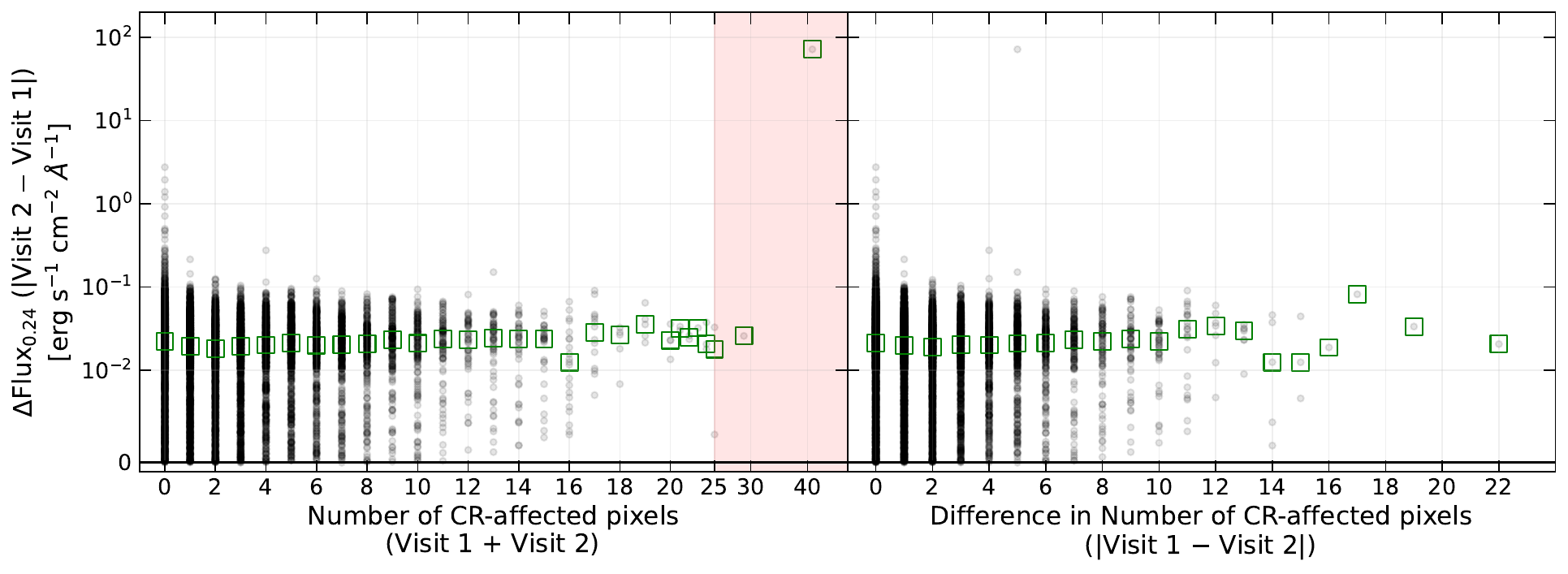}
    \caption{\tealnb{We analyze the potential effect of cosmic rays on the measured flux difference of variable objects in an 8$\times$8 pixel (0\farcs24$\times$0\farcs24) area centered on each object. A pixel is considered at risk of cosmic ray contamination if it encounters a cosmic ray in four or more contributing exposures. Plotted is a random selection of 10,000 galaxies (represented by black points), with median values indicated by green boxes. \textbf{Left:} The absolute value of the difference in measured F606W flux versus the total number of pixels (in Visit 1 $+$ Visit 2) possibly impacted by cosmic rays. The red shaded region shows objects that are removed for containing too many cosmic rays. The y-axis is logarithmically scaled except between -0.1 and 0.1, which is scaled linearly. The x-axis is linearly scaled up to 20, where it is then logarithmically scaled to 50. \textbf{Right:} The absolute value of the difference in measured F606W flux versus the difference in the number of pixels potentially affected by cosmic rays.}}
    \label{fig:cosmic_ray_numbers}
\end{figure}

\tealnb{In Figure \ref{fig:cosmic_ray_numbers}, we plot the measured change in flux of the objects (between two epochs) versus the combined number of CR-affected pixels. We also plot the difference in the number of pixels affected by cosmic rays between the two epochs, because if one epoch has significantly more CR-affected pixels, this could cause artificial variability. The exercise demonstrates that cosmic rays do not affect our variability detections because the measured flux difference does not depend on the number of CR-affected pixels or the difference in the number of CR-affected pixels. Moreover, \tsim31\% of the 10,000 galaxies sampled in Figure \ref{fig:cosmic_ray_numbers} have zero CR-affected pixels in both epochs.}

\tealnb{One object with very high measured differences in flux also have a large number of pixels with cosmic rays. We opt to reject and remove two objects with more than 25 pixels (\tsim25\% of the pixels in the aperture) affected by cosmic rays. There were no variable candidates that met this criteria.}

\begin{figure}
    \centering
    \includegraphics[width=0.8\linewidth]{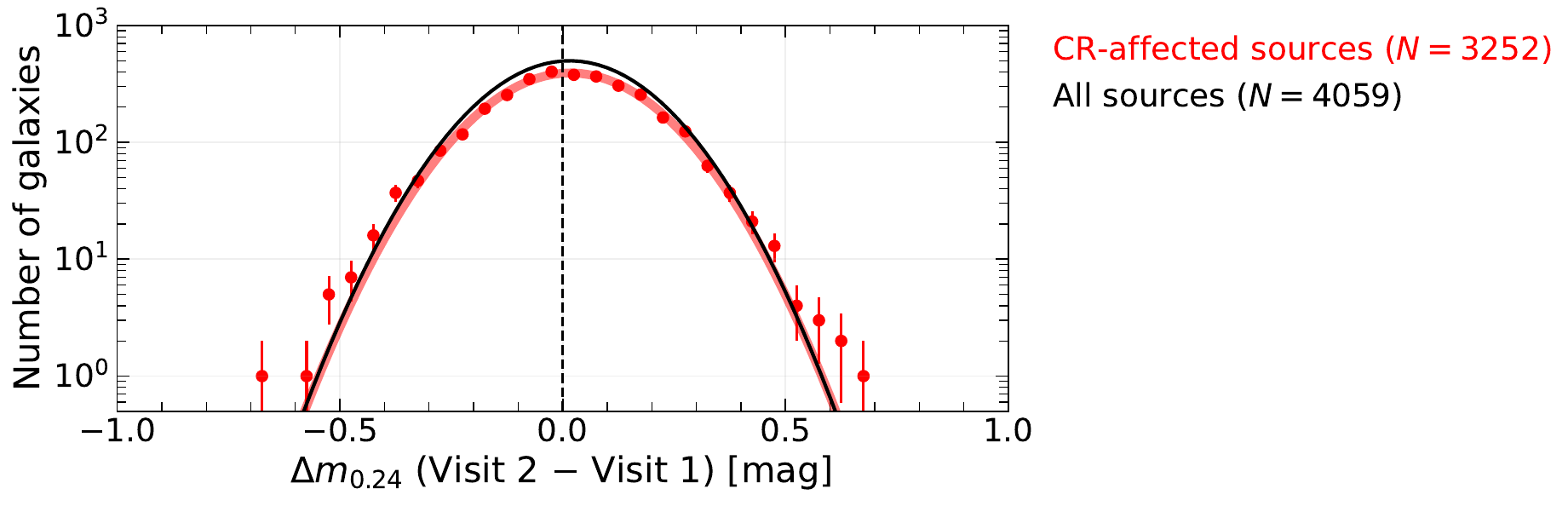}
    \caption{\tealnb{Histogram of the distribution of magnitude differences measured between two epochs ($\Delta\mTF$) for 4,059 galaxies between $28.0 < \mAB < 28.5$. We compare the sample of CR-affected galaxies (red) to the full sample (black). The points show the fraction of CR-affected galaxies ($N$) for each $\Delta\mTF$ bin, where the error bar is $\sqrt{N}$. The solid lines are Gaussian fits to the data. The black solid line represents the Gaussian fit to the full sample. Our empirical uncertainties estimated in Figure \ref{fig:variable_candidates} already account for cosmic ray effects, as shown by the black line closely following the red line.}}
    \label{fig:cosmic_ray_hist}
\end{figure}

\tealnb{We also confirm the distribution of $\Delta \mTF$ is Gaussian for sources potentially affected by cosmic rays, shown in Figure \ref{fig:cosmic_ray_hist}. More importantly, since a majority ($\tsim80$\%) of sources in our dataset are potentially affected by cosmic rays, our empirical uncertainties in Section \ref{sec:source_selection} already account for cosmic ray affects. This is specifically represented by the fact that the Gaussian fit to the CR-affected sources (red line) is within 0.01 mag to the Gaussian fit for the full data (black line). The latter (black line) was used to calibrate our empirical uncertainties in Section \ref{sec:source_selection}. We note that, by eye, there appears to be a slight divergence from the Gaussian fit at $\Dmagerr < -0.5$ and $\Dmagerr > 0.5$. However, only the outer-most points seem to contribute to this trend, corresponding to less than $5$ sources (or less than $0.2$\% of sources).}

\tealnb{In conclusion, this evidence suggests that cosmic rays rarely appear in the final drizzled images. The spurious transients (in Section \ref{sec:transient_methods}) were likely outliers, and any other spurious cosmic rays to this degree would have been identified during the transient search. In addition, if cosmic rays affect the brightness of sources regularly, our empirical estimation of uncertainties in Section \ref{sec:source_selection} already accounts for this. Nonetheless, we do not completely eliminate the possibility that some (albeit very few) of our variable candidates may be affected by cosmic rays that cause artificial variability. In other words, although we assume a majority of our sources are not affected by cosmic rays, follow-up observations remain necessary to confirm which sources are truly variable, as well as reveal other sources that were missed in this work.}

\end{document}